\documentclass[a4paper,useAMS,usenatbib]{mn2e}

\usepackage{times}
\usepackage{deluxetable}
\usepackage{longtable}

\usepackage{url}
\usepackage{epsfig}
\usepackage{amsmath}
\usepackage{natbib}
\usepackage{graphicx}
\usepackage{aas_macros}
\date{accepted to MNRAS}

\begin{document}

\title{Star Formation \& Dust Heating in the FIR Bright Sources of M83}

\author[Foyle et al.]{\parbox{\textwidth}{K. Foyle,$^{1}$$\dagger$ G. Natale,$^{2}$$\dagger$ C. D. Wilson,$^{1}$ C. C. Popescu,$^{2,3}$ 
 M.~Baes,$^{4}$ G. J.~Bendo,$^{5}$ M.~Boquien,$^{6}$ A.~Boselli,$^{6}$A.~Cooray,$^{7}$ D.~Cormier,$^{8}$ I.~De Looze,$^{4}$ 
 J. Fischera,$^{9}$ O. \L. Karczewski,$^{10}$ V.~Lebouteiller,$^{8}$ S.~Madden,$^{8}$ M.~Pereira-Santaella,$^{11}$ M.~W.~L.~Smith,$^{12}$ L.~Spinoglio,$^{13}$ 
R. J. Tuffs$^{3}$}\vspace{0.4cm}\\\\
\parbox{\textwidth}{$^{1}$Dept. of Physics \& Astronomy, McMaster University, Hamilton, Ontario, L8S 4M1, Canada\\
$^{2}$Jeremiah Horrocks Institute, University of Central Lancashire, Preston, PR1 2HE, UK\\
$^{3}$Max Planck Institute f\"{u}r KernPhysik, Saupfercheckweg 1, D-69117 Heidelberg, Germany\\
$^{4}$Sterrenkundig Observatorium, Universiteit Gent, Krijgslaan 281 S9,  B-9000 Gent, Belgium\\
$^{5}$UK ALMA Regional Centre Node, Jodrell Bank Centre for Astrophysics, School of Physics and Astronomy, University of Manchester, Oxford Road, Manchester M13 9PL, United Kingdom\\
$^{6}$Aix Marseille Universit\'e, CNRS, LAM (Laboratoire d'Astrophysique de Marseille) UMR 7326, 13388, Marseille, France\\
$^{7}$Department of Physics \& Astronomy, University of California, Irvine,CA 92697, USA\\
$^{8}$ CEA, Laboratoire AIM, Irfu/SAp, Orme des Merisiers, F-91191 Gif-sur-Yvette, France\\
$^{9}$Canadian Institute for Theoretical Astrophysics, University of Toronto, 60 Saint George Street, Toronto, ON, M5S 3H8, Canada\\
$^{10}$Department of Physics and Astronomy, University College London, Gower Street, London, WC1E 6BT, UK\\
$^{11}$Istituto di Astrofisica e Planetologia Spaziali, INAF, Via Fosso del Cavaliere 100, I-00133 Roma, Italy\\
$^{12}$School of Physics \& Astronomy, Cardiff University, Queen Buildings, The Parade, Cardiff CF24 3A\\
$^{13}$Istituto di Fisica dello Spazio Interplanetario, INAF, Via del Fosdso del Cavaliere 100, I-00133 Roma, Italy}}

\maketitle
\label{firstpage}

\begin{abstract}
We investigate star formation and dust heating in the compact FIR bright sources detected in the {\it Herschel} maps of M83. 
We use the source extraction code \textsc{getsources} to detect and extract sources in the FIR, as well as their photometry in the MIR and H$\alpha$.
By performing infrared SED fitting and applying an H$\alpha$ based star formation rate (SFR) calibration, we derive the dust masses and 
temperatures, SFRs, gas masses and star formation efficiencies (SFEs). The detected sources lie exclusively on the spiral 
arms and represent giant molecular associations (GMAs), with gas masses and sizes of 10$^{6}$-10$^{8}$M$_{\odot}$ and 200-300 pc, respectively.  
The inferred parameters show little to no radial dependence and there is only a weak correlation between the SFRs and gas masses, which suggests that more massive clouds are less efficient at forming stars.
Dust heating is mainly due to local star formation. However, although the sources are not optically thick, the total intrinsic young stellar 
population luminosity can almost completely account for the dust luminosity. This suggests that other radiation sources contribute to the dust heating as well 
and approximately compensate for the unabsorbed fraction of UV light.

\end{abstract}
\begin{keywords}
galaxies: individual ---, galaxies: ISM, galaxies: spiral, ISM: dust
\end{keywords}

\section{Introduction}
 The {\it  Herschel Space Observatory} (Pilbratt et al.\ 2010) has provided us with high sensitivity and angular resolution maps of nearby galaxies in the far-infrared, which has allowed us to spatially resolve their cold dust emission.   Dust plays a key role in the chemistry of the interstellar medium, acting as a catalyst for the formation of molecular gas, the fuel for star formation.  However, it complicates our view of galaxies by obscuring UV and optical photons from stars and then re-radiating this light in the infrared.  Thus, over a third of a galaxy's bolometric luminosity comes to us at these longer wavelengths  (e.g. Draine et al.\ 2003;
Bernstein et al.\ 2002).    \let\thefootnote\relax\footnote{$\dagger$ The first two lead authors have been co-equal contributors to the majority of the work presented in this paper.}
In the past it has been very difficult to detect the dust emission in galaxies between 200 and 850$\mu$m.  Ground-based telescopes have lacked the sensitivity and, prior to {\it Herschel}, space telescopes could not make detections at these wavelengths.

Apart from the measurement of dust luminosities and masses, the maps from {\it Herschel}, spanning 70 to 500~$\mu$m, can be used to estimate two 
important quantities: 
the gas mass and the average intensity of the radiation field heating dust within galaxies.  In the first case, since dust is generally well-mixed with gas,
we can use dust mass estimates inferred by SED fitting as a proxy for the total gas mass, assuming a gas-to-dust ratio (e.g. Hildebrand 1983, Boselli et al.\ 2002, Eales et al.\ 2010, Eales et al.\ 2012).  While atomic gas measurements are relatively well-known due to 21 cm line measurements, molecular gas measures are more challenging since we are forced to use an alternative tracer like CO rather than directly measure molecular hydrogen.  
This requires the calibration of this tracer, which is known to vary with  environment, metallicity and density (e.g. Shetty et al.\ 2011).  
{\it Herschel} maps have been used to help spatially resolve this calibration factor (i.e. the X-factor) improving our view of the molecular gas 
component in galaxies (Sandstrom et al.\, sub.\ ). Typically dust emission maps also have higher resolutions, at least at the shorter wavelengths, and superior sensitivity to CO maps.  

Dust emission SED fitting also provides a measure of the average radiation field energy density 
heating the dust or, alternatively, the average dust temperature. These latter quantities are connected to the luminosity of the heating sources and are 
therefore, in principle, useful to understand which radiation sources are heating the dust. In particular, 
in order to use dust emission to infer star formation rates, one would like to quantify the fraction of dust heating due to recent star formation.  
 However, the dust can also be powered by an older stellar population (e.g.\ Popescu \& Tuffs 2002, Groves et al.\ 2012, Bendo et al.\ 2012, Smith et al.\ 2012).

{\it Herschel} has improved our view of the morphological structure of the dust emission and has allowed us to better separate compact sources from diffuse emission in some of the far-infrared bands.  
We find both clumpy, compact regions associated with sites of recent star formation, detected also in H$\alpha$ emission, and a smoother, 
more diffuse component (e.g.\ Verley et al.\ 2010).  Although {\it Herschel} images have a resolution of 6$''$ at the shortest wavelengths (70$\mu$m), at 
longer wavelengths it increases to 36$''$ at 500~$\mu$m.  Therefore, to date most studies have been forced to degrade the high-resolution PACS (Photodetector Array Camera; Poglitsch et al.\ 2010) maps (70 and 160~$\mu$m) in order to match them to the longer 
wavelength SPIRE (Spectral Photometric Imaging REceiver; Griffin et al.\ 2010) data  (i.e.\ Bendo et al.\ 2010, Smith et al.\ 2012, Bendo et al.\ 2012, Boquien et al.\ 2012, Aniano et al.\ 2012).  This has meant that these studies have averaged over large physical areas (i.e. pixel-by-pixel analysis at 36$''$ resolution or 
azimuthal averages).  Thus, the structure evident in the PACS maps is mostly lost.  In this work, we attempt to capitalize on the 
highest resolution PACS maps, while still making use of the lower resolution SPIRE maps, in order to study the compact, clumpy regions in the FIR continuum emission. Our method relies on the
multi-wavelength extraction tool, \textsc{getsources} (Men'shchikov et al.\ 2012), which allows us to preserve the native resolution of the images.  

This paper uses this new technique on M83, a galaxy from the Very Nearby Galaxies Survey (PI: C.D. Wilson).  The close 
proximity of M83 (4.5 Mpc; Thim et al.\ 2003) affords us high spatial resolution (130 pc at 70~$\mu$m).  M83 also has a strong spiral structure and 
prominent bar, allowing us to investigate different regions within the galaxy.

The main goal of this work is the investigation of the properties of the compact sources detected in the FIR, in terms of gas masses, star formation 
and dust heating. Specifically, we address the following questions: 1) Which range of giant molecular cloud masses are associated with the FIR compact 
sources in M83?;  2) How efficient is the star formation within them?; and 3) Is the local star formation the predominant radiation source powering the dust 
emission?

To this purpose, we developed a procedure consisting of the following steps.  We use \textsc{getsources} to detect and extract compact regions from the FIR (70-350~$\mu$m), 
which we call ``clumps''.  We also measure the flux in the MIR (8 and 24~$\mu$m) and H$\alpha$ using the clump area and position.  In all cases a local 
background emission component is subtracted.  For each well-detected source, the corresponding dust luminosity, dust mass, average radiation field energy density
 and dust temperature are determined using a 
two-component dust emission SED fitting method of the MIR and FIR.   Using the dust mass we infer a gas mass for each source, using a constant gas-to-dust 
ratio.  We also derive the clump SFRs by applying the Calzetti et al.\ (2007) calibration on the measured H$\alpha$ and 24~$\mu$m fluxes.

We structure the paper in the following sections: background and
motivation (\S 2); observations and data reduction (\S3); compact
source extraction (\S 4); dust emission SED fitting (\S 5); star
formation rates and gas mass estimations; presentation of results (\S 7) and discussion (\S 8). 
We conclude with a summary of the main findings (\S 9).  

\section{Background \& Motivations}

\subsection{Gas \& Star Formation on sub-kpc scales}

During the last few years, high-resolution maps of nearby galaxies in several wavelengths including UV, IR, emission line and radio have become gradually available and opened the possibility to study the relation between gas mass surface density and star formation, known as Schmidt-Kennicutt law (hereafter S-K relation; 
 Kennicutt 1998; Schmidt 1959) on scales of kpc/sub-kpc within galaxies. The S-K relation is commonly expressed as
\begin{equation}
\Sigma_{\text{gas}} \text{ = A}\Sigma_{\text{SFR}}^{\text{N}}\text{.}
\end{equation}
The value of the exponent N is particularly important and remains debated.  N=1 implies that the star formation efficiency (SFE) or gas depletion timescale 
is constant, while N$>$1 implies that the SFE is greater for high-density regions, or that the timescale is shorter for high-density regions.  
While global averages across galaxies have shown that the total gas mass (HI + H$_{2}$) surface density has a power law relation with the SFR surface density 
(i.e. N=1.4; Kennicutt et al.\ 1989), spatially resolved studies of star formation have shown that the SFR surface density is  more tightly coupled to the 
molecular gas surface density (e.g.\ Wong \& Blitz 2002 and Bigiel et al.\ 2008).

Reliable measurements of SFR within galaxies can be performed by combining optical/UV and infrared flux measurements, according to calibrations such as Calzetti et al.\ (2007) or 
Zhu et al.\ (2008) for H$\alpha$+24$\mu$m and Bigiel et al.\ (2008) for UV+24$\mu$m. In these empirically based calibrations, dust emission is used to estimate the fraction of optical/UV light attenuated by dust. These calibrations are now considered the most accurate way to determine SFRs within galaxies, at least when used on galaxy regions which are bright enough (SFR $>=$ 0.001 M$_{\odot}$ yr$^{-1}$, Kennicutt \& Evans 2012).

Spatially resolved studies of the SFR typically involve aperture photometry of the SFR tracers.  The size of the aperture has been primarily dictated by the beam sizes.  Typically background emission is subtracted from the SFR tracers either locally, in annuli directly surrounding the apertures, or, in the case of crowded regions, more extended zones are used (Calzetti et al.\ 2005).  
Background subtraction is typically performed because the star formation tracers have contamination from other sources not directly 
associated with star formation$^1$\footnote{ $^1$The relative amount of background emission that needs to be removed is still the subject of much debate. See e.g. Leroy et al.\ (2012) for a discussion on the nature of the diffuse MIR emission.}.  
The 24$\mu$m emission includes cirrus emission from an older stellar population (Popescu \& Tuffs 2002), which may contribute as much 
as 30\% of the emission (Kennicutt et al.\ 2007).  H$\alpha$ emission may also include diffuse ionized gas (Ferguson et al.\ 1996).  Background subtraction has proven to have effects on the S-K relation.  Subtracting a diffuse component more readily suppresses the fainter star forming regions (Liu et al.\ 2011).  Thus, studies that have used aperture photometry on star forming regions and background subtraction have found a super-linear relation (e.g. Kennicutt et al.\ 2007).  Meanwhile pixel-by-pixel analyses that have not subtracted a diffuse component have recovered a linear relation (e.g. Bigiel et al.\ 2008).

 The measures of star formation tracers are quite advanced due to the relatively high resolutions available but tracing gas in extragalactic studies has 
proven to be more complicated.  While measurements of the atomic gas component are available by using the 21 cm line emission (e.g.\ THINGS; Walter et al.\ 2008), 
cold molecular hydrogen has proven more challenging.  
Due to the low mass of the molecule, it requires high temperatures to excite the rotational transitions of molecular hydrogen.  Thus, it is virtually 
impossible to measure the total amount of molecular hydrogen directly. 
Instead, an alternative tracer like CO is commonly used (Bolatto et al.\ 2013).  Typically conversions from CO are done using a constant X-factor.  However, this factor can vary considerably depending upon the metallicity, density and temperature (Kennicutt \& Evans 2012).  Accurate measures of the molecular component associated with star formation are particularly important, because it seems that star formation is more directly coupled to molecular gas than to total gas.

Due to the comparatively low resolution and sensitivity of the molecular gas maps, only few studies (e.g. Rahman et al.\ 2011) have subtracted diffuse emission from the gas. 
 However, diffuse gas unrelated to star formation sites is most likely present. Low mass clouds that are unresolved will present themselves as a diffuse 
component and may not host star formation.  Ideally, one should also account for a diffuse gas component. 
Furthermore, there is growing evidence that low mass GMCs may not have a corresponding star forming region (e.g. Hirota et al.\ 2011).   

In light of the issues surrounding the molecular gas measures we are
motivated to capitalize on the high resolution dust emission maps from
{\it Herschel}.  
By employing a gas-to-dust ratio we present an alternative way to probe the star formation efficiency on spatially resolved scales. 
Inspired by the aperture photometry studies of star forming regions, we employ the dust maps to isolate and extract clumpy regions which are likely associated with 
individual or groups of molecular clouds. 
For both the star formation tracers and the dust we can also subtract a diffuse background component, thus treating both the star formation and gas tracers in a 
similar fashion.

\subsection{Dust heating within galaxies} 
In general, dust emission within galaxies without AGNs is powered by
radiation coming from both sites of recent star formation and from 
more evolved stellar populations. However, there is a longstanding
debate about the exact fraction of dust heating contributed by each
stellar population (e.g. Law et al.\ 2011, Boquien
et al.\ 2011, Bendo et al.\ 2012, for recent references), which depends on several factors: the intrinsic emission spectral energy distribution 
of the stellar populations, the dust mass and optical properties, 
the relative dust-star geometry.

Recent observational works investigating the origin of the radiation
heating the dust in nearby galaxies have looked for correlations
between the source dust temperature (or, alternatively, FIR colour) and 
source stellar
populations luminosities, as traced by SFR for the young stars and NIR
luminosity for the old stars (e.g. Bendo et al.\ 2012, Boquien et al.\
2011, Smith et al.\ 2012, Foyle et al.\ 2012).$^2$\footnote{ {$^2$ The quantities typically considered are \textit{surface densities}
 of SFR and NIR luminosity rather than total values.  For pixel-by-pixel analyses this amounts to multiplication with a constant, 
thus it does not affect the nature of the correlation with dust temperature.}} 

The presence of an observed correlation of this kind has often been used as evidence that a
particular stellar population is the dominant source heating the
dust. Although it might seem intuitive that a higher dust temperature
should correspond to a stellar population with a higher intrinsic luminosity,
deducing the origin of the radiation heating the dust in this way can
be potentially misleading. The reason is that the intrinsic luminosity
of a stellar population and the average dust temperature are not
necessarily proportional to each other, even in the case where the
stellar population in question is the dominant source of dust heating. 

The luminosity of a stellar population gives the amount of total radiative energy per unit time injected by a stellar population locally in the 
interstellar medium. Therefore, one can expect at most a
proportionality between the observed dust luminosity and the intrinsic stellar
population luminosity for a set of sources, provided all the sources
are mainly heated locally by the same kind of stellar population and heating from radiation 
sources external to the areas considered is negligible$^3$\footnote{$^3$The latter assumption can easily break down when considering arbitrary 
galactic regions. For the compact FIR sources associated with star formation regions, the heating is usually thought to be dominated by the local young 
stellar populations but see further discussion in \S8.3.}. 

Specifically, one can express the
total dust luminosity L$_{\rm{dust}}$ for a source as: 
\begin{equation}
{L}_{\rm{dust}}={L}_{\rm{stars}}^{\rm{int}}\left(1-e^{-\tau}\right) 
\label{ldusteq}
\end{equation}
where $L_{\rm{stars}}^{\rm{int}}$ is the total intrinsic luminosity
from a certain stellar population and $\tau$ is the source luminosity-weighted 
optical depth.  The optical depth can be expressed as: 
\begin{equation}
 e^{-\tau}=\frac{L_{\rm{stars}}^{\rm{out}}}{L_{\rm{stars}}^{\rm{int}}}
\end{equation}
where $L_{\rm{stars}}^{\rm{out}}$ is the escaping unabsorbed source stellar
population luminosity. All the aforementioned luminosities are integrated over 
wavelength spanning the entire emission spectral range.  
From Eq. \ref{ldusteq} one can see that, if $\tau$ has similar values for a sample of sources, a correlation between L$_{\rm{dust}}$ 
and $L_{\rm{stars}}^{\rm{int}}$ will be found. Note also that
Eq.\ref{ldusteq} implies that $L_{\rm{dust}}$ cannot exceed
$L_{\rm{stars}}^{\rm{int}}$.  If this is 
observed for a sample of sources, it would mean that the
dust is significantly heated by radiation coming from another stellar
population beyond that associated with $L_{\rm{stars}}^{\rm{int}}$.  
         
In order to have a proportionality between the source intrinsic stellar luminosity and the source average dust temperature, an additional 
assumption is required: each of the detected sources should be
associated with a similar amount of dust mass, $M_{\rm{dust}}$.  Assuming that the dust emission can be described by a modified blackbody function such that $\kappa_{\nu}B_{\nu}(T_{\rm{dust}})$, 
where $\kappa_{\nu}\propto \nu^\beta$ is 
the absorption coefficient, $\beta$ is the emissivity, $B_{\nu}$ the Planck function and $T_{\rm{dust}}$ the dust temperature, it can be shown that 
$L_{\rm{dust}} \propto M_{\rm{dust}}\int{\kappa_{\nu}B_{\nu}(T_{\rm{dust}})d\nu} \propto M_{\rm{dust}}T_{\rm{dust}}^{4+\beta}$. 
By combining the latter relation with Eq.~\ref{ldusteq}, it follows that: 
\begin{equation}
 \frac{L_{\rm{stars}}^{\rm{int}}}{M_{\rm{dust}}} \propto \frac{T_{\rm{dust}}^{4+\beta}}{\left(1-e^{-\tau}\right)}
\label{lstarequ}
\end{equation}
 
If $M_{\rm{dust}}$ and $\tau$ have similar values for all the sources considered, the average intensity of the radiation heating the dust approximately 
scales only with the 
stellar luminosity and, as a consequence, the dust temperature $T_{\rm{dust}}$ is directly related only to $L_{\rm{stars}}^{\rm{int}}$ through 
Eq.~\ref{lstarequ}.    
In this case, it is likely that a specific stellar population is heating the dust, if one finds that the intrinsic 
luminosity of that stellar population correlates with the observed dust temperature.
However,  the amount of dust mass, the dust-star geometry and, therefore, the value of $\tau$, can be substantially different for each source. 
Thus, in general, it cannot be expected that the luminosity of a stellar population is correlated with the dust temperature, even if that stellar 
population is responsible for the dust heating.

Eq.~\ref{ldusteq} and \ref{lstarequ} are almost equivalent to each
other and they both express the relation between the dust luminosity and stellar 
population luminosity. The only difference is that $L_{\rm{dust}}$ in Eq.~\ref{ldusteq} 
includes all the dust and PAH emission throughout the entire infrared range. In Eq.~\ref{lstarequ}, we assumed that the dust emission can be 
modelled by a single modified blackbody curve.  This is a good approximation if one considers only the FIR region of the emission spectra. However, usually 
the FIR luminosity is the dominant contribution to $L_{\rm{dust}}$.

Thus, Eq.~\ref{lstarequ} can be used to probe the extent to which a given stellar population is powering the dust emission. 
We can do so by comparing the local, cold dust temperature with a measure of the local stellar population luminosity divided by the corresponding dust mass. 
As before, if the value of $\tau$ is varying in a relatively small range, a correlation between $L_{\rm{stars}}^{\rm{int}}/{M_{\rm{dust}}}$ and $T_{\rm{dust}}$ 
will be 
observed if the stellar population considered is responsible for
heating the dust. 

For example, one can use the SFR value as a tracer of the UV luminosity of the young stellar populations associated with each source. 
Therefore, if the radiation from young stars dominates the dust heating, the SFR/$M_{\rm{dust}}$ 
ratio is expected to be more tightly coupled to the dust temperature than the SFR alone.   
Thus, an observed correlation between SFR/$M_{\rm{dust}}$ and $T_{\rm{dust}}$ would suggest that star formation is powering the observed 
dust emission.

Given this, we are motivated to include a comparison of the measured SFR/$M_{\rm{dust}}$ (or, equivalently, SFE if one assumes a constant dust-to-gas mass ratio) 
with $T_{\rm{dust}}$ for the sources 
we detect, in addition to simply a comparison of the SFR and dust temperature, which is typically seen in the literature. 
The results of these comparisons are shown in Sect. \S7 and discussed in Sect. \S8.

\section{Observations}
\begin{figure*}
\centering
\includegraphics[trim=0mm 0mm 0mm 0mm,width=180mm]{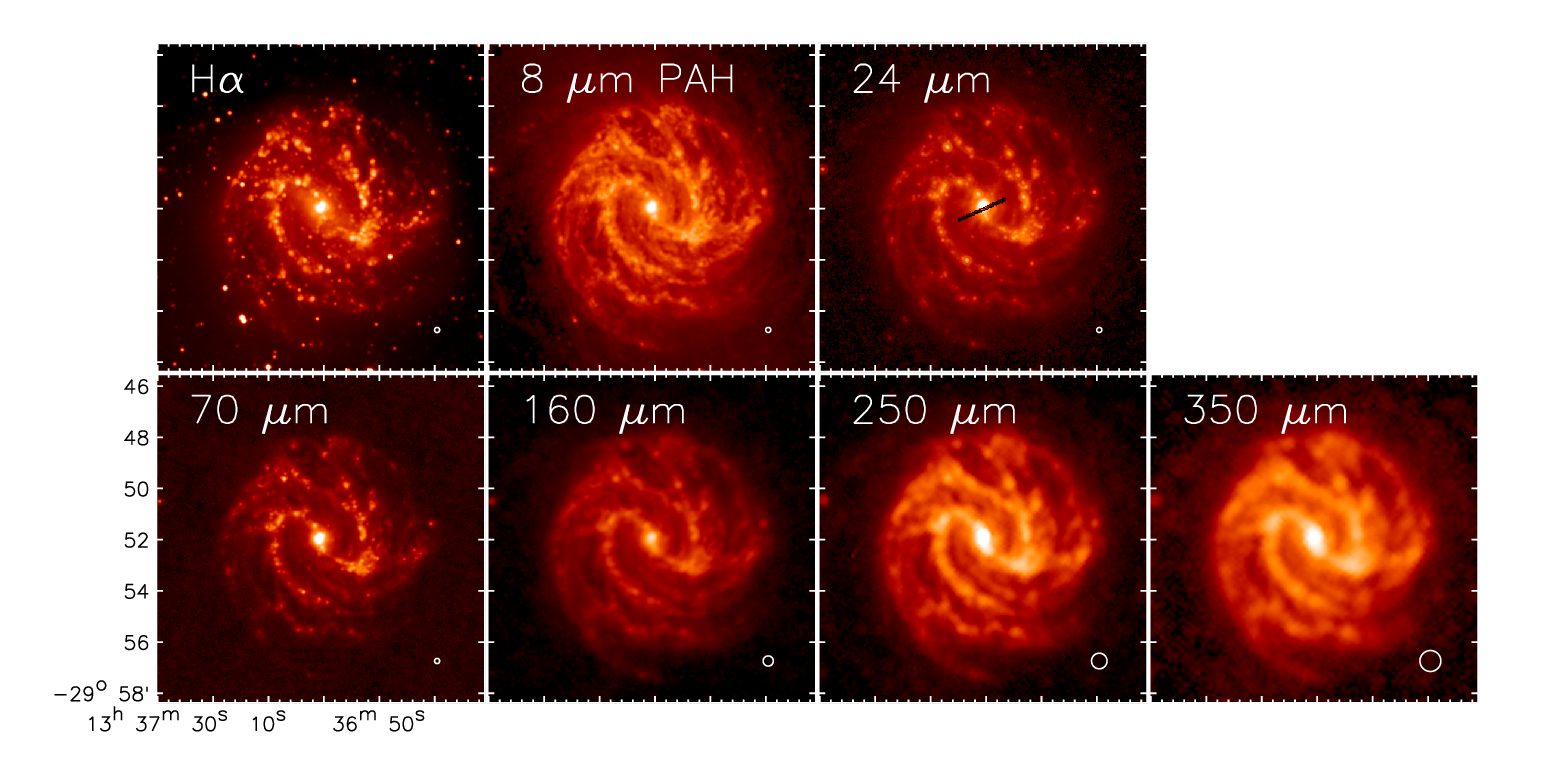}
\caption{M83 at different wavelengths at the resolution used in this study denoted by the white circle in the bottom right of each panel. Upper row: H$\alpha$ from SINGG survey and {\it Spitzer} IRAC 8 and MIPS 24~$\mu$m maps.  Lower row: {\it Herschel} 70, 160, 250 and 350~$\mu$m maps.} 
\label{maps}
\end{figure*}

In this work we use the {\it Herschel} far-infrared maps of M83 from the Very Nearby Galaxies Survey (PI: C. D. Wilson) as well as ancillary mid-infrared 
and H$\alpha$ maps to trace dust/PAH emission and star formation rate. Fig.~\ref{maps} shows M83 at each of the wavebands considered.  

\subsection{Far-Infrared Images}
We use far-infrared (FIR) images from the {\it {\it Herschel} Space Observatory} to trace cold dust emission.  
We use 70 and 160~$\mu$m maps taken with the Photodetector Array Camera (PACS; Poglitsch et al.\ 2010) and 250 and 350~$\mu$m maps taken with the 
Spectral Photometric Imaging REceiver (SPIRE; Griffin et al.\ 2010)$^{4}$\footnote[4]{$^{4}$We performed an initial test including the 500~$\mu$m map as well, but we found 
that the uncertainties on the source fluxes were so large that they were not useful to constrain the source SEDs. This was caused by the  
low resolution of the 500~$\mu$m map (PSF FWHM=36$''$). Thus, we decided to not include the 500~$\mu$m map in our analysis.}. 
The PACS images are processed using both \textsc{hipe} v5 and \textsc{scanamorphos} v8 
(Roussel 2012) and the SPIRE images are processed with \textsc{hipe} and \textsc{brigade} (Smith et al.\ 2012).  The PACS images were corrected from the v5 photometric calibration files to v6 with corrective factors of 1.119 and 1.174 for the 70- and 160-~$\mu$m maps, respectively.  The SPIRE images are 
multiplied by 0.9828 and 
0.9839 for the 250 and 350~$\mu$m maps respectively in order to convert from monochromatic intensities of point sources to monochromatic extended sources. The images are kept in their native 
resolution with a FWHM of the PSF of 6.0$''$, 12.0$''$, 18.2$''$ and 24.5$''$ for the 70, 160, 250 and 350~$\mu$m maps, respectively.  For more details on how the images were processed we refer to  
Bendo et al.\ (2012) and Foyle et al.\ (2012, hereafter F12).

\subsection{Mid-Infrared Images}
We trace the warm dust and PAH emission using mid-infrared maps (MIR) taken from the {\it Spitzer}   Local Volume Legacy Survey (Dale et al.\ 2009). 
Specifically, we use the 8 and 24~$\mu$m maps from IRAC and MIPS instruments. We  subtract the stellar component of the emission in the 8~$\mu$m map 
using a scaling of the IRAC 3.6~$\mu$m map, according to the relation provided by Helou et al.\ (2004): $F_{\nu}$(8 $\mu$m, dust) = $F_{\nu}$(8 $\mu$m)-0.232$F_{\nu}$(3.6 $\mu$m).  
In the 24$\mu$m map the nuclear region is saturated, so this region is excluded from our analysis. Because of reasons explained in \S4, we degrade the resolution of the MIR maps to 6$''$ in order to match the resolution of the 70~$\mu$m map. 

\subsection{H$\alpha$ Images}
We use continuum subtracted H$\alpha$ maps from the Survey for Ionization in Neutral Gas Galaxies (SINGG; Meurer et al.\ 2006). 
We correct the H$\alpha$ maps for Galactic extinction using a factor 1.167 from the NASA/IPAC Extragalactic Database that is based on Schlegel et al.\ (1998).  
As for the MIR maps, we degrade the H$\alpha$ map to a resolution of 6$''$ matching that of the 70~$\mu$m map. 
We use the H$\alpha$ maps in conjunction with the 24~$\mu$m MIR map, in order to measure the star formation rate (SFR) of the extracted compact sources 
(discussed in greater detail in \S6).


\section{Compact Source Extraction}
In order to extract compact sources in the FIR maps (70-350~$\mu$m), we use the new
multi-scale, multi-wavelength tool, \textsc{getsources} (Men'shchikov
et al.\ 2012).  \textsc{getsources} is specifically designed to work
with the FIR images of {\it Herschel}.  The data from {\it 
    Herschel} spans a range of angular resolutions from 6.0$''$ to 36.0$''$ and, thus, any source extraction code must be able to handle these extremes, which poses concerns for source blending at longer wavelengths.  

Rather than extracting sources directly from observed images (i.e. \textsc{gaussclump}; Stutzki \& Gusten 1990, \textsc{clumpfind}; Williams et al.\ 1994 or \textsc{sextractor}; Bertin \& Arnouts 1996), \textsc{getsources} analyzes spatial decompositions of the images across different scales and different wavelengths.  Wavelength independent images are generated to detect sources at each spatial scale.  The original images are then used to perform photometry on the detected sources. 
This procedure takes into account the different angular resolutions of the maps, background subtraction and source blending.  
We briefly describe the key steps of the process here, but refer the reader to Men'shchikov et al.\ (2012) for a detailed description on 
how \textsc{getsources} extracts and measures the properties of compact sources, including a comparison with other similar codes.    

After the images are aligned to the same spatial grid, \textsc{getsources} decomposes of the original maps into single scale detection images. 
This is done by using a 
process of successive unsharp masking, where the original images are convolved with Gaussians and subtracted successively, in order to enhance 
the visibility of emission on different scales.  The FWHM of the Gaussians 
varies between twice the pixel size to a maximum of 18 times the resolution of the image or the image size. The image resolution is the only information 
the user needs to provide.   The background and noise in the single scale 
detection images is then removed by intensity thresholding.  The clean single-scale detection images at each wavelength are then combined into single-scale 
wavelength-independent detection images, allowing for the use of all the information across all bands simultaneously for the source detection. Many detection codes rely on independent catalogues at each waveband which are then matched using an 
association radius, which can introduce large unknown errors.  By dividing the images into single scales, this process can be avoided.  

On the combined single-scale detection images, a given source will appear at a small scale and gradually get brighter until it is seen at a scale roughly the true 
size of the source.  Beyond this, the source begins to vanish again.  \textsc{getsources} tracks the evolution of the source through the spatial scales and 
creates source masks to identify the sources.  The scale where the source is brightest provides an initial estimate of the source {\it footprint}.  A source must have a signal-to-noise ratio of at least 3$\sigma$ in at least two bands in order to be considered detected.
Coordinates of the sources are determined using the moments of the intensities.  Once the sources have been detected in the combined images over all spatial 
scales, \textsc{getsources} performs source flux measurements on the observed images at each waveband and simultaneously subtracts a background by interpolating 
under the sources {\it footprints}. Partially overlapping sources are ``deblended'' using an iterative process.  

Upon completing the extraction and measurements, the user is supplied with a table giving the location and size of each source at each waveband.  
Due to the fact that the resolution decreases with increasing wavelength, the apparent source size increases with wavelength.  The measured properties also 
include total flux, peak flux, degree of source blending, monochromatic and global detection significance.  The detection significance at each wavelength is 
essentially a signal-to-noise ratio, which is determined by the ratio of the peak flux of the source to the standard deviation in an annulus surrounding the 
source on the detection maps. The global detection significance is determined by the square root of the sum of the 
squares of the detection significance at each wavelength.  As discussed in the following sections, we apply a minimum threshold for the global detection 
significance value to remove some sources not well detected.

We now briefly describe the parameters we defined in the source extraction process.
\subsection{Preparing Images}
We supply all images (H$\alpha$, MIR and FIR) to \textsc{getsources}, but only use the FIR images to detect the compact sources, since we seek to trace the location of cold compact clouds.  

The images are aligned to the same grid and converted to MJy sr$^{-1}$, with a pixel size of 1.4$''$.  
 For the SPIRE images, \textsc{getsources} uses the beam areas in order to convert the units from Jy/beam (423 $\pm$ 3, 751 $\pm$ 4, 1587 $\pm$ 9 
arcsec$^{2}$, for the 250, 350 and 500~$\mu$m images respectively).  The images are all aligned to the WCS of the 70$\mu$m maps.   Observational masks are 
created, which denote the image area over which \textsc{getsources} is meant to look for sources, which speeds the detection process.  The alignment and masks are visually checked.

Since \textsc{getsources} has been designed to work with FIR {\it Herschel} maps, we decided to degrade the MIR and H$\alpha$ maps to the angular resolution of the 70~$\mu$m map. 
This should reduce possible systematic effects due to the use of the program on a larger range of spatial resolutions than the one on which it has been designed and tested.  
Furthermore, we performed a test to examine the efficiency of \textsc{getsources} in recovering source fluxes at different resolutions, by running the program
on a set of convolved 70$\mu$m maps. In this test, source fluxes are expected to be the same at each resolution. We found the accuracy
of the source flux measurement decreases with poorer resolution, but the average systematic effect does not seem to be large enough to affect our 
results substantially (see Appendix A for more details).

\subsection{Compact Source Measurements}
The source detection has been performed using only the FIR maps, since we aimed to select sources due to their FIR brightness and not necessarily MIR and H$\alpha$ counterparts. 
However, MIR and H$\alpha$ maps have been provided to \textsc{getsources} for source photometry.     
\textsc{getsources} detects 186 compact sources across M83. However, in the analysis described in the following sections, 
we only include those sources with a detection significance above 20.   Furthermore, since a high global detection 
significance does not necessarily imply that the photometry
is accurate for each single waveband, we further removed all the sources with a photometric signal-to-noise ratio lower than 1$\sigma$ in at least one band 
between 8 and 250~$\mu$m (that is, we occasionally retained sources detected in all bands but the 350~$\mu$m band). After this source selection, 
our sample is reduced to 121 sources. 

\begin{figure*}
\centering
\includegraphics[trim=0mm 0mm 0mm 0mm,width=100mm]{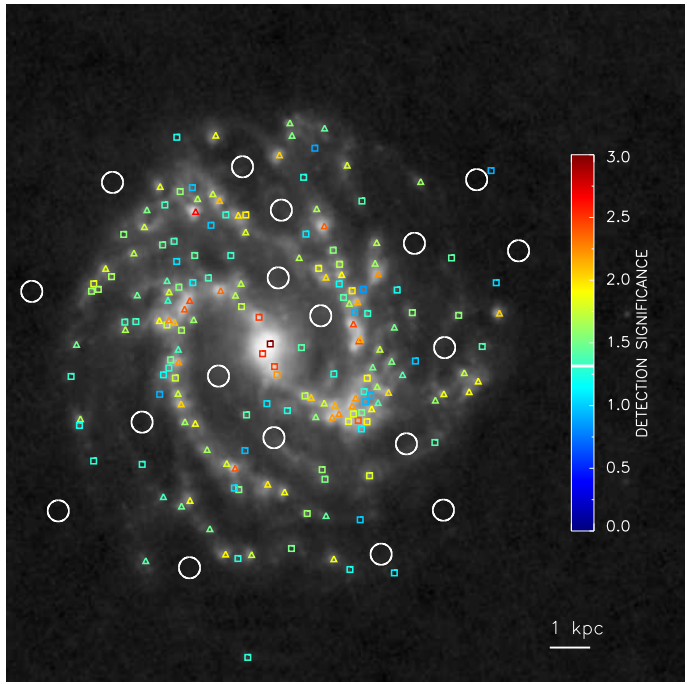}
\caption{M83 at 70~$\mu$m with the compact source locations marked with colors corresponding to the detection significance (Log 10 scale).  A detection significance of 20 is used as a cut-off for retaining sources for further analysis.  
Sources that are used in the study are denoted with triangles and excluded sources are denoted with squares (see text for criteria). Apertures showing the measurement locations of interarm emission are displayed in white.}
\label{sources_on_map}
\end{figure*}

Fig.~\ref{sources_on_map} shows the location of all the detected sources overlaid on the 70~$\mu$m map, with colours denoting the detection significance.  
Sources that are retained in the study are denoted with triangles, while sources that are excluded are denoted with squares. 
We note that the sources lie almost exclusively on the bar ends and the spiral arms.  Very few sources are detected in the interarm regions and along 
the bar.  This does not mean that compact sources do not exist in
these regions, but rather, the sources there are not bright enough to be detected.  
Thus, it seems that only the bar ends and spiral arms harbour bright compact sources that we can detect.

Our spatial resolution at 70~$\mu$m allows us to measure intrinsic sources sizes only for sources having radii larger than 130 pc.  
We find a median radius of 150 pc at this waveband. This implies that
the majority of our sources are unresolved or barely resolved. 
Thus, it is not possible to determine a physical radius for all the sources.  

\begin{figure}
\centering
\includegraphics[trim=0mm 0mm 0mm 0mm,angle=-90,width=80mm]{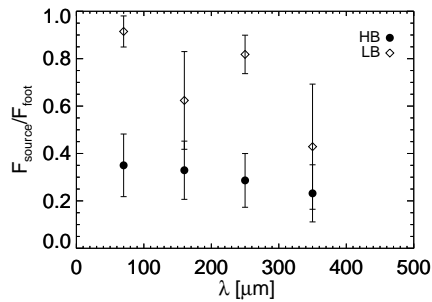}
\caption{Median value of the ratio of the measured source flux to the total flux in
  the source footprint for both the high (full circles) and low (open
  squares) source background subtractions.  At
  longer wavelengths the background is greater due to the increasing
  size of the footprint.}
\label{comp_bk}
\end{figure}

At each waveband we obtain the total flux of each of the extracted sources (hereafter, F$_{\text{source, HB}}$) and of the corresponding 
background emission, which has been interpolated and subtracted by \textsc{getsources}. 
Because \textsc{getsources} interpolates the background from regions very close to the sources, this background measurement is actually determined in the 
vicinity of the spiral arms.  We find that the background makes up
more than 50\% of the flux in the footprint.  Fig.~\ref{comp_bk} shows
the relative flux of the source to the total flux in the footprint.
The relative fraction decreases with wavelength mainly because, due to the
poorer resolution at longer wavelengths, the footprint size increases
and thus a relatively larger amount of background flux is included in
the footprint. However, part of this relative flux variation might be also due to an intrinsic higher fraction of diffuse background 
emission at longer wavelengths.
 We find that at all wavelengths the source makes up less than 50\% of the emission in the footprint.  For this reason, we 
refer to this way of performing the source photometry as ``high-background'' source flux measurement.     
In order to identify possible systematic uncertainties due to the way the background is 
subtracted, we also consider an alternative measure for the background, estimated in the interarm regions, as shown in the next subsection. 


\subsection{Low-Background Source Flux Measurement}
The background estimated by \textsc{getsources}  is 
basically a measure of the smooth emission component associated mainly with the spiral arms. 
However, typically the photometry of star forming regions is done by 
subtracting an 
average brightness measured in the interarm regions close to the sources (i.e. Calzetti et al.\ 2007 and Liu et al.\ 2011) and, as said before, it is 
important to check if there are substantial differences on the final results if the background measurement is performed differently.

For these reasons, we decided to perform aperture photometry in several interarm regions selected by eye (see Fig. \ref{sources_on_map}). 
Specifically, we selected 18 interarm regions,
which are all sufficiently far away from the detected sources at all wavelengths. The diameter size of the apertures is 25$''$ which is large enough to obtain 
a good estimate of the rather uniform interarm emission. For each aperture we measured the average surface brightness with an uncertainty derived from the standard deviation of the surface brightness within each aperture and the flux calibration uncertainty. 

Once we obtained the flux measurements for the interarm apertures, we performed a ``low background'' source flux measurement in the following way:  
for each source, we consider the flux in the 
full footprint determined by \textsc{getsources} and subtract the interarm surface brightness, estimated from the aperture that is 
closest to the source, multiplied by the area of the source footprint (typically, the distance between the centres of each source and the closest interarm 
region is about 1 kpc or less)$^5$\footnote{ $^5$We note that there are variety of possible approaches for subtracting background emission.  Here we contrast two techniques, one which utilizes an interpolation scheme in the vicinity of the source and one that relies on a more distant estimate in the interarm regions.}. Since the \textsc{getsources} background is usually larger than that determined in 
the interarm region, in general the ``low background'' (LB) source fluxes are higher than the ``high-background'' (HB) ones.  
Fig.~\ref{comp_bk} shows that the LB measurements have higher fluxes relative to their background at all wavelengths than HB flux measurements.

\section{SED Fitting}
In the absence of a strong contribution from AGNs, the galaxy dust/PAH emission SED on sub-kpc/kpc scales can be modelled as the sum of two 
emission components: 
a warm component largely emitting in the MIR, produced by dust in photo-dissociation regions (PDRs) and heated predominantly by young stellar populations; 
and a diffuse emitting component, emitting mostly in the FIR and MIR PAH line emission, which can be 
powered both
by older stellar populations and by the fraction of UV photons escaping from PDRs.     
This concept is at the base of the SED fitting method developed by 
Natale et al.\ (2010, NA10), which we used in this work to fit all the well-detected source emission SEDs derived in \S4.  
In the following we explain the main features of this SED fitting
method, together with minor updates, and its application to our data set. We refer the reader to NA10
for additional details.

\begin{figure*}
\centering
\includegraphics[trim=0mm 0mm 0mm 0mm,width=150mm]{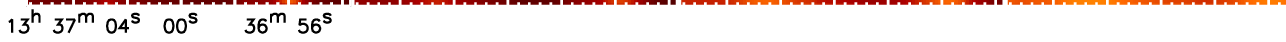}
\caption{Example of source flux measurement and SED fitting. The maps on the left show a small area of the disk of M83 around one of the detected sources. 
The ellipses overplotted on each map delineate the source footprint determined by \textsc{getsources}. On the right, we show the two-component SED fit of the 
same source with best fit parameter and SFR values listed aside. The dashed and the dotted line represent the diffuse dust emission component and the PDR emission component
respectively.}
\label{sedfit}
\end{figure*}

Using the mentioned fitting method, each observed source dust emission SED is fit by 
combining two infrared SED components (see Fig.\ref{sedfit}). The first component is a PDR SED template which has been selected by Popescu et al.\ (2011) 
among the models of Groves et al.\ (2008) because it provides a good fit to the dust 
emission from Milky Way star formation regions (specifically the chosen model is for compactness parameter $log(C)=6.5$, solar metallicity and hydrogen
column density $log(N)=22$, see \S 2.8 of Popescu et al.\ 2011 for more details). The second component, suitable to fit the diffuse dust emission,  
is taken from a grid of 
uniformly heated dust emission templates, obtained by using the dust emission code of Fischera \& Dopita (2008). The diffuse dust emission is calculated 
assuming 
a Milky Way 
dust+PAH composition (exact dust model parameters can be found in Table 2 of Fischera \& Dopita 2008) and taking into account the stochastic heating of grains following the 
method of Guhathakurta \& Draine (1989), combined with the step
wise analytical solution of Voit (1991). Since we assumed a 
fixed dust/PAH composition, each element of the grid is determined only by the parameters of the radiation field heating the dust. 
The adopted spectral shape of the radiation field is the classical Mathis et al.\ (1983) profile, which was derived for the local interstellar radiation field, 
scaled by two linear factors: $\chi_{\rm{UV}}$, which multiplies the whole curve, and $\chi_{\rm{col}}$ which multiplies only the optical part of the Mathis spectra 
(see appendix B.2 of NA10). Therefore, $\chi_{\rm{UV}}$ can be seen as the intensity of the UV radiation field and 
$\chi_{\rm{col}}$ as the optical to UV ratio in the units of the Mathis et al.\ (1983) profile. 

Compared to NA10, we extended the size of the 
 grid of the diffuse dust templates, in order to cover a larger range of possible radiation field parameters. Specifically, $\chi_{\rm{UV}}$ and $\chi_{\rm{col}}$ are both allowed to vary 
between 0.1 and 10, a range which is reasonably large to include  
all the plausible values of the diffuse radiation field intensity and colour within galaxies (note that even higher radiation fields and, therefore, 
warmer dust are associated with the PDR component in our SED fitting procedure). We also point out that there is a degeneracy 
between $\chi_{\rm{col}}$ and the relative abundance of PAH and solid dust grains, in the sense that they both affect the 8~$\mu$m/FIR ratio (see \S B.\ 2 of NA10). 

One of the main differences between our dust emission models and those of Draine \& Li (2007) is 
that in their models the PAH abundance is varied but the optical/UV intensity ratio is fixed to the Mathis et al.\ (1983) value.  
Therefore, this should be taken into account when considering results involving the $\chi_{\rm{col}}$ parameter. 

We performed a $\chi^2$ minimization fitting of the data with the two dust emission components by varying four free parameters: $\chi_{\rm{UV}}$ and $\chi_{\rm{col}}$, the 
radiation 
field parameters defined above; $M_{d}$ the dust mass of the 
diffuse dust component; F$_{24}$, the fraction of 24~$\mu$m emission provided by the PDR component. For each pair of $\chi_{\rm{UV}}$ 
and $\chi_{\rm{col}}$ values, the parameters $F_{\rm{24}}$ and $M_{\rm dust}$ correspond to linear scaling factors for the amplitudes of the PDR template and the 
diffuse dust component respectively. The fit takes into account colour corrections calculated for each SED template following the conventions adopted 
for each 
instrument (See also Sect. 5 of NA10 for details and the {\it Spitzer} and {\it Herschel} observer manuals for the colour correction definitions). 

The uncertainties on each parameter are calculated by analyzing the variation of 
 $\chi^{2}$ around the minimum found by the fitting procedure. The one sigma interval is defined as the minimum variation of a given fitting parameter 
around its best fit value, 
which produces a variation $\Delta \chi^{2}=\chi^2-\chi^2_{min}$ always higher than 1, independently of all the possible values of the other fitting parameters. 
The program checks also if ``islands'' of $\chi^{2}$ low values (such that $\Delta \chi^{2}<1$) are present, which are detached from the region where the minimum of $\chi^{2}$ has been found. 
In that case, 
a conservative 
uncertainty covering all the regions of low $\chi^{2}$ values is provided for the fitting parameter. 

Apart from the fitting parameters, our SED fitting procedure provides the total dust emission luminosity and the luminosities of each SED component.
However, in contrast to blackbody fits, our method does not provide a single average dust temperature, since the 
output total spectra of the 
dust emission code is determined by probability distributions of dust temperatures, which are different for each dust grain size and composition. 
Furthermore, in principle 
different combinations of UV and optical radiation field energy densities can cause the dust to have approximately the same average cold dust temperature (that is, similar 
FIR peak wavelength), 
which would not be 
immediately evident by comparing different pairs of $\chi_{\rm{UV}}$ and $\chi_{\rm{col}}$ values.  
However, for the purposes of comparisons with other works that have relied on blackbody fits, it is useful to quote an average dust temperature.  Taking advantage of the fact 
that the cold 
FIR part of the diffuse dust emission component ($\lambda$ $>$70~$\mu$m) can be well fit with a modified blackbody curve with a dust emissivity index of
$\beta$=2  (consistent with observational results, i.e. F12, Davies et al.\ 2012, Auld et al.\ 2013),  it is straightforward to associate to each 
diffuse dust template the dust temperature of the modified blackbody which 
best reproduces its FIR part. We will refer to this cold dust
temperature as T$_{\rm dust}$.  In Appendix B we compare the dust
masses and temperatures found using the SED fitting method outlined
here to those determined directly from a modified blackbody function
fit to the FIR wavelengths.


\section{Star Formation Rates and Gas Masses}
 Although the sources are detected in the FIR bands, measurements of the flux in the source footprint 
in the H$\alpha$ and MIR maps are also made. H$\alpha$ and 24$\mu$m emission can be used in conjunction to trace recent star formation - both unobscured and obscured (Calzetti et al.\ 2007).  
\textsc{getsources} also interpolates and subtracts a background for these images as well.  A visual comparison of the source measurements in H$\alpha$ confirmed that the source footprints are centred on bright emission peaks.  In this way, we feel confident that the flux measured in the footprints is directly related to an active star forming region.  

For each source, we combined the fluxes measured in H$\alpha$ and 24 $\mu$m emission to derive a star formation rate by using the calibration of 
Calzetti et al.\ (2007): 
\begin{equation}
\text{SFR} [M_\odot \text{ yr}^{-1}] = 5.3 \times 10^{-42} (L(\text{H}_{\alpha}) + 0.031L(\text{24} \mu\text{m}))\text{.}
\end{equation}

There are two key differences between our SFR and that derived by Calzetti et al.\ (2007).  We detect the compact sources in the FIR maps rather than 
the H$\alpha$. This means we essentially estimate the SFR in the region of a FIR compact source.  However, a visual check shows that almost all sources in 
the FIR are also present in H$\alpha$. A second difference is related to how the background is treated. Here, we rely on the background subtraction 
performed by \textsc{getsources}.  This background is interpolated in the region surrounding the sources and takes into consideration neighbouring sources 
and deblending. In Calzetti et al.\ (2007), the background is determined in 12 rectangular regions surrounding sources.  These regions cover a large number of 
pixels and extend well beyond the local neighbourhood of the sources. The mode of these regions is then used as a measure of the background. Our method, 
produces a more `local' background, which is greater than that which would be found by using the same method of Calzetti et al.\ (2007). 
In order to check the effect of choosing a different background level, we also determine 
 source star formation rates by using the H$\alpha$ and 24~$\mu$m emission in the total source footprint, after subtraction of the background estimated 
from the nearest interarm aperture. This is exactly the same approach as for the LB measure for the compact source dust emission.  

We note a potential problem with the SFR as outlined above.  During the source extraction, it is not known which sizes the sources will potentially have.  The size of the region plays an important role in deciding whether the calibration described above can be used.  Regions which are too small may have SFRs which are too low.  
It is known that SFRs below $\approx$ 0.001 M$_{\odot}$ yr$^{-1}$ can be problematic because there may be incomplete 
sampling of the IMF and the assumption of continuous star formation in the last few Myr may not be valid (Kennicutt \& Evans 2012; Leroy et al.\ 2012).  
Furthermore, the adopted calibration assumes that all the ionizing photons are absorbed by the gas in the HII regions. 
However, a fraction of ionizing photons could escape the star formation regions before ionizing the gas or, instead, could be 
absorbed by dust (see e.g. Boselli et al.\ 2009,
Calzetti 2012, Relano et al.\ 2012). Due to all these concerns, it is useful to consider the SFR measured in this process as a ``corrected'' H$\alpha$ 
luminosity. Whether this luminosity can accurately trace the SFR will depend on the source size and the value of the SFR. 

We estimate the source gas masses $M_{\text gas}$ by making use of the fact that the dust mass can be used as a proxy for the total gas mass (both molecular and atomic), 
provided one can translate the dust mass with a constant gas-to-dust ratio (GDR) to a gas mass.  Dust mass is known to be better correlated with the total gas mass than with the atomic or molecular component alone (i.e.\ Corbelli et al.\ 2012).

We assume a constant GDR of 100 for the compact regions.  While, there is some evidence that the GDR may vary within galaxies (e.\ g.\ Sandstrom et al. 2013), particularly with metallicity, given that our sources lie almost exclusively along the spiral arms, it is not obvious that a varying GDR should be used for these sources.  For example, the metallicity values for M83 are based on azimuthal averages, which include interarm regions. Thus, employing a varying GDR based on metallicity values would require introducing another assumption, namely that the GDR depends more on radius than arm and interarm regions.  In the pixel-by-pixel analysis of F12, the GDR was found to be relatively constant
on the spiral arms ($\approx$ 100) and, in contrast to the interarm regions, declined only slightly with radius with the inner regions of the spiral arm having a GDR of 130 and the outer tips of the spiral arm having a GDR of 120.  Given the current uncertainties and possible systematic effects, we simply adopt a value of 100, which is in the range found by F12 (mean value of 84 and standard deviation of 40).
   
By combining the SFR and gas mass measurement for each source, we estimate the star formation efficiency defined as SFE=SFR/$M_{\text gas}$ for both types of compact source flux measurement.  
We stress that this SFE is different from that presented in other works (Leroy et al.\ 2008; Bigiel et al.\ 2008 etc.) which describe the SFE using only the molecular gas component.  However, it is likely that the dust is mostly tracing molecular gas in these regions. M83 is known to be a galaxy with a dominant molecular component and, over the region we consider, the molecular gas constitutes 80\% of the total gas (Crosthwaite et al.\ 2002).  F12 found that on the spiral arms, where most sources lie, the molecular gas component has column densities on the order of 100 M$_{\odot}$ pc$^{-2}$, whereas the atomic gas had column densities of less than 10 M$_{\odot}$ pc$^{-2}$.

\section{Results}

Upon completing the SED fitting, as described in \S5, we further remove sources that have high (total) $\chi^{2}$ values, that is $\chi^{2}>10$,  or 
have a dust mass uncertainty greater than a factor of 2. In the first case we remove all the sources which are not well-fit
by our model (mainly because the observed SED shape is too irregular to be fitted with our two-component model), while in the second we remove those 
sources with flux uncertainties that are too large at some wavebands and, therefore, cause the fitting parameters to not be constrained within small ranges. 
After applying this selection criteria, we are left with 90 sources. Appendix D presents tables of the extracted flux and SED fitting parameters for both the 
HB and LB source measurements. In this section we present the results for this set of sources. Specifically we describe 
the distributions of the inferred source parameters in\S7.1, their radial variation in \S7.2 and their interdependence in \S7.3.

\begin{figure*}
\centering
\includegraphics[trim=0mm 0mm 0mm 0mm,width=100mm,angle=-90]{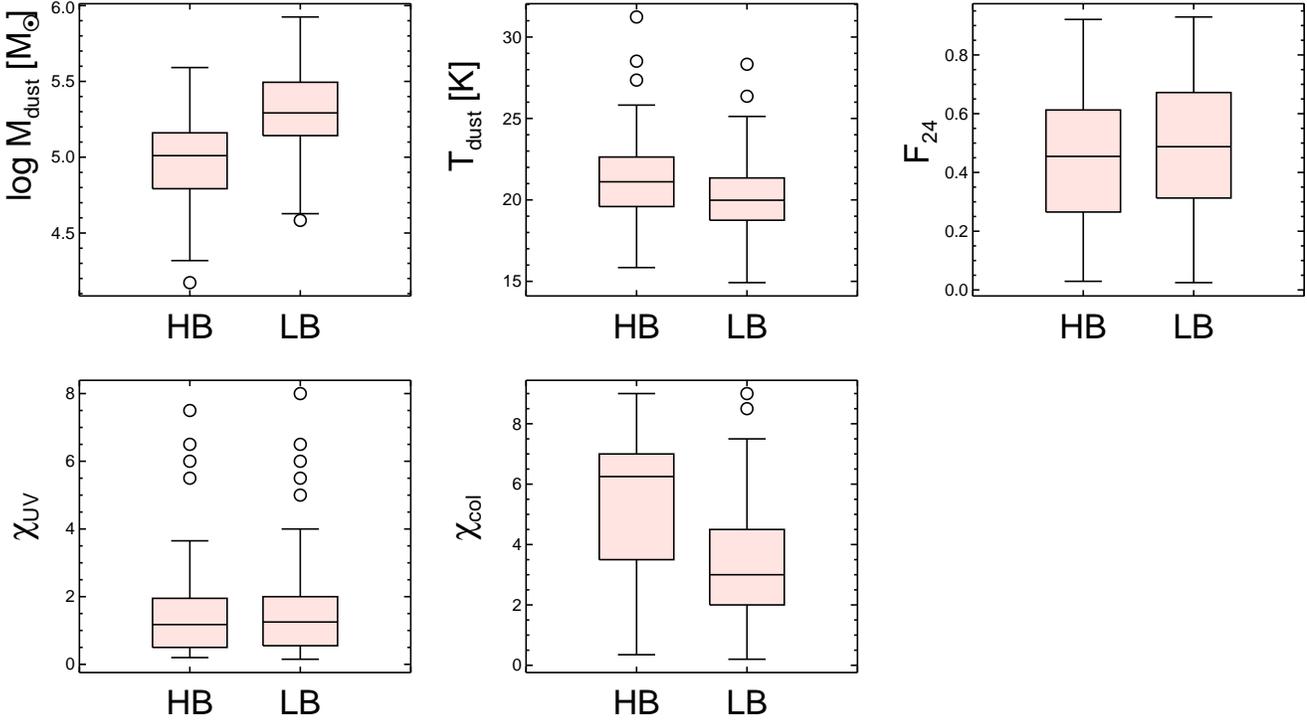}

\caption{Box-and-whisker plots showing the distribution of values for
  the compact source dust mass (top left), temperature (top middle),
  $F_{24}$ (top left)), $\chi_{\rm{UV}}$ (bottom right) and $\chi_{\rm{col}}$
  (bottom middle).  In each panel the source measurements with high
  background (left) and low background (right) subtractions are shown.  The solid horizontal line denotes the median value with the box delineating the 25th and 75th quartile range.  The lines extend to the maxima and minima with open circles marking outliers (see text). }
\label{bigbox}
\end{figure*}


\subsection{Inferred Source Parameter distributions}
In order to show the distributions of inferred parameters and compare them for the two types of source flux measurements, we created box-and-whisker plots
which are shown in Fig.~\ref{bigbox}, \ref{lumbox} and  \ref{bigbox_sfr_sfe}.  The solid line in the boxes shows the median value and the box delineates the 25th 
and 75th quartiles. The lines 
extend to either the maximum and minimum values or to 1.5 times the
75th and 25th quartiles.  If there are values beyond the later range,
they are denoted with open circles.
In the following we describe in detail the distributions for each set of parameters, that is, the SED fitting parameters, the source luminosities and the 
star formation rates and efficiencies.    

\subsubsection{SED fitting parameters}

As explained in \S5, the SED fitting has four free parameters $\chi_{\rm{UV}}$, $\chi_{\rm{col}}$, $M_{\rm{dust}}$ and $F_{24}$. The box-and-whisker plots of these 
parameters and of the dust temperature are shown in Fig.\ref{bigbox}. 

The $\chi_{\rm{UV}}$ values (see bottom left panel), representing the UV radiation energy density in the units of the Mathis et al.\ (1983) profile (hereafter, MMP profile), 
show a 
similar distribution for both types of source measurements with most values between 0.1 and 2. Therefore, the average UV radiation field energy density of 
the sources seems to be typically from a few tenth up to a factor 2 the intensity of the local Milky Way interstellar radiation field, 
as described by the MMP curve.    

The bottom right panel shows the distribution of $\chi_{\rm{col}}$. As explained in \S5, $\chi_{\rm{col}}$ expresses the ratio of the optical to UV radiation field 
energy density (relative to the standard MMP curve). Here the two measurements show some differences with the HB source measurement showing 
higher values and a greater range. This could mean that the average optical/UV ratio needed to fit the source emission is higher when one subtracts the 
higher \textsc{getsources} background. Alternatively, this could mean 
that the PAH abundance required for the HB fit is lower than for the LB fit (because of the $\chi_{\rm{col}}$-PAH abundance degeneracy described 
in \S5). In other words, a lower fraction of 8~$\mu$m emission relative to the FIR emission is removed by the background estimated in the interarm apertures.    

As explained in \S5, the cold dust temperature is not fit in our SED
fitting procedure but is determined by the values of both $\chi_{\rm{UV}}$ and $\chi_{\rm{col}}$. 
Both source measurements show similar temperature ranges (see top middle panel of Fig.~\ref{bigbox}), with the HB measurement only slightly higher 
(about 1 K higher).  

The source dust masses are shown in the upper left panel. In both cases, the compact sources have a quite narrow range in dust masses peaking close to  
10$^{5}$ M$_{\odot}$ and with most values between 10$^4$ and 10$^6$M$_\odot$. The small range of masses is likely due to our resolving power, which prevents us from detecting fainter sources which 
tend to also have smaller dust masses. The LB measurements for the sources have higher masses, because the background determined in the interarm aperture 
subtracts less flux than the \textsc{getsources} background. 
 
The distribution of $F_{24}$, the relative amount of 24~$\mu$m emission associated with PDR, is shown in the top right panel. Both source measurements show relatively 
similar values, mainly in the range 0.25-0.6, which suggests that a substantial part of the 24~$\mu$m emission is contributed by stochastically heated small 
grains in the diffuse dust component. However, we note that PDR SED dust emission spectra can vary substantially (Groves et al. 2008) and might not 
be accurately reproduced by our PDR template for each individual case.  

\subsubsection{Source Luminosities}
\begin{figure}
\centering
\includegraphics[trim=0mm 0mm 0mm 0mm,width=50mm,angle=-90]{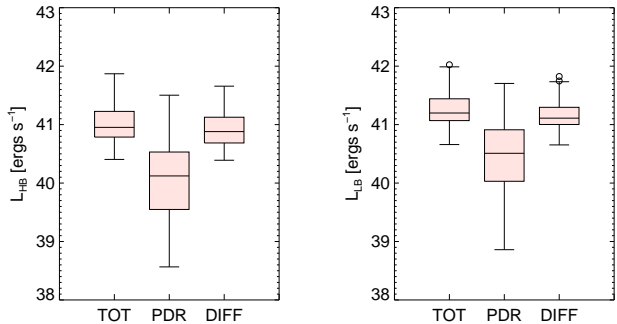}

\caption{Box-and-whisker plots showing the distribution of
  luminosities for the sources in the high background (left) and low
  background (right) subtractions.  We show the distributions of
  the total luminosity, PDR luminosity component and total diffuse
  component. See Fig.~\ref{bigbox} for details.}
\label{lumbox}
\end{figure}

The SED fitting procedure also provides us with a measure of the total infrared luminosity of the sources as well as the luminosities due to the PDR 
and the diffuse dust emission component.
Fig.~\ref{lumbox} shows the distribution of the inferred luminosity values for each component. The total luminosity values are found in a rather 
small interval around $10^{41}~\rm{ergs/s}$. This is likely due to the detection technique. 
We find that \textsc{getsources} does not detect sources below a minimum dust infrared luminosity of $\approx 0.5\times10^{41}~\rm{ergs/s}$.  Since there are not many sources with dust 
luminosities higher than a few times $10^{41}~\rm{ergs/s}$, the inferred dust luminosity range covered is rather small and this has important consequences
 on the inferred range of SED fitted parameters and SFR values (see discussion in \S8). Fig.~\ref{lumbox}  also shows that the diffuse dust emission component
dominates the dust luminosity. We point out that this does not
necessarily mean that the dust heating from star forming regions is not responsible for most of the dust 
emission, since the diffuse dust emission component can be powered by both radiation from older stellar populations and the fraction of UV photons escaping 
from PDRs. The origin of the dust heating will be further discussed in $\S8.3$.     

\subsubsection{Star formation rates and efficiencies}
\begin{figure}
\centering
\includegraphics[trim=0mm 0mm 0mm 0mm,width=45mm,angle=-90]{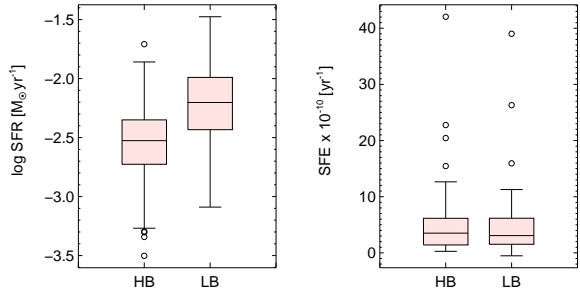}

\caption{Box-and-whisker plots showing the distribution of values for
  the SFR (left) and SFE (right).  In each panel the source
  measurements with high background (left) and low background (right)
  subtractions are shown.  See Fig.~\ref{bigbox} for details.}
\label{bigbox_sfr_sfe}
\end{figure}

Fig.~\ref{bigbox_sfr_sfe} shows the distribution of the SFR and SFE inferred values.  Due to the small areas considered here (i.e. average radius of sources 
is 150 pc), the SFRs are quite low. As outlined in \S6 if the SFRs are too low (i.e. below 0.001 M$_{\odot}$ yr$^{-1}$) then the calibration of H$\alpha$ and 24~$\mu$m fluxes may not adequately trace the SFR.  However, the majority of 
the sources have SFRs above this value and furthermore, we have restricted ourselves to regions associated with recent star formation.  
These regions should have a good correspondence between gas and star formation.   We note that the H$\alpha$ contribution represents half or slightly more than half of the SFR for the sources.  The median ratio of the H$\alpha$ luminosity, $L(\text{H}_{\alpha})$, to the scaled 24~$\mu$m luminosity, 0.031$L(\text{24} \mu\text{m}$), is 1.4.

 The SFEs of the source measurements have median values with median deviation values of 3.57 $\pm$ 3.45 $\times$ 10$^{-10}$ yr$^{-1}$ and 3.1 $\pm$ 3.1 $\times$ 10$^{-10}$ yr$^{-1}$ 
 in the case of the HB and LB measurement, respectively.  This is consistent with the  
findings of Leroy et al.\ (2008) who found that in terms of molecular gas, the SFE has a mean value of  
5.25 $\pm$ 2.5 $\times$ 10$^{-10}$ yr$^{-1}$.
  
   The LB and HB measurements show considerable differences between the inferred SFR and source gas masses, with the LB measurements having higher values in 
both cases.  However, the LB and HB measurements show roughly the same range of SFE values.  This means that roughly the same relative amount of background is being subtracted for both the SFR tracers and dust emission.  This suggests the amount of background subtraction does not affect the results of the SFE, provided the same technique is used for both the gas and SFR tracers.

\begin{figure}
\centering
\includegraphics[trim=0mm 0mm 0mm 0mm,width=70mm,angle=-90]{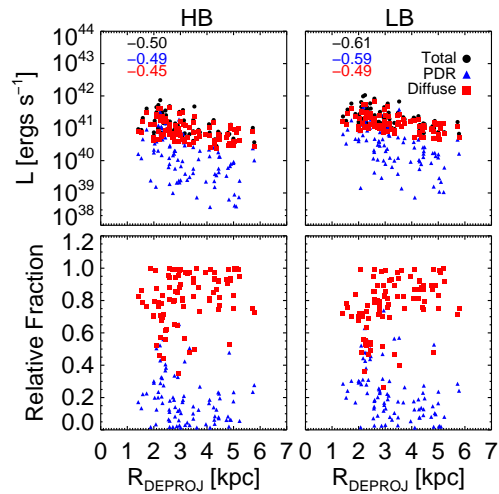}
\caption{The radial variation of the luminosity components including
  the total (black circles), PDR component (blue triangles) and
  diffuse component (red squares), for the compact sources with high
  background subtraction (left) and low background subtraction (right).   In the upper left corner we list the Spearman rank correlation coefficient for each component and radius.  The relative fraction of the PDR component to the total (blue triangles) and diffuse component to the total (red squares) are shown on the bottom panel for both cases. }
\label{lumrad}
\end{figure}


\subsection{Radial Variations}

We now turn to the radial variations of the SED fitting parameters, the SFR and SFE and of the dust luminosities. 
 Previous works which have 
used azimuthal averages (e.g. Mu\~noz-Mateos et al.\ 2009, Pohlen et al.\ 2010, Engelbracht et al.\ 2010, Boquien
et al.\ 2011) have found that the parameters
describing the dust, 
including temperature and surface mass density, and the SFR  tend to decrease with radius particularly for late-type spirals. These studies have averaged over both the compact regions and the 
more diffuse emission. By separating the compact regions, we can better disentangle how and if the radial position affects the dust properties 
of these sources.
 
In order to determine the deprojected radial position, we assume an inclination and position angle for M83 of 24$^{\circ}$ and 225$^{\circ}$ respectively  (Tilanus \& Allen 1993).  
\begin{figure}
\centering
\begin{tabular}{cc}
\includegraphics[trim=0mm 0mm 0mm 70mm,width=80mm,angle=-90]{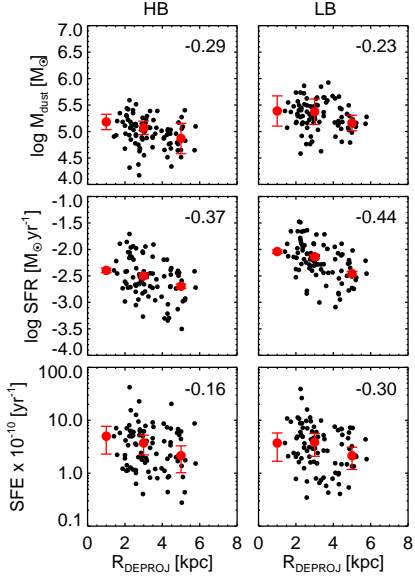}&

\end{tabular}

\caption{The radial variation of the mass, SFR and SFE for the two compact source measures.  The upper right hand corner lists the Spearman rank correlation coefficient.  The error bars in red denote the median uncertainties of the points in three bins between 0-2 kpc, 2-4 kpc and 4-6 kpc.}
\label{rad1}
\end{figure}


\begin{figure}
\centering
\includegraphics[trim= 0mm 0mm 0mm 20mm,width=80mm,angle=-90]{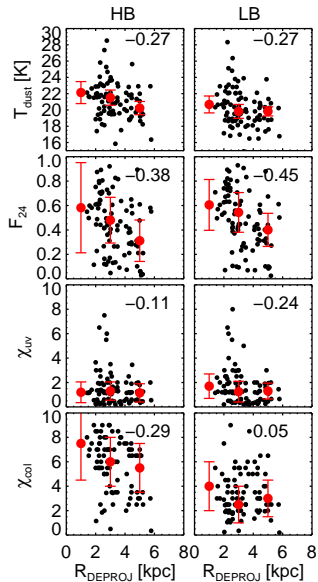}
\caption{The radial variation of $\chi_{\rm{col}}$, $\chi_{\rm{UV}}$, $F_{24}$
  and temperature for the two compact source measures.  The upper right hand corner lists the Spearman rank correlation coefficient.  The error bars in red denote the median uncertainties of the individual measurements in three bins between 0-2 kpc, 2-4 kpc and 4-6 kpc.}
\label{rad2}
\end{figure}


\subsubsection{Source infrared luminosities}
 Fig.~\ref{lumrad} shows how the luminosity of the sources (upper panels) and the relative fraction of PDR and diffuse dust emission luminosity 
(lower panels) 
varies with radius. In the upper left corner we list the Spearman rank correlation coefficient for the correlation between the luminosity components and 
radius.   Values of the correlation coefficient approaching $+$1 or $-$1 reflect stronger correlations or anti-correlations respectively.  Values close or 
less than 0.5 represent mild or weak correlations.  In both the HB and LB measurements, we find only a mild correlation between the source 
luminosities and radius.  The source luminosities show a slight decrease with radial position with scatter.  
We also note that the regions with the highest PDR luminosity fraction are found in the inner regions, at a radial distance of 2-3~kpc 
from the galaxy centre (roughly the end of the bar). However, the diffuse dust component dominates the emission for all the sources, except in a few cases.   

\subsubsection{Radial Variations of the SED Fitting Parameters, SFR and SFE}
Fig.~\ref{rad1} and \ref{rad2} show the SED fitting parameters and the SFR and SFE values plotted versus the source deprojected radial positions for both the flux 
measurements of the compact sources. The median uncertainties of the individual measurements are shown by the red error bars and the Spearman rank coefficient is listed in the upper right of each panel.

For both the HB and LB measurements, we find little to no variations with radius for any of the SED parameters including dust mass, temperature, 
F$_{24}$, $\chi_{\rm{UV}}$, $\chi_{\rm{col}}$.  There is a slight trend showing that the SFR decreases with radius, but this is far less than what is typically 
seen in studies on  late-type spiral galaxies that employ azimuthal averages.  It seems that the properties of the compact regions are quite uniform 
and do not vary much with 
location in the galaxy.  This is perhaps not too surprising, since our source sample covers only a relatively small range of dust luminosities, 
as shown in \S7.1. Therefore, the sources we consider likely represent similar types of objects. Although the low spatial 
resolution did not allow us to detect a larger range of source luminosities, it is striking that the detectable bright sources lie almost exclusively on 
the spiral arms or at the ends of the bar in M83, which means that the local environment may in fact be quite similar as well.  
This suggests that azimuthal averages of dust properties could well mask important differences in environment within an azimuthal bin.  The radial decreases 
that are typically  seen in other galaxies and for M83 in F12 using azimuthal averages might simply reflect the increasing contribution of the interarm regions at larger radii.  
We note, however, that the SFE is quite constant with radius.  This finding is consistent with studies that use azimuthal averages and pixel-by-pixel analyses 
(e.g.\ Leroy et al.\ 2008, Blanc et al.\ 2009, Foyle et al.\ 2010, Bigiel et al.\ 2011).

\subsection{Correlations between source properties}

\begin{figure}
\centering

\includegraphics[trim=0mm 0mm 0mm 50mm,width=100mm,angle=-90]{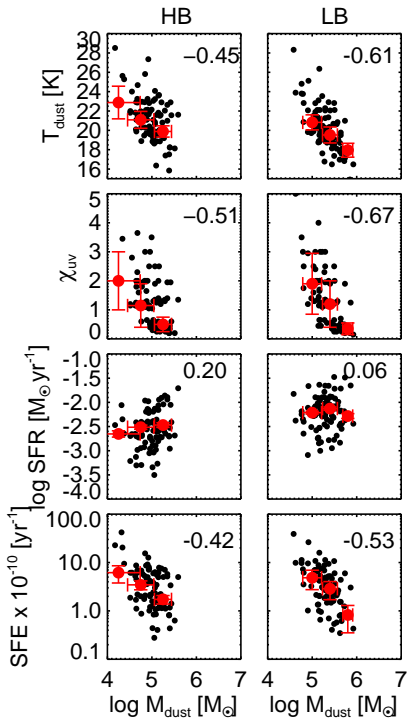} 

\caption{Correlation plots showing dust temperature, $\chi_{\rm{UV}}$, SFR and SFE versus dust mass for the HB (left) and LB source flux measurements (right).  
The Spearman rank coefficient is shown in the upper right of each panel.  The red uncertainties show the median uncertainties in bins of 10$^{0.4}$ M$_{\odot}$. }
\label{masscorr}
\end{figure}

\begin{figure}
\centering

\includegraphics[trim=0mm 0mm 0mm 100mm,width=100mm,angle=-90]{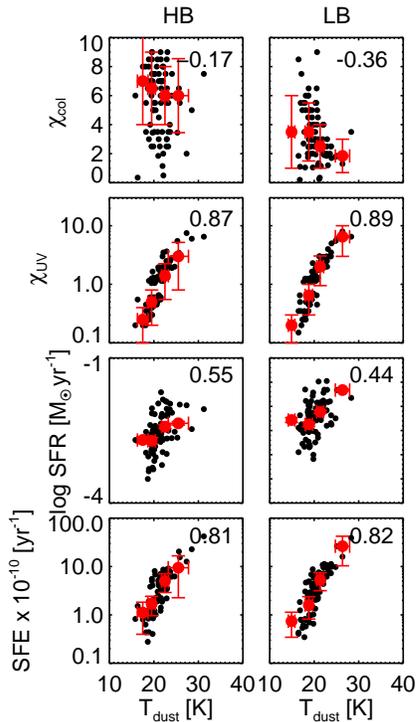}

\caption{Correlation plots showing $\chi_{\rm{col}}$, $\chi_{\rm{UV}}$, the SFR and SFE versus dust temperature for the HB (left) and LB source flux measurement (right).
  The Spearman rank coefficient is shown in the upper right of each panel.  The red uncertainties show the median uncertainties in bins of 2 K. }
\label{tempcorr}
\end{figure}

Having seen how the dust and star formation parameters of the compact sources vary with radius, we now turn to how they are inter-related and what 
types of correlations may exist between some of these parameters. In particular, we examine how the source dust mass and temperature are related 
to the radiation field parameters $\chi_{\text UV}$ and $\chi_{\text col}$ and with the star formation rate and efficiency. 
Fig.~\ref{masscorr} 
shows the dust temperature, $\chi_{\text UV}$, SFR and SFE plotted versus the dust mass, while Fig.~\ref{tempcorr} shows $\chi_{\text UV}$, $\chi_{\text col}$, SFR and SFE values plotted against the dust temperature$^{6}$\footnote[6]{$^{6}$In Appendix C we present how these parameters vary as a function of colour.}.  As before, the red bars show the median uncertainties in equally spaced bins
and the Spearman correlation coefficients are shown in the upper-left of each panel. In general, we find that both LB and HB measurements of the compact sources 
present quite similar results. 

We observe a strong 
correlation of $\chi_{\text UV}$ with the dust temperature, while the optical-UV intensity ratio $\chi_{\text col}$ is not as strongly correlated to the dust temperature measured for the sources. These findings reflect the fact the UV radiation field intensity 
is the fundamental parameter that determines the average dust temperature, while the optical-UV intensity ratio varies randomly among sources 
having same average dust temperatures. We point out that this does not mean that optical photons are not an important source of dust heating. 
Actually, the optical part of the radiation field accounts for more than 50\% of the dust heating for the diffuse dust SED component when 
$\chi_{\text col} > 1$.

 The upper left panel of Fig.~\ref{masscorr} reveals a mild anti-correlation between the dust temperature and mass for the sources.  The typical error bars are much smaller than the inferred range of dust mass and dust temperature.  Thus, 
this anti-correlation cannot be explained by the uncertain determination of dust temperature and mass, which are connected in the SED fitting since 
approximately $L_{\text dust}\propto M_{\text dust}T_{\text dust}^{4+\beta}$.  Therefore, we are confident that the sources with higher 
dust masses tend to have lower dust temperatures in the sample we considered. Given the strong correlation between $\chi_{\text UV}$ and temperature, 
we also find an anti-correlation between $\chi_{\text UV}$ and dust mass in the second row from the top of  Fig.~\ref{masscorr}. 
We will discuss the significance of the dust mass - dust temperature anti-correlation in \S8.3. 

Fig.~\ref{masscorr} (third row, left) shows that for both measures of the compact sources the dust mass is only weakly correlated with the SFR.  
If the dust traces gas, this plot should be similar to the S-K relation (Kennicutt 1998), except that typically surface densities are 
plotted rather than total values. The S-K relation for nearby galaxies typically shows a tight correlation between gas surface densities 
and the SFR surface densities, at least when the scales considered are large enough (e.g.\ Schruba et al.\ 2010, Bigiel et al.\ 2008, see other references in \S 2.1). 
The relatively constant SFR with mass suggests that more massive sources are less efficient at forming stars and will be further discuss in \S8.2.
We also find that for both measures of the compact sources, the SFR is only mildly correlated with the dust temperature (see third row of Fig.~\ref{tempcorr}).
This is not unexpected since, as explained in Sect. \S2.2, the SFR is connected to the dust temperature through the dust mass and the total luminosity-weighted 
optical-depth.    

In contrast to the lack of correlations for the SFR, we find a mild and strong correlation respectively of dust mass and dust 
temperature with the SFE.  
While the SFE was roughly constant with radius, 
we find that the SFE is anti-correlated with dust mass and correlated with dust temperature. These findings have not been seen before, as most studies 
have found that the SFE does not vary much with other properties in terms of pixel-by-pixel and azimuthal averages (i.e.\ Leroy et al.\ 2008). This will be further discussed in \S8.2 and \S8.3.  We note that that the anti-correlation between the SFE and mass is primarily due to the fact that the SFR does not vary much with the inferred gas mass.  Thus, the correlated axes produce the anti-correlation.  To check this, we performed Monte-Carlo simulation assuming the SFR was roughly constant with a Gaussian distribution and a mean and standard deviation equivalent to that found in our measurements.  We assumed a similar distribution for the simulated source masses.  The simulation naturally produced a linear plot of SFE versus mass with a slope of -1.05 $\pm$ 0.15.  The compact sources in our analysis produced a relation with a slope of -0.71 $\pm$ 0.14. The slopes while not identical within the uncertainties, are similar, suggesting that the correlated axes are the primary reason for the relation.

\begin{figure}
\centering
\includegraphics[trim=20mm 20mm 0mm 20mm,width=40mm,angle=-90]{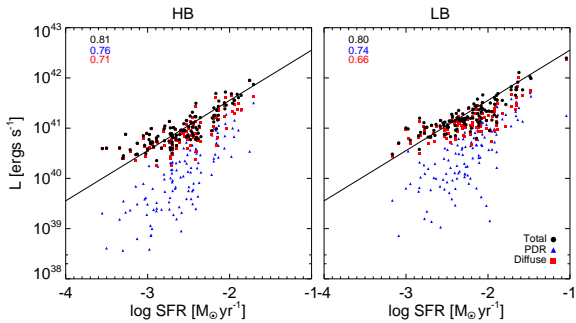}
\caption{Dust luminosity versus the SFR for both measures of compact
  sources including the total luminosity (black circles), PDR
  component (blue triangles) and diffuse component (red squares).  The
correlation coefficients for each component are shown in the upper
right. The solid line denotes the relation of
SFR=2.8$\times$10$^{-44}\times$ $L_{\text{dust}}$ (Calzetti 2012), which assumes that all the young stellar luminosity is absorbed and re-emitted by dust.}
\label{lumsfr}
\end{figure}

We also examined the relation between dust luminosity and the SFR, as shown in Fig.\ref{lumsfr}. We see that not only is the total dust luminosity 
but also the PDR and diffuse dust emission luminosities are well-correlated with the SFR. As explained in section $\S2.2$, this is what is expected
 if a young stellar population is responsible for the dust heating and the total luminosity-weighted opacity of the sources, $\tau$, does not vary arbitrarily among the sources.   

\section{Discussion}
In this section we discuss the properties of the detected compact sources of M83 in terms of gas masses and locations (\S8.1),  star formation (\S8.2) 
and dust heating (\S8.3). 

\subsection{Source Gas Masses and Locations} 
The inferred dust masses of the compact sources lie within the range of 10$^{4}$-10$^{6}$ M$_{\odot}$.  These dust masses can be used to estimate the 
source gas masses as discussed in \S6.  We use a gas-to-dust ratio of 100, which means the gas masses of the sources have values in the range 
10$^{6}$-10$^{8}$ M$_{\odot}$. These values correspond to the high end of the molecular cloud mass distribution function derived both observationally (see e.g.\ Solomon et al.\ 1987, Rosolowski et al.\ 2005, Gratier et al.\ 2012) and by numerical simulations (e.g.  Nimori et al.\ 2012). Furthermore, previous CO studies of nearby galaxies have highlighted the presence of very massive clouds with gas masses of order of 10$^{7}$ M$_{\odot}$ or higher, which are usually referred to as Giant Molecular Associations (GMA) (e.g. Vogel et al.\ 1988, Koda et al.\ 2009, Muraoka et al.\ 2009).  The high masses of our compact sources suggests that they are GMAs rather than less massive GMCs. 

However, because dust continuum emission does not provide kinematical information, it is not possible to check whether the detected sources are gravitationally bound clouds.  The smallest cloud radius we can measure is 130 pc due the 70~$\mu$m beam size (6$''$).  The majority of the detected sources are found on the spiral arms.  Studies of the molecular gas in M83 have shown that most of the bright compact 
sources on the spiral arms are bound clouds in GMAs.  Muraoka et al.\ (2009) examined CO in M83 and measured the virial parameter, $\alpha$, defined as the ratio of the virial mass to the CO luminosity mass.  They found that $\alpha$ is almost equal to unity in on-arm clouds, suggesting that they are in a gravitationally bound state (see also Rand et al.\ 1999, Lundgren et al.\ 2004).  The resolution of their study was 7$''$.5, which is comparable to our resolution.  Thus, it is quite likely that our sources are bound GMAs.

The spiral arms are the natural location where ISM gas is brought to higher densities, thus leading to the formation of massive clouds. 
Therefore, it is not surprising, given our resolution and sensitivity, that we only detect compact sources on the spiral structure.   We note that the bar-spiral arm 
transition region harboured the most sources and that just few were detected along the length of the bar.  Due to shear motions in bars, GMCs can be 
easily pulled apart (Downes et al.\ 1996).  In contrast to spiral arms, bars exhibit lower star formation rates and in some cases lower SFE 
(Momose et al.\ 2010).  Meanwhile, studies of atomic and molecular gas have shown that the most massive complexes are found in the bar-spiral arm 
transition region. This can be explained by orbit crowding when gas on highly elliptical orbits in the bar converges with the gas orbiting in the spiral 
structure (e.g.\ Kenney \& Lord 1991).  The location of the compact sources detected in the FIR is consistent with these findings in the gas.
 
\subsection{Star Formation Rates and Efficiencies}

One of the goals of this work has been to infer cloud gas masses using the dust emission of compact sources. Combined with a measure of the SFR, we can then examine the SFE in these regions.

There have been many recent studies that have examined star formation on spatially-resolved scales. Typically these studies have employed one of 
three methods:  pixel-by-pixel analyses, where the SFR tracers and gas tracers are compared in individual pixels; azimuthal average analyses, where average 
values in radial bins are compared; or aperture photometry on compact sources bright in some SFR tracer such as H$\alpha$ emission. 

Our work differs in several key ways and, before discussing the results, we summarize these differences here:
1) We use the dust emission to detect the compact sources.  While most studies have focused on aperture photometry of active star forming regions 
(e.g.\ Calzetti et al.\ 2005, Kennicutt et al.\ 2007, etc.), few have selected compact sources as seen in dust emission maps, which may be more akin to 
locating peaks in the gas (e.g. Schruba et al.\ 2010).   However, there is a good correspondence between the detected FIR sources and star formation regions, which is demonstrated by the tight correlation between total infrared luminosities of the sources and their SFRs (see Fig. 13); 2) 
we perform a diffuse background subtraction of the dust emission using both a local background defined by \textsc{getsources} and one determined from 
nearby apertures in the interarm regions. Previous studies have performed a background subtraction on the star formation tracers but not on the gas mass 
(though see Rahman et al.\ 2011); 3) it is important to keep in mind, when comparing this work with others, that we analysed the total SFR and 
dust masses for individual sources while most studies consider the surface densities of the gas mass and SFR when examining trends between the two. The reason why we did not 
consider surface densities is that it is not clear what the source area should be, since the apparent size of the sources 
varies on each map depending on the map resolution. In addition, since most of the sources are not well-resolved, it is not possible to determine the 
physical source area which is typically smaller than the PSF beam area.   

As discussed in \S2,  previous works have found that, on relatively large spatial scales ($>$ 500 pc) clear correlations are found between the gas, SFR 
and dust as well as a tendency for these parameters to decline with radius.  However, as we saw in \S7, the compact regions detected in this study show a relatively constant 
SFR and dust mass with radius, albeit with large scatter and there is little to no correlation between the SFR and dust mass.  Due to the small spatial 
scales of these sources ($<$ 300 pc), one does not expect to recover the relations found by averaging over much larger areas.  Recently, it has become clear 
that on small spatial scales (less than 400 pc ) the S-K relation between the gas surface density and SFR surface density breaks down (Calzetti et al.\ 2012, Feldmann \& Gnedin 2011 and Schruba et al.\ 2010). This has been found to be even more prominent in the case where 
gas peaks are selected (Schruba et al.\ 2010). The scatter is due to the fact that individual GMCs and HII regions will show varying gas mass to star 
formation rates depending on their evolutionary state. Averaging over large areas means that many objects in different states are averaged and the 
relation between SFR and gas mass is recovered.

The scales on which the SFR and gas mass surface density are measured play an important role in defining the relation (Liu et al.\ 2011).  Typical SFR tracers suitable for relatively large galaxy regions are not straightforwardly applicable to small regions of sub-kpc scales. In fact, in this case, the intrinsic assumption that the stochastic characteristics of star formation are averaged out from the integration on large areas can easily break down (i.e.\ Calzetti et al.\ 2012). This happens because, one ideally needs a complete sampling of the stellar IMF.  Small regions may not encompass a large enough stellar population to do so.  
Furthermore, beyond the issue of sampling the stellar IMF, there is the problem of time-averaging.  The SFR is calibrated based on the assumption of a constant SFR over 100 Myr (Calzetti et al.\ 2007).  While this is true for entire galaxies, on small scales this assumption may no longer hold and can create a large scatter in the relation.

Our sources have sizes of roughly 300 pc and thus, we find large scatter between the inferred gas mass and the SFR. We also note that our sources populate 
only a narrow range in FIR luminosity and dust mass (see Fig.\ref{bigbox} and \ref{lumbox}).      
Thus, we see a relatively constant SFR for sources with such similar characteristics.  If we could populate the plot with lower mass regions, 
an S-K relation with considerable scatter might be recovered.  Our results reflect stochasticity and the fact that individual star forming regions can exhibit a range of SFRs for a given dust or gas mass.  However, we should note a few other caveats that might introduce scatter into the relation.

First, it is possible that dust emission does a poor job of tracing molecular gas especially in light of our use of a constant GDR. There may be large variations in the GDR (i.e. due to metallicity variations), which means that our gas mass estimates might not be reliable.   However, the compact regions all lie on the spiral arms and beyond the nuclear region of the galaxy.  In F12, the largest variations in the GDR were seen in the central regions and, beyond, the GDR was quite constant.   It is also possible that some of the peaks found in the dust emission are not associated with active star forming regions.  However, we found a correlation between the dust luminosity and the SFR and studies like Verley et al.\ (2010) have found a tight correlation between compact regions in M33 and star formation rate tracers.  We should also note that all of the sources were detected in the 70$\mu$m map, which is also known as a good tracer of the SFR (Li et al.\ 2010).

A second possibility is that our background emission subtraction has introduced scatter. Rahman et al.\  (2011) showed that as one increases the amount of diffuse emission subtracted from the gas, the scatter in the S-K relation increases. However, particularly given that we are using dust emission, which may trace gas of different phases, it is important to account for a diffuse component.

Despite the large scatter, we find that the SFR varies little with the inferred gas mass and is roughly constant.  This implies that the more massive clouds are less efficient at forming stars.   A physical reason for this is that the more massive clumps might be more extended and thus have lower gas surface densities. Studies of GMCs have revealed that only the densest gas is directly associated with star formation.  In this way, the more massive and thus extended regions may have roughly comparable SFRs (or lower SFEs).   We note that we also found that there is a strong anti-correlation between the SFE and the inferred gas mass.  However, as discussed in \S 7.3, this is is largely a product of the correlated axis since mass appears both in the y and x-axes (see Murray (2011) who examined galactic GMCs and found an anti-correlation which was attributed to a similar effect).    We also note that we find no radial variation of the SFE of the compact regions. Previous works,  which have examined kpc-sized regions, have found that the SFE is relatively constant regardless of the variable considered (i.e. Leroy et al.\ 2008). This is consistent with our findings with radius.   

\subsection{Dust Heating}
We showed in \S2.2 that the simultaneous presence of a correlation between the 
SFR and source dust luminosity as well as a correlation between the SFE and the dust temperature suggests that the radiation impinging on the dust 
mass is mainly produced by a local, young stellar population.   
  Contrary to common lore, a strong correlation between SFR and dust temperature should not be necessarily expected in this case.   
Indeed we saw in \S7 (see Fig.~\ref{lumsfr} and \ref{tempcorr} ) that while the dust temperature was only mildly correlated with SFR, it was strongly 
correlated with the SFE.  We also saw that the dust luminosity was correlated with the SFR (see Fig.~\ref{lumsfr}).  This supports a scenario where a young, 
local stellar population is powering the dust emission of the FIR bright sources of M83. 

In Fig.~\ref{lumsfr}  the solid line illustrates the maximum level of dust luminosity that can be powered by the young stellar population by assuming that all
the luminosity from this population is absorbed and re-emitted by dust (optically thick case) using the following relation by Calzetti (2012):
\begin{equation}
SFR=2.8\times10^{-44}L_{\text dust}
\label{sfr_ltir}
\end{equation}
 This relation has been derived using the same assumptions on the star formation 
history, metallicity and IMF as the H$\alpha$ based calibration we used to infer SFR of the sources.  

This line should delineate the maximum value, yet we find many sources lie close to or above this limit.  Specifically we find that the percentage of 
sources that lie above or only within 2$\sigma$ below this limit is 60\% or 20\% in the HB or LB case respectively.  
While it is surprising to find sources that lie above or close to this maximum limit, these findings can be explained in one of two ways.  

Firstly, it is possible that some regions are indeed very close to being completely optically thick and
therefore the total dust luminosity will be approximately equal to the intrinsic young stellar population luminosity in those cases. However, this is 
unlikely because these regions are also H$\alpha$ emitters, which means, by definition, that at least some fraction of the young stellar population 
luminosity is able to escape
unabsorbed (at least through re-emission in H$\alpha$). Furthermore, we checked that the differences between the observed dust luminosities and the dust 
luminosities for an optically thick case, predicted by the relation in Eq.~\ref{sfr_ltir}, do not depend on the ratio F(H$\alpha$)/SFR[H$\alpha-24\mu m$].  
This suggests that the scatter of the observed points is not driven by differences in attenuation.   

Secondly, it is possible that there is a local older stellar population or stellar populations outside the projected source area contributing 
substantially to the heating. In this
case, the approximate equality between dust and young stellar population luminosity would be reached through this extra-heating from other radiation sources.
Thus, the infrared total luminosity could still be used as a star formation rate indicator in those cases despite the fact that the dust heating
is not powered exclusively by a local, young stellar population  (see Bendo et al.\ 2012 also). This might explain why the recent work of Verley et al.\ (2010) and Boquien et al.\ (2010) 
showed that the single {\it Herschel} 100, 160 and 250 $\mu$m band luminosities of individual sources in M33 are good tracers of SFR, since most of the dust 
luminosity is emitted at FIR wavelengths.     

We should note that the SFE vs $T_{\rm{dust}}$ correlation is at least partially driven by the $M_{\rm{dust}}$ vs $T_{\rm{dust}}$ anti-correlation 
(more evident for the LB measurement), since the SFR alone does not
correlate with $M_{\rm{dust}}$ and only mildly with $T_{\rm{dust}}$. There are multiple potential explanations for an anti-correlation between 
mass and temperature.  
One possible explanation is that it is due to the different
evolutionary stages of the sources, with the more massive sources having a relatively lower number of young, recently formed stars.  
In this way, the radiation field would be less intense for massive sources and the dust would be cooler. Alternatively, it could be that the more massive 
sources are more efficient at shielding radiation originating from outside the local region. As outlined in the previous paragraph, there is likely some 
heating due to other stellar populations other than just the local, young population, and thus such shielding could keep more massive sources cooler.
  
\begin{figure}
\centering
\includegraphics[trim=0mm 0mm 0mm 0mm,width=80mm]{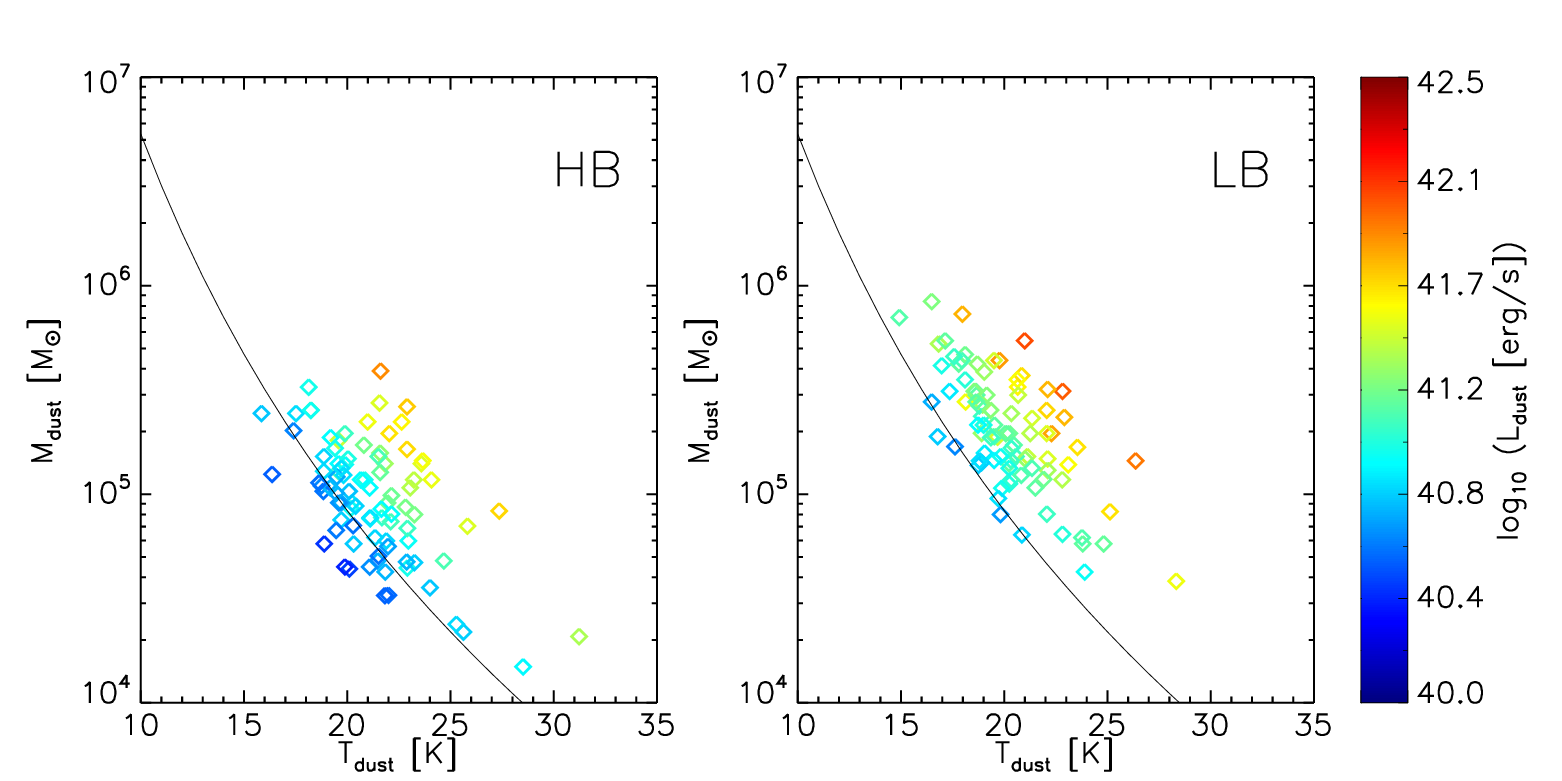}
\caption{Source dust masses versus dust temperature for the HB (left) and LB (right) measurements.  The points are colour-coded based on their total dust luminosity. The solid line represents the relation $L_{\text dust}$ $\propto$ $M_{\text dust}$ $T_{\text dust}$$^{6}$ for $L_{\text dust}$=6$\times$10$^{40}$ergs/s.}
\label{mass_vs_temp}
\end{figure}

The anti-correlation between dust temperature and mass could be in part due to the lower limit in dust luminosity for the sources.  
Fig.\ref{mass_vs_temp} shows the inferred dust masses versus dust temperatures with the plotted points colour coded depending 
on the source total dust luminosity. Since most of the dust luminosity is emitted at FIR wavelengths where $L_{\rm{dust}} \propto M_{\rm{dust}}T_{\rm{dust}}^{4+\beta}$, 
the points correspond to higher total dust luminosities while moving from the bottom-left to the upper-right region of the plot. From this figure one 
can see that there is a lower limit in luminosity ($\approx 5-6\times10^{40}$erg/s), determined by the minimum fluxes for source detection. 
This explains why there are no sources populating the lower-left part of the diagram. The steepness of the observed anti-correlation is consistent with 
the $4+\beta$ exponent (with $\beta$=2), 
as shown by the line in Fig.~\ref{mass_vs_temp}.  However, higher mass sources do not populate uniformly the entire range of temperatures in the 
upper-right part of the diagram, where source detection would easily be possible. Therefore,  while the low-luminosity limit may contribute to the 
anti-correlation between dust temperature and mass, it cannot fully account for it.  Thus, there should be at least some physical restrictions at play.
 
We note that the lower limit to the luminosity 
range explored in our analysis does introduce a bias that should be taken into account. Essentially, our sources are the brightest FIR compact sources in the 
disk of M83 and they are associated with the high-end of the molecular cloud mass distribution. As discussed in \S8.1, the sources are associated with gas 
masses in the range $\approx$10$^{6}$-10$^{8}$~M$_{\odot}$, while their total dust luminosities are in the range $\approx$0.5-5$\times$10$^{41}$erg/s. 
The results of this work should be considered valid only within these parameter ranges. 

We found
a mild correlation between SFR and $T_{\rm{dust}}$. However, as shown in Appendix C, a better correlation is found when plotting the inferred 
$70/160~\rm{\mu m}$ flux ratio versus the SFR. The difference between the two cases is due to the fact that $T_{\rm{dust}}$ in our SED fitting 
procedure refers to the cold dust temperature of the diffuse dust component.  Meanwhile the observed 70 and 160$\mu$m fluxes have contributions from both the diffuse component and the PDR component.   Nonetheless, in Appendix C we also show that the correlation between SFE and $70/160~\rm{\mu m}$ flux ratio is clearly tighter 
and this is consistent with the arguments presented in \S2.2 and very similar to the strong correlation found by plotting SFE versus $T_{\rm{dust}}$.           

Pixel-by-pixel analyses have tended to find varying results with no strong consensus 
for a correlation between SFR and modified black body dust temperatures or color temperatures. We should note that the pixel-by-pixel analyses have 
considered the star formation rate surface density (hereafter SFRSD) as 
opposed to the SFR used here. For M83, F12
found a correlation between the SFRSD and $T_{\rm{dust}}$. A similar result for M83 has been found by Bendo et al.\ (2012)
when comparing the $70/160~\rm{\mu m}$ flux ratio with the observed H$\alpha$
brightness, uncorrected for internal dust attenuation. However, Bendo et al.\ (2012) found a weaker correlation for M81 and NGC 2403.$^{7}$\footnote[7]{$^{7}$The weaker correlation is attributed to the dust heating by bulge stars in
M81 and artefacts in the images of NGC 2403; see the
discussion in that paper.} 
Boquien et al.\ (2011) found a mild correlation between SFRSD
and the $70/160\rm{\mu m}$ flux ratio for M33. In addition, Smith et al.\ (2012) for the Andromeda galaxy and
Skibba et al.\ (2012) for the Large and Small Magellanic Clouds found no or only a weak correlation between SFRSD and $T_{\rm{dust}}$. Therefore,        
it seems that in general the correlation between SFR vs tracers of dust temperature is not observed to be very strong when one considers either 
total SFRs for individual FIR sources or SFR averages on areas equivalent to pixel sizes. As explained in \S2.2, this is expected even for the case where 
young stellar populations are responsible for the dust heating because the dust temperature also depends on the dust mass and the total luminosity-weighted 
optical depth (see Eq.~\ref{lstarequ}), which are not uniform throughout a galaxy disk.

As we have seen, while isolating FIR bright sources has highlighted the presence of interesting correlations between the inferred source parameters, 
the narrow range of luminosities of the sources has introduced a potentially important bias. In order to elucidate the influence of this bias on the 
inferred relationships, we need to expand this work to include sources within a 
wider range of luminosities, particularly fainter sources. To do so, we plan a follow-up study which will include other spiral galaxies that are even 
closer (i.e. NGC 2403 and M33). The closer proximity will allow for an even better spatial resolution, which is necessary to detect sources in a wider range 
of luminosities and determine if the results for the GMAs in M83 are also valid for GMCs in other galaxies.

\section{Conclusions}

The main aim of this paper has been the investigation of the star formation and dust heating properties of the compact FIR bright sources as observed 
on the {\it Herschel} maps of the nearby spiral galaxy M83. By combining the source detection and photometry algorithm \textsc{getsources}, the dust emission SED fitting method of Natale et al.\ (2010)
and the [H$\alpha$ - 24~$\mu$m] star formation rate calibration by Calzetti et al.\ (2007), we have developed a new procedure to determine gas masses, radiation field intensities, 
cold dust temperatures, dust luminosities, star formation rates and star formation efficiencies associated with those sources.\\

The main results of our analysis are the following:
\begin{itemize}
\item We have found that the well-detected compact FIR sources are mostly associated with giant molecular associations, with gas masses in the range 
10$^{6}$-10$^{8}$ M$_{\odot}$ and dust total infrared luminosities in the range 0.5$\times 10^{41}$-$10^{42}$ergs/s. The majority of the sources are 
located on the spiral arms of M83 with only few sources found in the interarm and within the bar in the central region of M83.
\item None of the inferred physical quantities for the sources shows a strong variation with radius, including SFR, gas masses and dust temperature. 
Previous studies have usually found radial variation for these quantities, although only after averaging on larger areas including interarm regions and 
without subtracting any local background.
\item The SFR does not seem to correlate strongly with the gas mass of the sources. The lack of correlation is most likely due to the small spatial 
scales considered ($\approx$200-300 pc) and/or the relatively small range of inferred gas masses.
\item The star formation efficiency SFE, defined as SFR/M$_{\text{gas}}$, 
shows an anti-correlation with source gas mass. This finding suggests that the more massive GMAs are less efficient 
in forming stars in the last few Myr.  However, we note that this anti-correlation is a consequence of the roughly constant SFR with inferred gas mass.   
\item We found that the SFR correlates well with total dust luminosity, which is consistent with a scenario where dust is predominantly heated by the local 
young stellar population. However, between 20-60\% of the sources show dust luminosities which are greater than or only 2$\sigma$ below those predicted when 
 the heating is due only to a local young stellar population, embedded in a optically thick dust distribution. It is unlikely that the 
sources are completely optically thick, as they are also bright in H$\alpha$.  Thus, it seems that there must be some extra-heating by either a local older 
stellar population or by an external radiation field.  
\item We found a correlation between the SFE and $T_{\rm{dust}}$ which is tighter than the mild correlation we found between SFR and $T_{\rm{dust}}$. 
 This is expected if the dust is heated primarily by recent star formation. 
\item We found a mild anti-correlation between dust and mass temperature. While our sources have a low-luminosity limit, which may contribute to this 
anti-correlation, we find that this can not fully account for it.  Thus, we speculate that the more massive sources are more efficient at shielding from an 
impinging radiation field or that more massive sources are in an earlier stage of star formation. 
\end{itemize}

We plan to use the same procedure presented in this pilot work on a set of nearby galaxies observed by {\it Herschel}. This will help to further clarify 
the origin of the observed correlations and their implications for star formation and dust heating associated with FIR bright sources on galactic scales.
\\
\\
We are grateful to the anonymous referee for their careful reading of this work and their comments.  K. Foyle acknowledges helpful conversations with S.\ J.\ Kiss, V. K\"onyves, P.\ G.\ Martin, A.\ Men'shchikov, M.\ Reid and R.\ Skibba.  This research was supported by grants from the Canadian Space Agency and the Natural Science and Engineering Research Council of Canada (PI: C. D. Wilson).  MPS has been funded by the Agenzia Spaziale Italiana (ASI) under contract I/005/11/0.  PACS has been developed by
a consortium of institutes led by MPE (Germany) and including
UVIE (Austria); KU Leuven, CSL, IMEC (Belgium); CEA, LAM
(France); MPIA (Germany); INAF-IFSI/OAA/OAP/OAT, LENS,
SISSA (Italy); IAC (Spain). This development has been supported
by the funding agencies BMVIT (Austria), ESA-PRODEX (Belgium), CEA/CNES (France), DLR (Germany), ASI/INAF (Italy),
and CICYT/MCYT (Spain). SPIRE has been developed by a consortium of institutes led by Cardiff University (UK) and including
Univ. Lethbridge (Canada); NAOC (China); CEA, LAM (France);
IFSI, Univ. Padua (Italy); IAC (Spain); Stockholm Observatory
(Sweden); Imperial College London, RAL, UCL-MSSL, UKATC,
Univ. Sussex (UK); and Caltech, JPL, NHSC, Univ. Colorado
(USA). This development has been supported by national funding agencies: CSA (Canada); NAOC (China); CEA, CNES, CNRS
(France); ASI (Italy); MCINN (Spain); SNSB (Sweden); STFC
(UK); and NASA (USA). HIPE is a joint development by the Herschel Science Ground Segment Consortium, consisting of ESA, the
NASA Herschel Science Center, and the HIFI, PACS and SPIRE
consortia. This research has made use of the NASA/IPAC Extragalactic Database (NED) which is operated by the Jet Propulsion
Laboratory, California Institute of Technology, under contract with
the National Aeronautics and Space Administration.


\begin{thebibliography}{}

\bibitem[Aniano et al.(2012)]{2012ApJ...756..138A} Aniano, G., Draine, B.~T., Calzetti, D., et al.\ 2012, \apj, 756, 138
\bibitem[Aniano et al.(2011)]{2011PASP..123.1218A} Aniano, G., Draine, B.~T., Gordon, K.~D., \& Sandstrom, K.\ 2011, \pasp, 123, 1218 
\bibitem[Auld et al.(2013)]{2013MNRAS.428.1880A} Auld, R., Bianchi, S., Smith, M.~W.~L., et al.\ 2013, \mnras, 428, 1880
\bibitem[Bendo et al.(2010)]{2010A&A...518L..65B} Bendo, G.~J., Wilson, C.~D., Pohlen, M., et al.\ 2010, \aap, 518, L65 
\bibitem[Bendo et al. (2012)]{2012MNRAS.419.1833B}  Bendo, G. J., Boselli, A., Dariush, A., Pohlen, M., Roussel, H. et al.  2012, MNRAS, 419, 1833
\bibitem[Bernstein et al.(2002)]{2002ApJ...571..107B} Bernstein, R.~A., Freedman, W.~L., \& Madore, B.~F.\ 2002, \apj, 571, 107 
\bibitem[Bertin \& Arnouts(1996)]{1996A&AS..117..393B} Bertin, E., \& Arnouts, S.\ 1996, \aaps, 117, 393
\bibitem[Bigiel et al. (2008)]{2008AJ....136.2846B} Bigiel, F., Leroy, A., Walter, F., Brinks, E. et al. 2008, AJ, 136, 2846
\bibitem[Bigiel et al.(2011)]{2011ApJ...730L..13B} Bigiel, F., Leroy, A.~K., Walter, F., et al.\ 2011, \apjl, 730, L13 
\bibitem[Blanc et al.(2009)]{2009ApJ...704..842B} Blanc, G.~A., Heiderman, A., Gebhardt, K., Evans, N.~J., II, \& Adams, J.\ 2009, \apj, 704, 842 
\bibitem[Bolatto et al.(2013)]{2013arXiv1301.3498B} Bolatto, A.~D., Wolfire, M., \& Leroy, A.~K.\ 2013, arXiv:1301.3498 
\bibitem[Boquien et al.(2010)]{2010A&A...518L..70B} Boquien, M., Calzetti, D., Kramer, C., et al.\ 2010, \aap, 518, L70 
\bibitem[Boquien et al. (2011)]{2011AJ....142..111B} Boquien, M., Calzetti, D., Combes, F., Henkel, C., Israel, F., Kramer, C. et al.  2011, AJ, 142, 111
\bibitem[Boquien et al.(2012)]{2012A&A...539A.145B} Boquien, M., Buat, V., Boselli, A., et al.\ 2012, \aap, 539, A145 
\bibitem[Boselli et al.(2002)]{2002Ap&SS.281..127B} Boselli, A., Lequeux, J., \& Gavazzi, G.\ 2002, \apss, 281, 127 
\bibitem[Boselli et al.(2009)]{2009ApJ...706.1527B} Boselli, A., Boissier, S., Cortese, L., et al.\ 2009, \apj, 706, 1527
\bibitem[Calzetti et al.(2005)]{2005ApJ...633..871C} Calzetti, D., Kennicutt, R.~C., Jr., Bianchi, L., et al.\ 2005, \apj, 633, 871 
\bibitem[Calzetti et al. (2007)]{2007ApJ...666..870C} Calzetti, D., Kennicutt, R. C., Engelbracht, C. W., Leitherer, C., Draine, B. T., et al. 2007, ApJ, 666, 870
\bibitem[Calzetti(2012)]{2012arXiv1208.2997C} Calzetti, D.\ 2012, arXiv:1208.2997 
\bibitem[Calzetti et al.(2012)]{2012ApJ...752...98C} Calzetti, D., Liu, G., \& Koda, J.\ 2012, \apj, 752, 98
\bibitem[Corbelli et al.(2012)]{2012A&A...542A..32C} Corbelli, E., Bianchi, S., Cortese, L., et al.\ 2012, \aap, 542, A32 
\bibitem[Crosthwaite et al.(2002)]{2002AJ....123.1892C} Crosthwaite, L.~P., Turner, J.~L., Buchholz, L., Ho, P.~T.~P., \& Martin, R.~N.\ 2002, \aj, 123, 1892
\bibitem[Dale et al.(2009)]{2009ApJ...703..517D} Dale, D.~A., Cohen, S.~A., Johnson, L.~C., et al.\ 2009, \apj, 703, 517 
\bibitem[Dale et al.(2012)]{2012ApJ...745...95D} Dale, D.~A., Aniano, G., Engelbracht, C.~W., et al. 2012, \apj, 745, 95 
\bibitem[Davies et al.(2012)]{2012MNRAS.419.3505D} Davies, J.~I., Bianchi, S., Cortese, L., et al.\ 2012, \mnras, 419, 3505 
\bibitem[Downes et al.(1996)]{1996ApJ...461..186D} Downes, D., Reynaud, D., Solomon, P.~M., \& Radford, S.~J.~E.\ 1996, \apj, 461, 186 
\bibitem[Draine(2003)]{2003ARA&A..41..241D} Draine, B.~T.\ 2003, \araa, 41, 241 
\bibitem[Draine \& Li(2007)]{2007ApJ...657..810D} Draine, B.~T., \& Li, A.\ 2007, \apj, 657, 810D
\bibitem[Eales et al.(2010)]{2010A&A...518L..62E} Eales, S.~A., Smith, M.~W.~L., Wilson, C.~D., et al.\ 2010, \aap, 518, L62
\bibitem[Eales et al.(2012)]{2012ApJ...761..168E} Eales, S., Smith, M.~W.~L., Auld, R., et al.\ 2012, \apj, 761, 168  
\bibitem[Engelbracht et al.(2010)]{2010A&A...518L..56E} Engelbracht, C.~W., Hunt, L.~K., Skibba, R.~A., et al.\ 2010, \aap, 518, L56 
\bibitem[Feldmann \& Gnedin(2011)]{2011ApJ...727L..12F} Feldmann, R., \& Gnedin, N.~Y.\ 2011, \apjl, 727, L12
\bibitem[Ferguson et al.(1996)]{1996AJ....111.2265F} Ferguson, A.~M.~N., Wyse, R.~F.~G., Gallagher, J.~S., III, \& Hunter, D.~A.\ 1996, \aj, 111, 2265 
\bibitem[Fischera \& Dopita(2008)]{2008ApJS..176..164F} Fischera, J., \& Dopita, M.~A.\ 2008, \apjs, 176, 164 
\bibitem[Foyle et al.(2010)]{2010ApJ...725..534F} Foyle, K., Rix, H.-W., Walter, F., \& Leroy, A.~K.\ 2010, \apj, 725, 534 
\bibitem[Foyle et al. (2012)]{2012MNRAS.421.2917F} Foyle, K., Wilson, C. D., Mentuch, E., Bendo, G., Dariush, A., et al. \ 2012, MNRAS, 421, 2917
\bibitem[Gratier et al.(2012)]{2012A&A...542A.108G} Gratier, P., Braine, J., Rodriguez-Fernandez, N.~J., et al.\ 2012, \aap, 542, A108
\bibitem[Griffin et al.(2010)]{2010A&A...518L...3G} Griffin, M.~J., et al.\ 2010, A\&A, 518, L3 
\bibitem[Groves et al. (2008)]{2008ApJS..176..438G} Groves, Brent, Dopita, Michael A., Sutherland, Ralph S., Kewley, Lisa J., Fischera, Jörg, et al. 2008, ApJS, 176, 438
\bibitem[Groves et al.(2012)]{2012MNRAS.426..892G} Groves, B., Krause, O., Sandstrom, K., et al.\ 2012, \mnras, 426, 892
\bibitem[Guhathakurta \& Draine (1989)]{1989ApJ...345..230G} Guhathakurta, P., Draine, B. T. \ 1989, ApJ, 345, 230
\bibitem[Helou et al. (2004)]{2004ApJS..154..253H} Helou, G., Roussel, H., Appleton, P., Frayer, D. et al. 2004, ApJS, 154, 253
\bibitem[Hildebrand(1983)]{1983QJRAS..24..267H} Hildebrand, R.~H.\ 1983, \qjras, 24, 267 
\bibitem[Hirota et al.(2011)]{2011ApJ...737...40H} Hirota, A., Kuno, N., Sato, N., et al.\ 2011, \apj, 737, 40 
\bibitem[Kenney \& Lord(1991)]{1991ApJ...381..118K} Kenney, J.~D.~P., \& Lord, S.~D.\ 1991, \apj, 381, 118 
\bibitem[Kennicutt(1989)]{1989ApJ...344..685K} Kennicutt, R.~C., Jr.\ 1989, \apj, 344, 685
\bibitem[Kennicutt(1998)]{1998ApJ...498..541K} Kennicutt, R.~C., Jr.\ 1998, \apj, 498, 541 
\bibitem[Kennicutt et al.(2007)]{2007ApJ...671..333K} Kennicutt, R.~C., Jr., Calzetti, D., Walter, F., et al.\ 2007, \apj, 671, 333
\bibitem[Kennicutt \& Evans (2012)]{2012ARA&A..50..531K} Kennicutt, Robert C., Evans, Neal J. 2012, ARA\&A, 50, 531K
\bibitem[Koda et al. (2009)]{2009ApJ...700L.132K} Koda, Jin, Scoville, Nick, Sawada, Tsuyoshi, La Vigne, Misty A., et al. \ 2009, ApJ, 700,132
\bibitem[Law et al.(2011)]{2011ApJ...738..124L} Law, K.-H., Gordon, K.~D., \& Misselt, K.~A.\ 2011, \apj, 738, 124
\bibitem[Leroy et al.(2008)]{2008AJ....136.2782L} Leroy, A.~K., Walter, F., Brinks, E., Bigiel, F., de Blok, W.~J.~G., Madore, B., \& Thornley, M.~D.\ 2008, AJ, 136, 2782 
\bibitem[Leroy et al.(2012)]{2012AJ....144....3L} Leroy, A.~K., Bigiel, F., de Blok, W.~J.~G., et al.\ 2012, \aj, 144, 3
\bibitem[Lee et al.(2009)]{2009ApJ...706..599L} Lee, J.~C., Gil de Paz, A., Tremonti, C., et al.\ 2009, \apj, 706, 599 
\bibitem[Li et al.(2010)]{2010ApJ...725..677L} Li, Y., Calzetti, D., Kennicutt, R.~C., et al.\ 2010, \apj, 725, 677
\bibitem[Liu et al. (2011)]{2011ApJ...735...63L} Liu, Guilin, Koda, Jin, Calzetti, Daniela, Fukuhara, Masayuki, Momose, Rieko \ 2011, \apj, 735, 63
\bibitem[Lundgren et al.(2004)]{2004A&A...413..505L} Lundgren, A.~A., Wiklind, T., Olofsson, H., \& Rydbeck, G.\ 2004, \aap, 413, 505 
\bibitem[Markwardt(2009)]{2009ASPC..411..251M} Markwardt, C.~B.\ 2009, Astronomical Data Analysis Software and Systems XVIII, 411, 251 
\bibitem[Mathis et al. (1983)]{1983A&A...128..212M} Mathis, J. S., Mezger, P. G., Panagia, N. \ 1983, A\&A, 128, 212
\bibitem[Men'shchikov et al. (2012)]{2012A&A...542A..81M} Men'shchikov, A., Andr\'e, Ph., Didelon, P., Motte, F., Hennemann, M., Schneider, N. \ 2012, A\&A, 542, 81
\bibitem[Meurer et al.(2006)]{2006ApJS..165..307M} Meurer, G.~R., et al.\ 2006, ApJS, 165, 307
\bibitem[Momose et al.(2010)]{2010ApJ...721..383M} Momose, R., Okumura, S.~K., Koda, J., \& Sawada, T.\ 2010, \apj, 721, 383 
\bibitem[Mu{\~n}oz-Mateos et al.(2009)]{2009ApJ...701.1965M} Mu{\~n}oz-Mateos, J.~C., et al.\ 2009, ApJ, 701, 1965 
\bibitem[Muraoka et al. (2009)]{2009ApJ...706.1213M} Muraoka, Kazuyuki, Kohno, Kotaro, Tosaki, Tomoka, Kuno, Nario, et al. \ 2009, ApJ, 706,1213
\bibitem[Murray(2011)]{2011ApJ...729..133M} Murray, N.\ 2011, \apj, 729, 133 
\bibitem[Natale et al. (2010)]{2010ApJ...725..955N} Natale, G., Tuffs, R. J., Xu, C. K., Popescu, C. C., Fischera, J., Lisenfeld, U., et al. \ 2010, \apj, 725, 955
\bibitem[Nimori et al.(2012)]{2012MNRAS.tmp..393N} Nimori, M., Habe, A., Sorai, K., et al.\ 2012, \mnras, 393
\bibitem[Pilbratt et al. (2010)]{2010A&A...518L...1P} Pilbratt, G. L., Riedinger, J. R., Passvogel, T., Crone, G., Doyle, D., et al. 2010, A\&A, 518,1
\bibitem[Poglitsch et al.(2010)]{2010A&A...518L...2P} Poglitsch, A., et al.\ 2010, A\&A, 518, L2 
\bibitem[Pohlen et al.(2010)]{2010A&A...518L..72P} Pohlen, M., Cortese, L., Smith, M.~W.~L., et al.\ 2010, \aap, 518, L72 
\bibitem[Popescu \& Tuffs(2002)]{2002MNRAS.335L..41P} Popescu, C.~C., \& Tuffs, R.~J.\ 2002, \mnras, 335, L41
\bibitem[Popescu et al. (2011)]{2011A&A...527A.109P} Popescu, C. C., Tuffs, R. J., Dopita, M. A., Fischera, J., Kylafis, N. D., Madore, B. F. \ 2011, A\&A,527,109
\bibitem[Rahman et al.(2011)]{2011ApJ...730...72R} Rahman, N., Bolatto, A.~D., Wong, T., et al.\ 2011, \apj, 730, 72 
\bibitem[Rand et al. (1999)]{1999ApJ...513..720R} Rand, Richard J., Lord, Steven D., Higdon, James L. \ 1999, \apj, 513, 720
\bibitem[Rebolledo et al. (2012)]{2012ApJ...757..155R}	Rebolledo, David, Wong, Tony, Leroy, Adam, Koda, Jin, Donovan Meyer, Jennifer 2012, ApJ,757, 155
\bibitem[Rela{\~n}o et al.(2012)]{2012MNRAS.423.2933R} Rela{\~n}o, M., Kennicutt, R.~C., Jr., Eldridge, J.~J., Lee, J.~C., \& Verley, S.\ 2012, \mnras, 423, 2933 
\bibitem[Roslolowski (2005)]{2005PASP..117.1403R} Rosolowsky, E. \ 2005, PASP,117,1403
\bibitem[Roussel(2012)]{2012arXiv1205.2576R} Roussel, H.\ 2012, arXiv:1205.2576
\bibitem[Sandstrom et al.(2012)]{2012arXiv1212.1208S} Sandstrom, K.~M., Leroy, A.~K., Walter, F., et al.\ 2012, arXiv:1212.1208
\bibitem[Schlegel et al.(1998)]{1998ApJ...500..525S} Schlegel, D.~J., Finkbeiner, D.~P., \& Davis, M.\ 1998, \apj, 500, 525
\bibitem[Schmidt(1959)]{1959ApJ...129..243S} Schmidt, M.\ 1959, \apj, 129, 243 
\bibitem[Schruba et al.(2010)]{2010ApJ...722.1699S} Schruba, A., Leroy, A.~K., Walter, F., Sandstrom, K., \& Rosolowsky, E.\ 2010, \apj, 722, 1699
\bibitem[Shetty et al.(2011)]{2011MNRAS.412.1686S} Shetty, R., Glover, S.~C., Dullemond, C.~P., \& Klessen, R.~S.\ 2011, \mnras, 412, 1686
\bibitem[Skibba et al.(2012)]{2012ApJ...761...42S} Skibba, R.~A., Engelbracht, C.~W., Aniano, G., et al.\ 2012, \apj, 761, 42 
\bibitem[Smith et al. (2012)]{2012ApJ...756...40S} Smith, M. W. L., Eales, S. A., Gomez, H. L., Roman-Duval, J., Fritz, J., et al. 2012, \apj, 756,40
\bibitem[Solomon et al. (1987)]{1987ApJ...319..730S} Solomon, P. M., Rivolo, A. R., Barrett, J., Yahil, A. \ 1987, \apj, 319, 730
\bibitem[Stutzki \& Guesten(1990)]{1990ApJ...356..513S} Stutzki, J., \& Guesten, R.\ 1990, \apj, 356, 513
\bibitem[Thim et al.(2003)]{2003ApJ...590..256T} Thim, F., Tammann, G.~A., Saha, A., Dolphin, A., Sandage, A., Tolstoy, E., \& Labhardt, L.\ 2003, ApJ, 590, 256 
\bibitem[Tilanus \& Allen(1993)]{1993A&A...274..707T} Tilanus, R.~P.~J., \& Allen, R.~J.\ 1993, \aap, 274, 707
\bibitem[Verley et al. (2010)]{2010A&A...518L..68V} Verley, S., Relaño, M., Kramer, C., Xilouris, E. M., Boquien, M., et al. 2010, A\&A, 518, 68
\bibitem[Vogel et al. (1988)]{1988Natur.334..402V} Vogel, S. N., Kulkarni, S. R., Scoville, N. Z. \ 1988, Nature, 334, 402
\bibitem[Voit (1991)]{1991ApJ...379..122V} Voit, G. M. \ 1991, ApJ, 379,122
\bibitem[Walter et al.(2008)]{2008AJ....136.2563W} Walter, F., Brinks, E., de Blok, W.~J.~G., et al.\ 2008, \aj, 136, 2563 
\bibitem[Williams et al.(1994)]{1994ApJ...428..693W} Williams, J.~P., de Geus, E.~J., \& Blitz, L.\ 1994, \apj, 428, 693
\bibitem[Wong \& Blitz(2002)]{2002ApJ...569..157W} Wong, T., \& Blitz, L.\ 2002, \apj, 569, 157
\bibitem[Zhu et al.(2008)]{2008ApJ...686..155Z} Zhu, Y.-N., Wu, H., Cao, C., \& Li, H.-N.\ 2008, \apj, 686, 155 
\end{thebibliography}

\appendix

\section{Resolution Tests}
\begin{figure}
\centering
\includegraphics[trim=0mm 0mm 0mm 0mm,width=40mm,angle=-90]{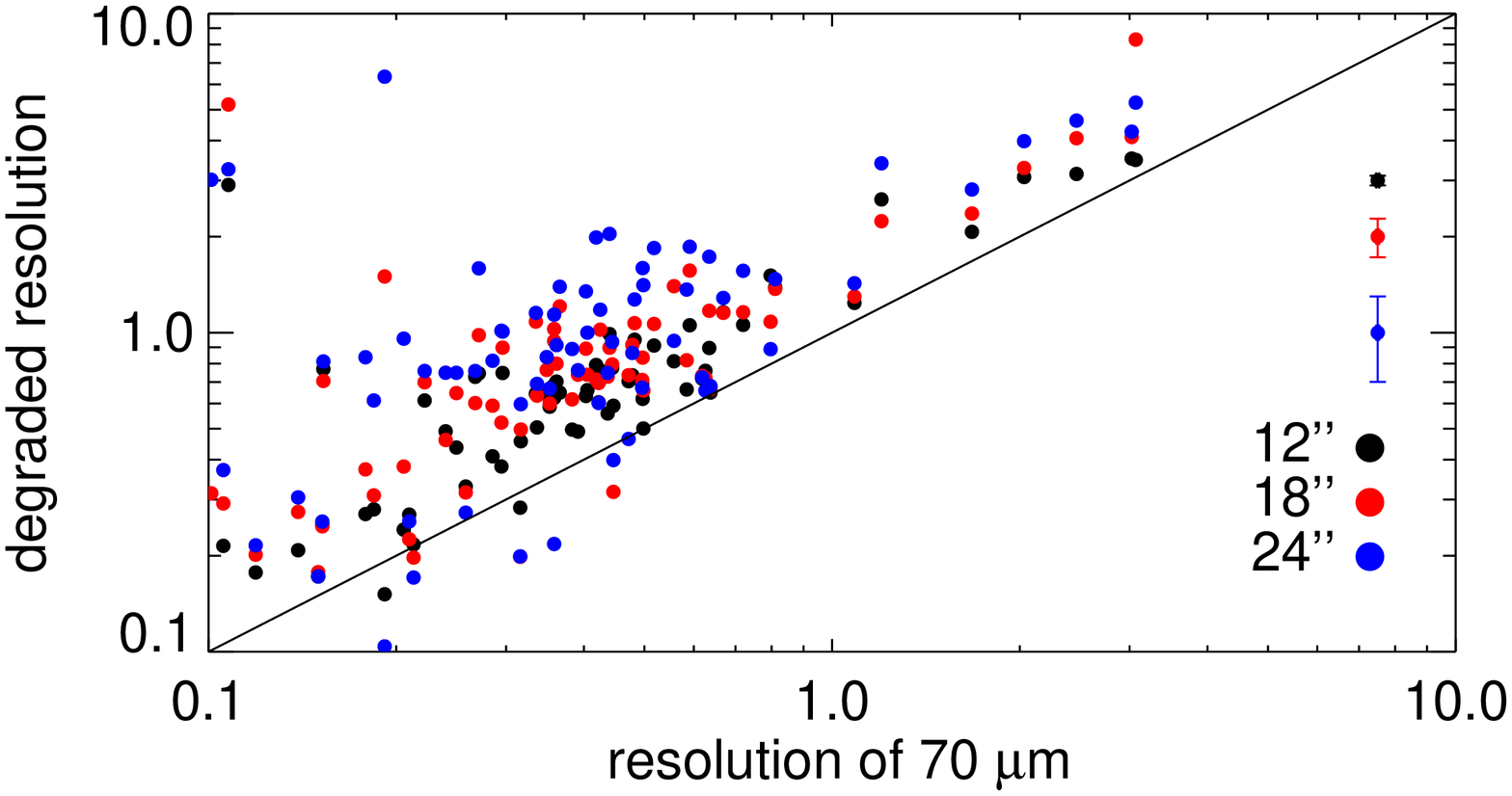}
\caption{Total flux of the sources measured by \textsc{getsources}
  using the degraded 70~$\mu$m maps versus the total flux of those
  sources measured in the 70~$\mu$m map in its native resolution. The median uncertainties for each resolution are shown on the right.  At 
  poorer resolutions, the source flux is increasingly overestimated.}
\label{deg70}
\end{figure}

We test for any systematic errors in the process of detecting and measuring fluxes of the compact sources using \textsc{getsources}.  
To do so, we take the 70~$\mu$m map with 6$''$ resolution and degrade its resolution to match that of 160~$\mu$m map (12$''$), the 250~$\mu$m map (18.2$''$) and the 
350~$\mu$m (24.5$''$) using the convolution kernels of Aniano et al. (2011). We then rerun the extraction process using these series of maps.  We can then compare how the flux of the sources is affected by the
 degradation of the resolution.  In principle, at each resolution, the flux should be the same for each source.  However, particularly in crowded regions, 
the source flux is more complicated to determine as the resolution is degraded.  Fig.~\ref{deg70} compares how the source flux at the degraded 
resolution compares to the original.  In general, we find that the flux is somewhat overestimated as the resolution is degraded.  Half of the sources 
fall within the uncertainties on the measurements and the majority lie within a factor of 2 of the flux determined at the 6$''$ resolution.  We also tested how 
such a systematic uncertainty could affect the results of the inferred dust temperature and mass. By applying the average correction factors to the source FIR 
flux measurements (see \S4.2) and performing SED fitting as described in \S5, we found an average relative difference in dust mass of $\approx 50\%$ and 
an average difference of 2K for the dust temperatures. Although non negligible, the possible systematic effect highlighted by this test is not large enough to 
affect the main conclusions of this work.    


\section{Comparison with Modified Blackbody Fits}
We compare the results of our two component SED fitting method, by determining the dust temperature and mass using a modified blackbody fit to the FIR 
emission (70-350~$\mu$m).  
  
We use a modified blackbody of the form:
\begin{equation}
S_{\lambda}=N B_{\lambda}(T) \lambda^{-\beta},
 \end{equation}
 where S$_{\lambda}$ is the flux density, B$_{\lambda}$(T) is the Planck function, N is a constant related to the column density of the material and $\beta$ is the dust emissivity index. We hold $\beta$ at a constant value of 2, based on the average value found by F12.   We fit the points using MPFIT, a least-squares curve fitting routine for IDL (Markwardt 2009). We refer to the temperature and dust masses determined by this technique as T$_{dust,MB}$ and M$_{dust,MB}$ 
respectively.  
 
\begin{figure}
\centering
\includegraphics[trim=0mm 0mm 0mm 0mm,width=40mm,angle=-90]{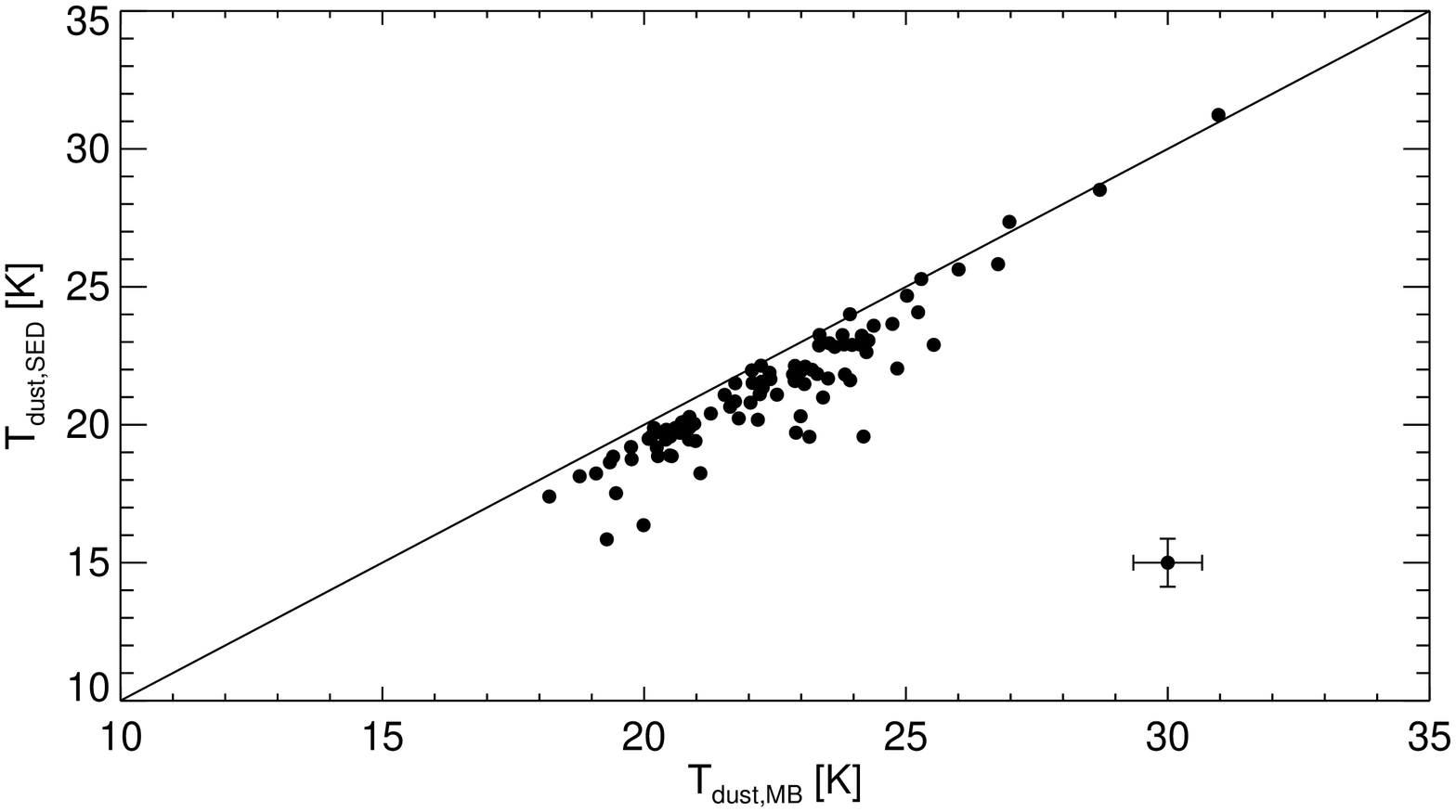}
\caption{Comparison of the dust temperature derived by the SED fitting to that obtained by a modified blackbody fit of the FIR wavelengths.  The median uncertainties are shown on the bottom right.}
\label{tmb}
\end{figure}

\begin{figure}
\centering
\includegraphics[trim=0mm 0mm 0mm 0mm,width=40mm,angle=-90]{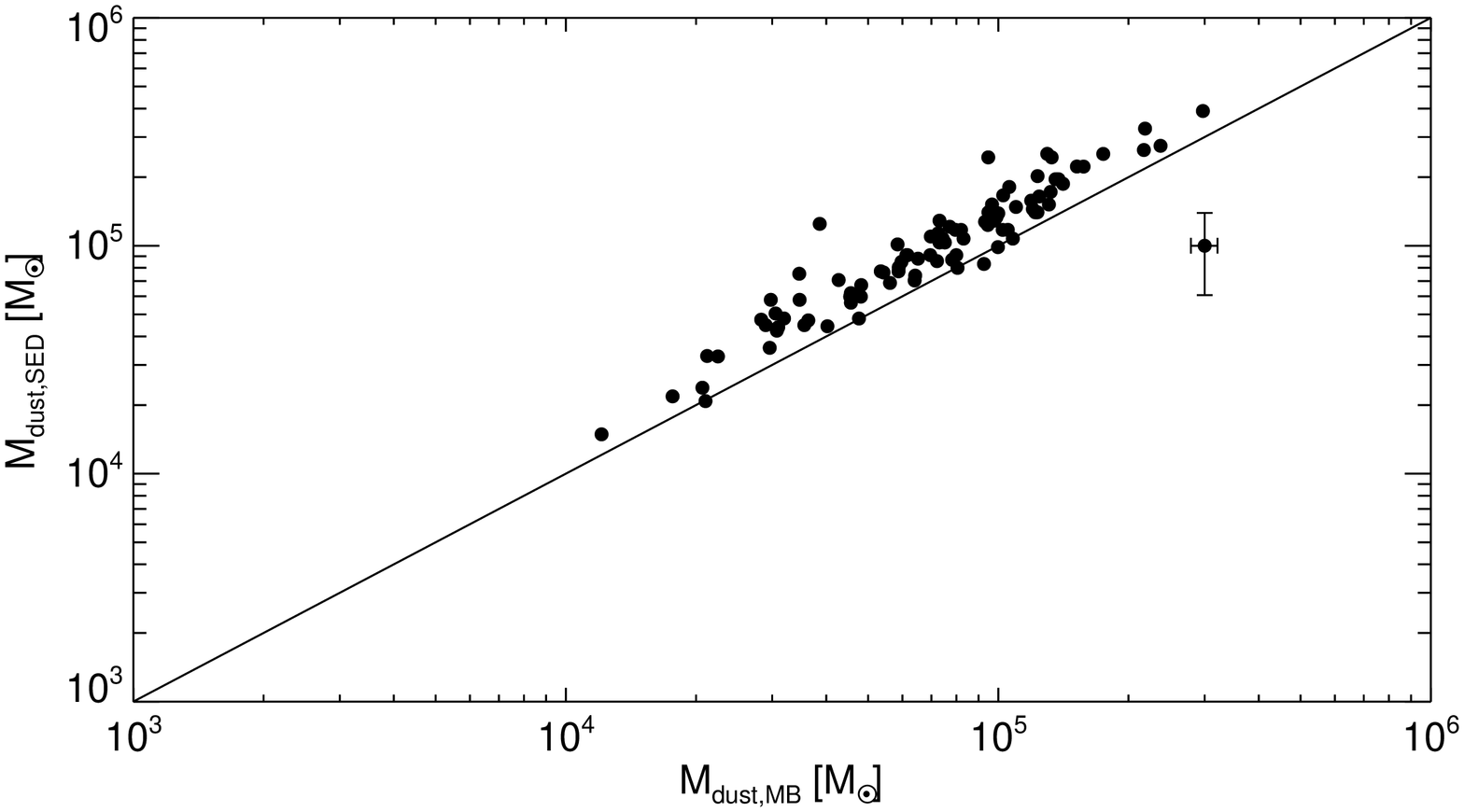}
\caption{Comparison of the dust mass derived by the SED fitting to that obtained by a modified blackbody fit of the FIR wavelengths.}
\label{mmb}
\end{figure}

Fig.~\ref{tmb} and \ref{mmb} compares the temperatures and masses derived from the SED fitting with those from a modified blackbody fit.  There is close agreement with the temperatures, although the modified blackbody dust temperatures tend to be slightly higher. The masses found by the SED fitting are larger than those from the modified blackbody fits with a median ratio of a factor of 1.4.  Dust emission models typically have higher masses than modified blackbody fits, because the latter do not encompass the full range of temperatures of the dust as they use only the FIR maps (Dale et al.\ 2012).    In F12, they found that fitting the SED using the dust emission models of Draine \& Li (2007) the total mass of M83 was greater by a factor of 1.3 than those found using a modified blackbody fit.  


\section{Color Correlations}   

\begin{figure}
\centering
\includegraphics[trim=0mm 20mm 0mm 20mm,width=60mm,angle=-90]{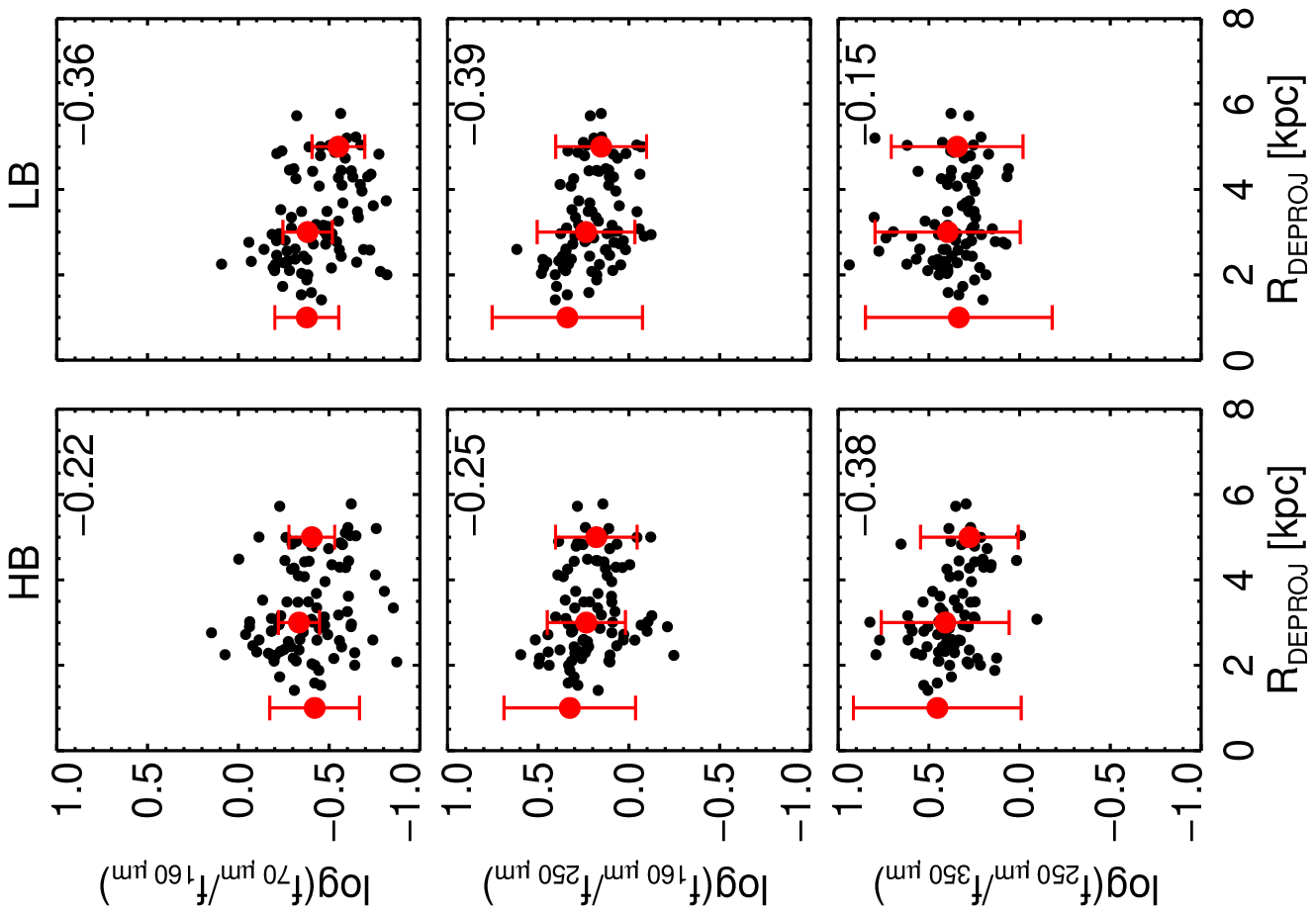}

\caption{Ratio of 70-to-160~$\mu$m emission (top), 160-to-250~$\mu$m
  emission (second from top) and 250-to-350~$\mu$m emission (bottom)
  for the compact source measurements with high background subtraction
  (left) and low background subtraction (right) versus deprojected radial position.}
\label{colrad}
\end{figure}


\begin{figure}
\centering
\includegraphics[trim=0mm 20mm 0mm 20mm,width=60mm,angle=-90]{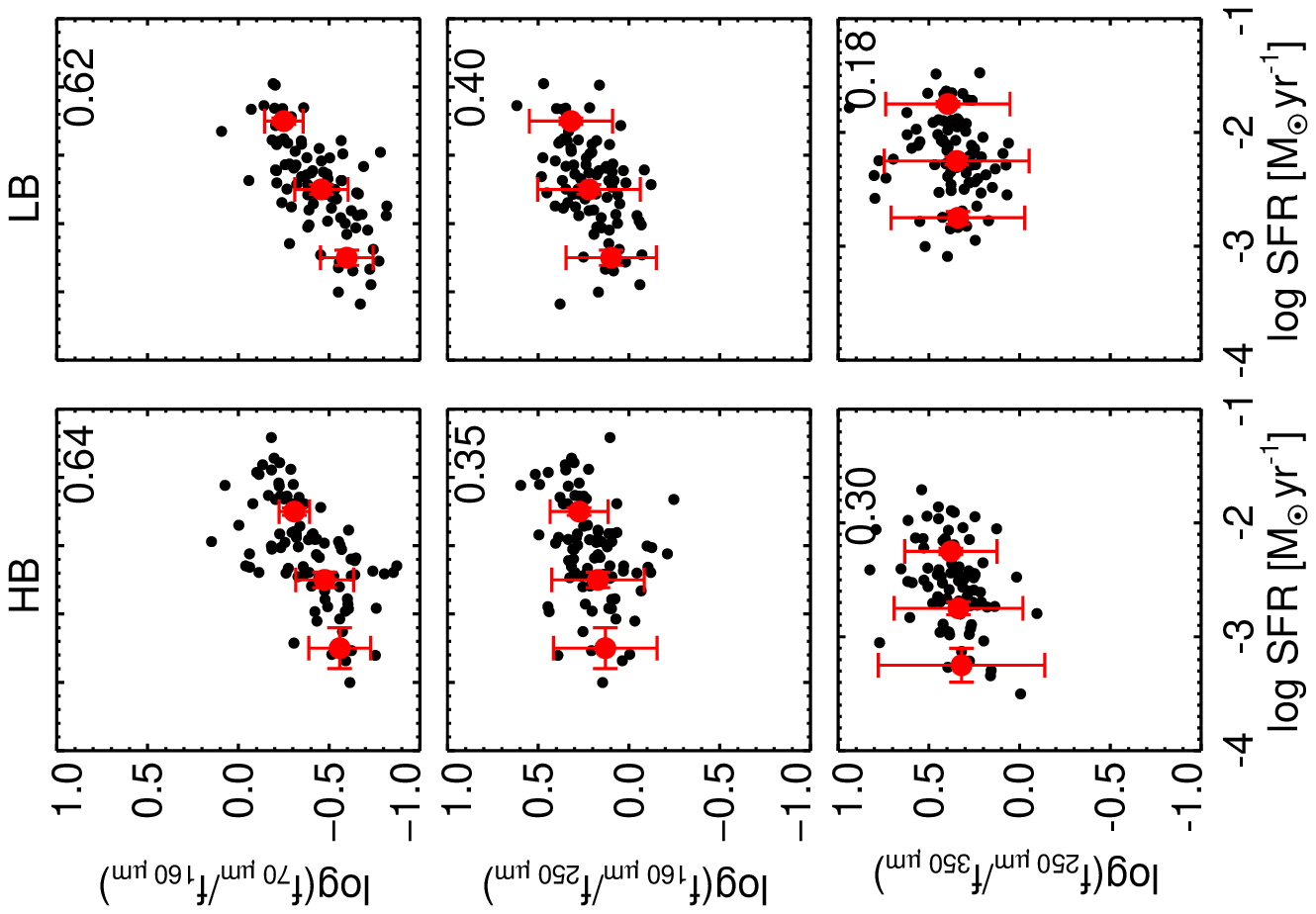}
\caption{Ratio of 70-to-160~$\mu$m emission (top), 160-to-250~$\mu$m
  emission (second from top) and 250-to-350~$\mu$m emission (bottom)
  for the compact sources with high background subtraction (left) and
  low background subtraction (right) versus the star formation rate measured in each case.}
\label{colsfr}
\end{figure}


\begin{figure}
\centering
\includegraphics[trim=0mm 20mm 0mm 20mm,width=60mm,angle=-90]{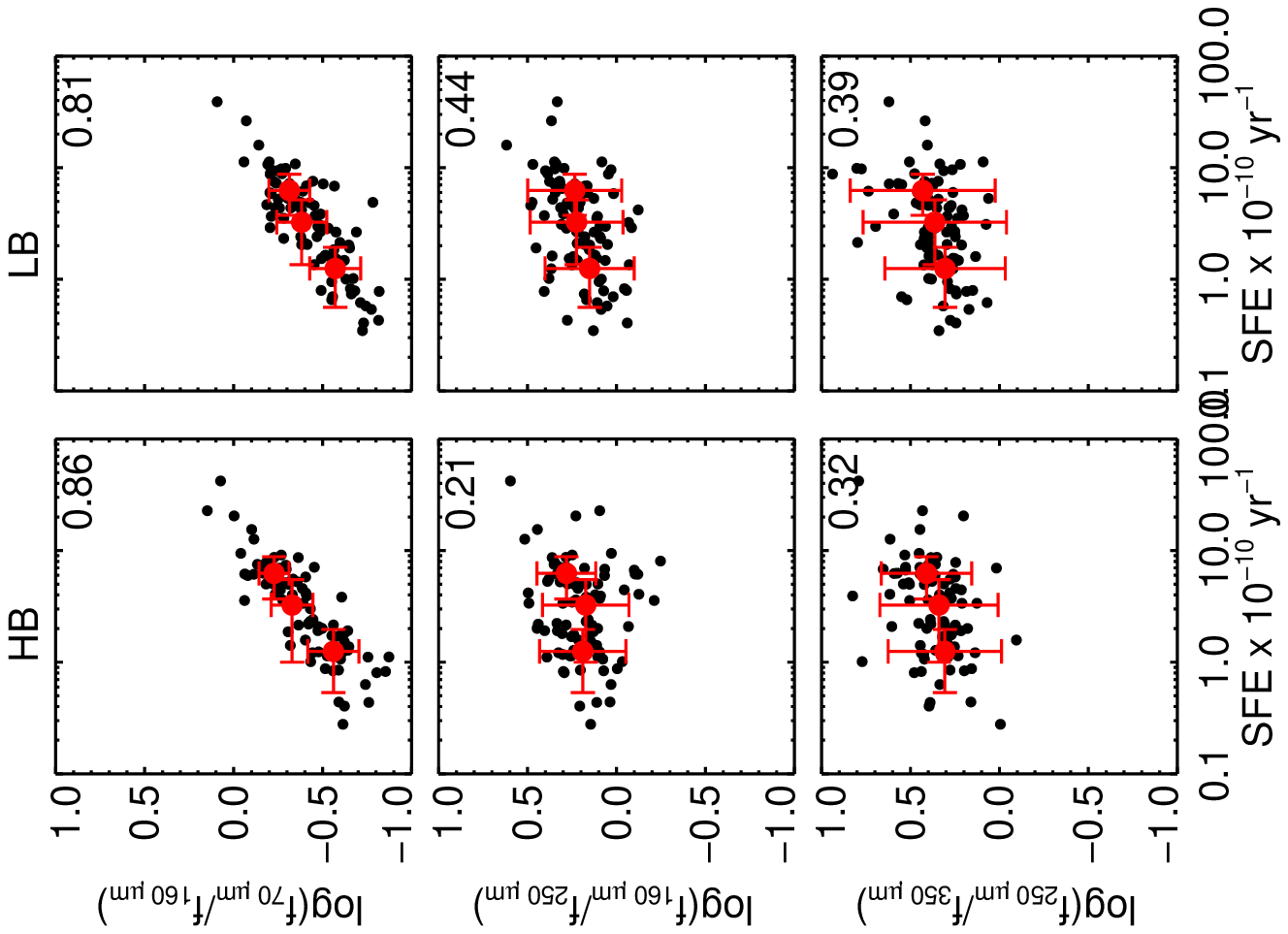}
\caption{Ratio of 70-to-160 $\mu$m emission (top), 160-to-250 $\mu$m emission (second from top) and 250-to-350 $\mu$m emission (bottom) for the compact source measurements from \textsc{getsources} (left) and alternative compact source measurements (right) versus the SFE measured in each case.}
\label{colsfe}
\end{figure}

Many previous works that have examined dust heating mechanisms and correlations with recent star formation rate tracers have used dust emission colors as 
opposed to SED fitting parameters (i.e.\ Bendo et al.\ 2012, Boquien et al.\ 2011).  For the purposes of comparison, 
we also examine how the dust emission colors vary with radius and with the star formation parameters SFR and SFE.

We consider the color ratios of 70-to-160~$\mu$m, 160-to-250~$\mu$m and 250-to-350~$\mu$m emission.  In following what we have done in the previous sections,
 we consider both measures of the compact sources.
 
Fig.~\ref{colrad}  shows how the colors vary with radius.  As we saw with the SED fitting parameters, for both measures of the compact sources, 
we find no or just a weak trend in the colors with radius.  Both Bendo et al.\ 2012 and Boquien et al.\ 2011 found in their pixel-by-pixel analysis 
that the FIR colours present decreasing radial trends. Our measurements show that, when compact regions are separated, they do not seem to behave in the same
 way.

Fig.~\ref{colsfr} shows how the colours correlate with the SFR. Bendo et al.\ 2012 and Boquien et al.\ 2011 highlighted the presence of correlations between 
 SFR and FIR colors, especially the 70-to-160 $\mu$m intensity ratio. This finding was used to imply that at shorter FIR wavelengths the dust heating is 
powered mostly by recent star formation. We find a mild correlation between the SFR and 70-to-160 $\mu$m intensity ratio. This correlation is 
stronger than that seen between the SFR and the dust temperature $T_{\rm{dust}}$ inferred by SED fitting.  The difference in the results when using 
the 70-to-160 $\mu$m intensity ratio and $T_{\rm{dust}}$ is due to the fact that the latter is the dust temperature associated with the cold diffuse dust 
component in the SED fitting. At both 70 and 160$\mu$m, a part of the emission is contributed by the PDR component as well, thus the FIR colors fitted by 
the diffuse dust component are usually slightly different than the observed ones.  

However, we note that there is a strong correlation between the SFE and the  70-to-160 $\mu$m intensity ratio (see Fig.~\ref{colsfe}), which is very similar 
with what we saw in \S7.3 for the correlation observed between SFE and $T_{\rm{dust}}$. On the other hand, we did not find any strong correlation between 
160/250 $\mu$m or 250/350 $\mu$m flux ratios with SFR or SFE. This might indicate that the dust emission at these wavelengths is more affected by heating 
from older stellar populations. However, one should also note that the flux uncertainties are quite large at longer wavelengths and, as a consequence, there 
could also be a stronger scatter hiding the presence of intrinsic correlations. 

\section{Source fluxes}

\onecolumn
\begin{center}
\small
\begin{longtable}{cccccccccc}
\caption{HB measurement source fluxes and positions}\\
\label{HBfluxes}\\

\hline
 & RA & $\delta$ & F$_{8\mu m}$ & F$_{24\mu m}$ & F$_{70\mu m}$ & F$_{160\mu m}$ & F$_{250\mu m}$ & F$_{350\mu m}$ & F$(H_\alpha)$\\
 & [deg] & [deg]& [mJy] & [mJy] & [mJy] & [mJy] & [mJy] & [mJy] & [$\mu$Jy] \\
\hline
\endfirsthead
\multicolumn{3}{c}%
{\tablename\ \thetable{} -- continued from previous page}\\
\hline
 & RA & $\delta$ & F$_{8\mu m}$ & F$_{24\mu m}$ & F$_{70\mu m}$ & F$_{160\mu m}$ & F$_{250\mu m}$ & F$_{350\mu m}$ & F$(H_\alpha)$\\
 & [deg] & [deg]& [mJy] & [mJy] & [mJy] & [mJy] & [mJy] & [mJy] & [$\mu$Jy] \\
\hline
\endhead

       1 &        204.27867 &       -29.826725 & 79 $\pm$ 6 & 190 $\pm$ 10 & 
2710 $\pm$ 150 & 3690 $\pm$ 410&1640 $\pm$ 410&720 $\pm$ 220&51 $\pm$ 2\\
       2 &        204.26433 &       -29.900421 & 35 $\pm$ 5 & 139 $\pm$ 9 & 
1495 $\pm$ 92 & 2490 $\pm$ 290&1320 $\pm$ 310&650 $\pm$ 210&36 $\pm$ 2\\
       3 &        204.22128 &       -29.858999 & 89 $\pm$ 8 & 180 $\pm$ 10 & 
2410 $\pm$ 160 & 3790 $\pm$ 680&1830 $\pm$ 480&650 $\pm$ 220&82 $\pm$ 3\\
       4 &        204.28271 &       -29.854724 & 68 $\pm$ 7 & 107 $\pm$ 7 & 
2090 $\pm$ 130 & 3060 $\pm$ 500&1540 $\pm$ 450&410 $\pm$ 200&34 $\pm$ 3\\
       5 &        204.23187 &       -29.830846 & 88 $\pm$ 7 & 73 $\pm$ 6 & 
1860 $\pm$ 120 & 3120 $\pm$ 370&1440 $\pm$ 410&580 $\pm$ 210&84 $\pm$ 3\\
       6 &        204.22652 &       -29.884745 & 70 $\pm$ 6 & 100 $\pm$ 10 & 
2010 $\pm$ 140 & 2530 $\pm$ 600&910 $\pm$ 360&330 $\pm$ 180&98 $\pm$ 3\\
       7 &        204.22894 &       -29.885796 & 48 $\pm$ 6 & 80 $\pm$ 9 & 
1840 $\pm$ 140 & 1550 $\pm$ 560&390 $\pm$ 310&--&80 $\pm$ 3\\
       8 &        204.22172 &       -29.882527 & 36 $\pm$ 5 & 81 $\pm$ 8 & 
1580 $\pm$ 120 & 1900 $\pm$ 500&1620 $\pm$ 430&710 $\pm$ 220&40 $\pm$ 3\\
       9 &        204.21930 &       -29.863860 & 120 $\pm$ 10 & 140 $\pm$ 10 & 
2840 $\pm$ 200 & 5690 $\pm$ 960&1830 $\pm$ 490&1370 $\pm$ 290&30 $\pm$ 4\\
      10 &        204.22046 &       -29.880233 & 102 $\pm$ 8 & 140 $\pm$ 10 & 
1630 $\pm$ 120 & 3160 $\pm$ 610&1890 $\pm$ 460&590 $\pm$ 210&79 $\pm$ 3\\
      11 &        204.21971 &       -29.852660 & 53 $\pm$ 6 & 62 $\pm$ 9 & 
1180 $\pm$ 110 & 2110 $\pm$ 340&1230 $\pm$ 420&460 $\pm$ 200&65 $\pm$ 4\\
      12 &        204.26957 &       -29.849437 & 87 $\pm$ 8 & 106 $\pm$ 9 & 
2260 $\pm$ 160 & 3820 $\pm$ 730&1900 $\pm$ 470&800 $\pm$ 230&127 $\pm$ 3\\
      13 &        204.28487 &       -29.869891 & 84 $\pm$ 8 & 71 $\pm$ 6 & 
1427 $\pm$ 98 & 2430 $\pm$ 380&1280 $\pm$ 400&500 $\pm$ 210&63 $\pm$ 3\\
      14 &        204.22800 &       -29.882334 & 62 $\pm$ 6 & 58 $\pm$ 8 & 
1620 $\pm$ 120 & 3370 $\pm$ 710&2610 $\pm$ 530&940 $\pm$ 240&16 $\pm$ 2\\
      15 &        204.28634 &       -29.858535 & 31 $\pm$ 5 & 49 $\pm$ 6 & 
531 $\pm$ 60 & 1970 $\pm$ 370&980 $\pm$ 390&370 $\pm$ 200&8 $\pm$ 3\\
      16 &        204.28083 &       -29.852275 & 150 $\pm$ 10 & 270 $\pm$ 20 & 
3850 $\pm$ 250 & 5850 $\pm$ 840&4590 $\pm$ 850&1330 $\pm$ 300&104 $\pm$ 4\\
      17 &        204.16814 &       -29.855863 & 29 $\pm$ 2 & 30 $\pm$ 2 & 
554 $\pm$ 31 & 940 $\pm$ 160&490 $\pm$ 160&216 $\pm$ 74&40.2 $\pm$ 0.9\\
      18 &        204.22554 &       -29.844806 & 20 $\pm$ 4 & 60 $\pm$ 5 & 
710 $\pm$ 67 & 1530 $\pm$ 290&640 $\pm$ 370&340 $\pm$ 190&70 $\pm$ 3\\
      19 &        204.21463 &       -29.883342 & 25 $\pm$ 5 & 37 $\pm$ 7 & 
666 $\pm$ 89 & 770 $\pm$ 280&1250 $\pm$ 350&390 $\pm$ 190&15 $\pm$ 3\\
      20 &        204.28397 &       -29.881946 & 37 $\pm$ 6 & 38 $\pm$ 4 & 
820 $\pm$ 74 & 1790 $\pm$ 280&1010 $\pm$ 320&430 $\pm$ 200&42 $\pm$ 3\\
      21 &        204.20849 &       -29.878634 & 33 $\pm$ 4 & 42 $\pm$ 5 & 
814 $\pm$ 92 & 2420 $\pm$ 330&960 $\pm$ 300&530 $\pm$ 210&20 $\pm$ 4\\
      22 &        204.26997 &       -29.823576 & 50 $\pm$ 6 & 44 $\pm$ 6 & 
986 $\pm$ 91 & 1830 $\pm$ 320&1030 $\pm$ 350&300 $\pm$ 180&86 $\pm$ 3\\
      23 &        204.29186 &       -29.858028 & 19 $\pm$ 4 & 12 $\pm$ 4 & 
383 $\pm$ 58 & 580 $\pm$ 260&730 $\pm$ 340&220 $\pm$ 180&43 $\pm$ 4\\
      24 &        204.25244 &       -29.905010 & 46 $\pm$ 6 & 23 $\pm$ 3 & 
451 $\pm$ 50 & 1610 $\pm$ 230&1130 $\pm$ 290&570 $\pm$ 190&38 $\pm$ 2\\
      25 &        204.21223 &       -29.844608 & 106 $\pm$ 8 & 99 $\pm$ 8 & 
2040 $\pm$ 140 & 3110 $\pm$ 410&1390 $\pm$ 410&790 $\pm$ 230&108 $\pm$ 3\\
      26 &        204.21877 &       -29.863393 & 16 $\pm$ 4 & 39 $\pm$ 8 & 
478 $\pm$ 84 & 760 $\pm$ 560&1350 $\pm$ 420&--&83 $\pm$ 3\\
      27 &        204.28831 &       -29.849784 & 23 $\pm$ 4 & 25 $\pm$ 4 & 
633 $\pm$ 65 & 450 $\pm$ 260&360 $\pm$ 310&--&36 $\pm$ 4\\
      28 &        204.24629 &       -29.907226 & 34 $\pm$ 3 & 18 $\pm$ 2 & 
453 $\pm$ 50 & 1220 $\pm$ 210&820 $\pm$ 250&380 $\pm$ 170&45 $\pm$ 2\\
      29 &        204.18835 &       -29.879014 & 16 $\pm$ 2 & 20 $\pm$ 2 & 
350 $\pm$ 33 & 810 $\pm$ 200&590 $\pm$ 210&330 $\pm$ 110&16 $\pm$ 3\\
      30 &        204.18091 &       -29.872982 & 12 $\pm$ 2 & 25 $\pm$ 3 & 
317 $\pm$ 39 & 630 $\pm$ 180&540 $\pm$ 200&119 $\pm$ 76&46 $\pm$ 4\\
      31 &        204.23124 &       -29.845585 & 69 $\pm$ 7 & 55 $\pm$ 5 & 
730 $\pm$ 64 & 1860 $\pm$ 430&590 $\pm$ 370&310 $\pm$ 180&18 $\pm$ 2\\
      32 &        204.29158 &       -29.819528 & 16 $\pm$ 4 & 17 $\pm$ 2 & 
447 $\pm$ 61 & 450 $\pm$ 250&270 $\pm$ 140&170 $\pm$ 110&61 $\pm$ 4\\
      33 &        204.22354 &       -29.813422 & 13 $\pm$ 2 & 15 $\pm$ 2 & 
300 $\pm$ 26 & 540 $\pm$ 160&360 $\pm$ 180&340 $\pm$ 140&44 $\pm$ 2\\
      34 &        204.29732 &       -29.831076 & 14 $\pm$ 4 & 16 $\pm$ 2 & 
360 $\pm$ 47 & 770 $\pm$ 280&580 $\pm$ 230&270 $\pm$ 150&39 $\pm$ 3\\
      35 &        204.27795 &       -29.823031 & 12 $\pm$ 3 & 21 $\pm$ 5 & 
241 $\pm$ 44 & 1530 $\pm$ 300&780 $\pm$ 290&260 $\pm$ 170&15 $\pm$ 2\\
      36 &        204.24819 &       -29.810471 & 59 $\pm$ 5 & 47 $\pm$ 3 & 
908 $\pm$ 57 & 1780 $\pm$ 240&820 $\pm$ 260&330 $\pm$ 150&68 $\pm$ 2\\
      37 &        204.26738 &       -29.899060 & 13 $\pm$ 4 & 17 $\pm$ 4 & 
381 $\pm$ 46 & 1590 $\pm$ 250&790 $\pm$ 270&410 $\pm$ 190&28 $\pm$ 2\\
      38 &        204.23665 &       -29.880081 & 46 $\pm$ 7 & 52 $\pm$ 6 & 
1040 $\pm$ 110 & 2730 $\pm$ 810&1260 $\pm$ 380&440 $\pm$ 190&15 $\pm$ 2\\
      39 &        204.21163 &       -29.866269 & 23 $\pm$ 5 & 21 $\pm$ 5 & 
423 $\pm$ 83 & 460 $\pm$ 250&430 $\pm$ 260&150 $\pm$ 150&20 $\pm$ 4\\
      40 &        204.31140 &       -29.843049 & 22 $\pm$ 3 & 14 $\pm$ 2 & 
329 $\pm$ 33 & 810 $\pm$ 250&400 $\pm$ 160&250 $\pm$ 120& -- \\
      41 &        204.25832 &       -29.925232 & 25 $\pm$ 3 & 16 $\pm$ 2 & 
313 $\pm$ 25 & 790 $\pm$ 170&410 $\pm$ 150&230 $\pm$ 100&47 $\pm$ 2\\
      42 &        204.28258 &       -29.887080 & 11 $\pm$ 4 & 19 $\pm$ 4 & 
314 $\pm$ 52 & 720 $\pm$ 220&340 $\pm$ 240&--&22 $\pm$ 3\\
      43 &        204.21262 &       -29.876963 & 34 $\pm$ 4 & 32 $\pm$ 5 & 
884 $\pm$ 92 & 1350 $\pm$ 310&660 $\pm$ 310&170 $\pm$ 160&23 $\pm$ 3\\
      44 &        204.27939 &       -29.857758 & 10 $\pm$ 4 & 20 $\pm$ 4 & 
312 $\pm$ 60 & 1360 $\pm$ 540&630 $\pm$ 320&390 $\pm$ 200&25 $\pm$ 3\\
      45 &        204.28841 &       -29.857795 & 170 $\pm$ 10 & 120 $\pm$ 10 & 
3080 $\pm$ 210 & 4000 $\pm$ 650&1220 $\pm$ 470&300 $\pm$ 200&72 $\pm$ 6\\
      46 &        204.27239 &       -29.821631 & 17 $\pm$ 4 & 17 $\pm$ 5 & 
424 $\pm$ 62 & 1140 $\pm$ 270&770 $\pm$ 290&370 $\pm$ 180&32 $\pm$ 2\\
      47 &        204.23190 &       -29.881713 & 27 $\pm$ 5 & 12 $\pm$ 5 & 
724 $\pm$ 84 & 2010 $\pm$ 680&950 $\pm$ 350&690 $\pm$ 220&21 $\pm$ 2\\
      48 &        204.17584 &       -29.875390 & 37 $\pm$ 4 & 12 $\pm$ 2 & 
266 $\pm$ 38 & 1530 $\pm$ 230&1180 $\pm$ 270&480 $\pm$ 110&8 $\pm$ 4\\
      49 &        204.26047 &       -29.832599 & 25 $\pm$ 3 & 20 $\pm$ 2 & 
374 $\pm$ 42 & 1350 $\pm$ 280&750 $\pm$ 380&350 $\pm$ 190&10 $\pm$ 2\\
      50 &        204.26316 &       -29.827777 & 92 $\pm$ 9 & 37 $\pm$ 4 & 
674 $\pm$ 82 & 1740 $\pm$ 370&1410 $\pm$ 490&210 $\pm$ 170&34 $\pm$ 2\\
      51 &        204.26779 &       -29.925076 & 28 $\pm$ 3 & 19 $\pm$ 2 & 
428 $\pm$ 34 & 1550 $\pm$ 190&810 $\pm$ 190&400 $\pm$ 120&38 $\pm$ 3\\
      52 &        204.28075 &       -29.909675 & 47 $\pm$ 4 & 37 $\pm$ 3 & 
605 $\pm$ 44 & 1400 $\pm$ 220&610 $\pm$ 180&250 $\pm$ 140&77 $\pm$ 3\\
      53 &        204.32062 &       -29.886212 & 14 $\pm$ 2 & 12 $\pm$ 1 & 
223 $\pm$ 19 & 410 $\pm$ 110&450 $\pm$ 170&270 $\pm$ 110&23.5 $\pm$ 0.8\\
      54 &        204.23495 &       -29.885641 & 15 $\pm$ 5 & 18 $\pm$ 6 & 
294 $\pm$ 64 & 780 $\pm$ 580&280 $\pm$ 260&--&--\\
      55 &        204.28509 &       -29.878952 & 26 $\pm$ 5 & 17 $\pm$ 4 & 
539 $\pm$ 63 & 1010 $\pm$ 250&600 $\pm$ 310&240 $\pm$ 190&22 $\pm$ 4\\
      56 &        204.27136 &       -29.804832 & 41 $\pm$ 3 & 27 $\pm$ 2 & 
695 $\pm$ 43 & 1450 $\pm$ 290&590 $\pm$ 180&250 $\pm$ 110&39 $\pm$ 2\\
      57 &        204.24438 &       -29.801254 & 8 $\pm$ 2 & 10 $\pm$ 2 & 
213 $\pm$ 31 & 280 $\pm$ 140&360 $\pm$ 150&200 $\pm$ 110&26 $\pm$ 2\\
      58 &        204.22839 &       -29.926435 & 22 $\pm$ 2 & 9 $\pm$ 1 & 
287 $\pm$ 20 & 1120 $\pm$ 170&700 $\pm$ 190&370 $\pm$ 110&10 $\pm$ 0.7\\
      59 &        204.17916 &       -29.878426 & 25 $\pm$ 3 & 9 $\pm$ 2 & 
311 $\pm$ 42 & 1380 $\pm$ 240&890 $\pm$ 240&500 $\pm$ 110&36 $\pm$ 3\\
      60 &        204.21423 &       -29.871364 & 20 $\pm$ 4 & 15 $\pm$ 4 & 
296 $\pm$ 66 & 800 $\pm$ 330&860 $\pm$ 340&--&--\\
      61 &        204.26419 &       -29.850760 & 17 $\pm$ 4 & 20 $\pm$ 7 & 
515 $\pm$ 90 & 1050 $\pm$ 570&710 $\pm$ 340&220 $\pm$ 170&51 $\pm$ 3\\
      62 &        204.26217 &       -29.847377 & 20 $\pm$ 5 & 18 $\pm$ 6 & 
432 $\pm$ 88 & 1230 $\pm$ 560&640 $\pm$ 340&190 $\pm$ 150&86 $\pm$ 6\\
      63 &        204.30113 &       -29.838620 & 15 $\pm$ 3 & 9 $\pm$ 1 & 
265 $\pm$ 40 & 800 $\pm$ 280&640 $\pm$ 240&350 $\pm$ 170&14 $\pm$ 2\\
      64 &        204.21281 &       -29.835625 & 24 $\pm$ 3 & 16 $\pm$ 2 & 
365 $\pm$ 45 & 890 $\pm$ 200&540 $\pm$ 260&310 $\pm$ 160&46 $\pm$ 2\\
      65 &        204.23465 &       -29.825908 & 14 $\pm$ 3 & 6 $\pm$ 4 & 
201 $\pm$ 50 & 800 $\pm$ 240&670 $\pm$ 310&250 $\pm$ 170&17 $\pm$ 2\\
      66 &        204.27451 &       -29.895053 & 24 $\pm$ 6 & 11 $\pm$ 3 & 
423 $\pm$ 62 & 1180 $\pm$ 260&820 $\pm$ 260&330 $\pm$ 180&35 $\pm$ 2\\
      67 &        204.20733 &       -29.869728 & 17 $\pm$ 4 & 13 $\pm$ 4 & 
387 $\pm$ 68 & 450 $\pm$ 240&570 $\pm$ 240&200 $\pm$ 150&26 $\pm$ 4\\
      68 &        204.28385 &       -29.911307 & 9 $\pm$ 2 & 6 $\pm$ 1 & 
194 $\pm$ 20 & 390 $\pm$ 110&290 $\pm$ 140&150 $\pm$ 120&5 $\pm$ 2\\
      69 &        204.27882 &       -29.887547 & 23 $\pm$ 6 & 11 $\pm$ 3 & 
290 $\pm$ 49 & 1590 $\pm$ 270&1480 $\pm$ 360&690 $\pm$ 220&26 $\pm$ 2\\
      70 &        204.29993 &       -29.850675 & 10 $\pm$ 3 & 8 $\pm$ 2 & 
195 $\pm$ 39 & 410 $\pm$ 270&330 $\pm$ 270&180 $\pm$ 150&25 $\pm$ 2\\
      71 &        204.21230 &       -29.881552 & 19 $\pm$ 4 & 7 $\pm$ 5 & 
285 $\pm$ 70 & 1200 $\pm$ 280&740 $\pm$ 300&300 $\pm$ 190&4 $\pm$ 3\\
      72 &        204.22276 &       -29.853089 & 21 $\pm$ 4 & 24 $\pm$ 7 & 
407 $\pm$ 74 & 1360 $\pm$ 380&750 $\pm$ 380&440 $\pm$ 190&12 $\pm$ 3\\
      73 &        204.21339 &       -29.841575 & 18 $\pm$ 3 & 26 $\pm$ 4 & 
308 $\pm$ 46 & 530 $\pm$ 220&700 $\pm$ 320&170 $\pm$ 140&30 $\pm$ 2\\
      74 &        204.29638 &       -29.826215 & 11 $\pm$ 4 & 7 $\pm$ 2 & 
157 $\pm$ 40 & 570 $\pm$ 270&480 $\pm$ 180&300 $\pm$ 140&9 $\pm$ 4\\
      75 &        204.24362 &       -29.804870 & 14 $\pm$ 3 & 9 $\pm$ 2 & 
215 $\pm$ 30 & 680 $\pm$ 160&530 $\pm$ 200&350 $\pm$ 140&23 $\pm$ 2\\
      76 &        204.28500 &       -29.866274 & 12 $\pm$ 4 & 16 $\pm$ 3 & 
271 $\pm$ 49 & 1180 $\pm$ 330&680 $\pm$ 350&300 $\pm$ 190&19 $\pm$ 2\\
      77 &        204.30424 &       -29.860591 & 12 $\pm$ 2 & 5 $\pm$ 1 & 
185 $\pm$ 28 & 750 $\pm$ 200&600 $\pm$ 260&220 $\pm$ 150&14 $\pm$ 1\\
      78 &        204.28876 &       -29.852195 & 13 $\pm$ 3 & 4 $\pm$ 3 & 
286 $\pm$ 49 & 890 $\pm$ 280&320 $\pm$ 300&--&16 $\pm$ 3\\
      79 &        204.28907 &       -29.846556 & 12 $\pm$ 2 & 6 $\pm$ 3 & 
161 $\pm$ 33 & 480 $\pm$ 260&560 $\pm$ 350&--&20 $\pm$ 3\\
      80 &        204.19144 &       -29.880882 & 10 $\pm$ 2 & 5 $\pm$ 1 & 
135 $\pm$ 20 & 530 $\pm$ 190&480 $\pm$ 200&330 $\pm$ 120&3 $\pm$ 2\\
      81 &        204.32204 &       -29.864861 & 10 $\pm$ 1 & 4.6 $\pm$ 0.8 & 
110 $\pm$ 12 & 410 $\pm$ 150&230 $\pm$ 150&110 $\pm$ 72&9 $\pm$ 1\\
      82 &        204.20639 &       -29.859928 & 14 $\pm$ 4 & 13 $\pm$ 3 & 
224 $\pm$ 47 & 570 $\pm$ 270&350 $\pm$ 290&430 $\pm$ 170&16 $\pm$ 3\\
      83 &        204.19682 &       -29.818118 & 15 $\pm$ 1 & 8.5 $\pm$ 0.7 & 
146 $\pm$ 12 & 584 $\pm$ 92&340 $\pm$ 120&180 $\pm$ 73&12.8 $\pm$ 0.5\\
      84 &        204.29686 &       -29.926976 & 16 $\pm$ 2 & 13 $\pm$ 1 & 
95 $\pm$ 11 & 400 $\pm$ 120&290 $\pm$ 130&145 $\pm$ 81&21 $\pm$ 1\\
      85 &        204.29044 &       -29.908584 & 10 $\pm$ 2 & 7 $\pm$ 0.8 & 
94 $\pm$ 14 & 309 $\pm$ 93&310 $\pm$ 140&220 $\pm$ 130&3 $\pm$ 1\\
      86 &        204.24101 &       -29.839948 & 10 $\pm$ 4 & 5 $\pm$ 2 & 
113 $\pm$ 35 & 840 $\pm$ 260&660 $\pm$ 370&340 $\pm$ 180&34 $\pm$ 2\\
      87 &        204.27353 &       -29.917803 & 18 $\pm$ 2 & 12 $\pm$ 2 & 
207 $\pm$ 30 & 840 $\pm$ 170&570 $\pm$ 180&260 $\pm$ 120&60 $\pm$ 3\\
      88 &        204.23122 &       -29.914030 & 8 $\pm$ 1 & 3.4 $\pm$ 0.7 & 
96 $\pm$ 14 & 540 $\pm$ 170&220 $\pm$ 160&--&5.5 $\pm$ 0.7\\
      89 &        204.23179 &       -29.802769 & 13 $\pm$ 3 & 3 $\pm$ 1 & 
134 $\pm$ 29 & 550 $\pm$ 180&390 $\pm$ 140&400 $\pm$ 120&2 $\pm$ 2\\
      90 &        204.20338 &       -29.873538 & 11 $\pm$ 4 & 30 $\pm$ 4 & 
148 $\pm$ 57 & 1050 $\pm$ 260&530 $\pm$ 250&190 $\pm$ 150&9 $\pm$ 3\\

\end{longtable}
\end{center}

\begin{center}
\small
\begin{longtable}{cccccccc}
\caption{HB measurement SED fitting parameters}\\
\label{HB_sed_table}\\

\hline
 & $\chi_{\text{UV}}$ & $\chi_{\text {col}}$ & T$_{\text {dust}}$ & M$_{\text {dust}}$ & F$_{24}$ & L$_{\text {dust}}$ & $\chi^2_{\text{FIT}}$ \\
 &                   &                     &    [K]   &       [$10^4~M_\odot$]  &     &   [$10^{40}$erg/s]  &   \\     
\hline
 \endfirsthead
\multicolumn{3}{c}%
{\tablename\ \thetable{} -- continued from previous page}\\
\hline
& $\chi_{\text{UV}}$ & $\chi_{\text {col}}$ & T$_{\text {dust}}$ & M$_{\text {dust}}$ & F$_{24}$ & L$_{\text {dust}}$ & $\chi^2_{\text{FIT}}$ \\
 &                   &                     &    [K]   &       [$10^4~M_\odot$]  &     &   [$10^{40}$erg/s]  &   \\    
\hline
\endhead

       1 & 1.3$\pm$0.7 & 9$\pm$2 & 22.9$\pm$0.8 & 16$\pm$5 & 0.8$\pm$0.2 & 47
$\pm$1&   0.6\\
       2 & 0.3$\pm$0.1 & 9$\pm$1& 19.6$\pm$0.1 & 18$\pm$3 & 0.9$\pm$0.1
 & 28.7$\pm$0.6 &   5.0\\
       3 & 1.4$\pm$0.6 & 6$\pm$2 & 22.0$\pm$0.6 & 20$\pm$5 & 0.9$\pm$0.2 & 45
$\pm$2 &   1.2\\
       4 & 2$\pm$1& 8$\pm$3 & 24$\pm$1& 12$\pm$5 & 0.7$\pm$0.2 & 35$\pm$2 & 
  2.0\\
       5 & 2$\pm$1& 4.$\pm$1& 23.6$\pm$0.7 & 14$\pm$3 & 0.4$\pm$0.2 & 32
$\pm$2 &   1.0\\
       6 & 3$\pm$2 & 6$\pm$2 & 26$\pm$1& 7$\pm$3 & 0.6$\pm$0.1 & 31$\pm$1& 
  0.6\\
       7 & 7$\pm$4 & 8$\pm$3 & 31$\pm$3 & 2$\pm$1& 0.6$\pm$0.2 & 24$\pm$1& 
  1.1\\
       8 & 1.3$\pm$0.7 & 9$\pm$1& 23.1$\pm$0.6 & 11$\pm$4 & 0.8$\pm$0.2 & 25
$\pm$1&   6.6\\
       9 & 2$\pm$1& 6$\pm$3 & 22.9$\pm$0.7 & 30$\pm$10 & 0.6$\pm$0.1 & 52
$\pm$2 &   4.5\\
      10 & 2$\pm$1& 2$\pm$1& 21$\pm$1& 22$\pm$9 & 0.7$\pm$0.2 & 35$\pm$1
 &   2.5\\
      11 & 1.4$\pm$0.6 & 6$\pm$2 & 21.8$\pm$0.4 & 14$\pm$3 & 0.7$\pm$0.2 & 23
$\pm$1&   0.5\\
      12 & 1.4$\pm$0.6 & 8$\pm$2 & 22.6$\pm$0.4 & 22$\pm$6 & 0.6$\pm$0.1 & 40.6
$\pm$0.9 &   0.4\\
      13 & 3$\pm$2 & 3$\pm$2 & 23.2$\pm$0.3 & 12$\pm$3 & 0.4$\pm$0.1 & 27.1
$\pm$0.8 &   0.4\\
      14 & 0.6$\pm$0.2 & 9$\pm$2 & 21.6$\pm$0.6 & 27$\pm$8 & 0.5$\pm$0.2 & 31
$\pm$2 &   1.9\\
      15 & 0.4$\pm$0.3 & 6$\pm$4 & 18.2$\pm$0.8 & 30$\pm$10 & 0.7$\pm$0.2 & 14.2
$\pm$0.8 &   2.3\\
      16 & 1.4$\pm$0.6 & 6$\pm$2 & 21.6$\pm$0.2 & 39$\pm$7 & 0.8$\pm$0.1 & 
74.1$\pm$0.5 &   2.9\\
      17 & 2$\pm$1& 3.$\pm$1& 22.9$\pm$0.7 & 4.4$\pm$0.8 & 0.6$\pm$0.1 & 
10.3$\pm$0.3 &   0.9\\
      18 & 0.3$\pm$0.2 & 9$\pm$2 & 20$\pm$1& 13$\pm$6 & 0.9$\pm$0.1 & 14.6
$\pm$0.8 &   1.5\\
      19 & 1.2$\pm$0.8 & 7$\pm$3 & 22$\pm$1& 8$\pm$4 & 0.7$\pm$0.3 & 12$\pm$1
 &   5.8\\
      20 & 1.3$\pm$0.8 & 7$\pm$3 & 21.6$\pm$0.7 & 13$\pm$4 & 0.6$\pm$0.2 & 16.2
$\pm$0.8 &   0.2\\
      21 & 0.5$\pm$0.2 & 9$\pm$2 & 20.8$\pm$0.5 & 17$\pm$4 & 0.6$\pm$0.2 & 17
$\pm$1 &   1.7\\
      22 & 2$\pm$1& 4$\pm$2 & 23.3$\pm$0.3 & 8$\pm$2 & 0.4$\pm$0.2 & 18
$\pm$1 &   0.9\\
      23 & 2$\pm$2 & 6$\pm$4 & 23$\pm$2 & 5$\pm$3 & 0.4$\pm$0.4 & 7$\pm$1 & 
  1.8\\
      24 & 1.3$\pm$0.7 & 3$\pm$2 & 19.9$\pm$0.4 & 20$\pm$5 & 0.13$\pm$0.1 & 12.9
$\pm$0.8 &   0.9\\
      25 & 3$\pm$2 & 4$\pm$2 & 23.7$\pm$0.7 & 14$\pm$4 & 0.5$\pm$0.1 & 37
$\pm$1 &   1.3\\
      26 & 1.0$\pm$1.0 & 7$\pm$4 & 20$\pm$2 & 9$\pm$6 & 0.8$\pm$0.2 & 10$\pm$1
 &   5.4\\
      27 & 6$\pm$4 & 5$\pm$3 & 29$\pm$3 & 1.5$\pm$1.0 & 0.5$\pm$0.2 & 8.8$\pm$
0.8 &   0.6\\
      28 & 1.4$\pm$0.6 & 4$\pm$2 & 20.8$\pm$0.4 & 12$\pm$3 & 0.13$\pm$0.1 & 10.6
$\pm$0.5 &   0.7\\
      29 & 0.5$\pm$0.3 & 8$\pm$3 & 20.2$\pm$0.8 & 9$\pm$4 & 0.7$\pm$0.2 & 7.8
$\pm$0.6 &   1.1\\
      30 & 0.6$\pm$0.4 & 7$\pm$4 & 20$\pm$1& 6$\pm$3 & 0.9$\pm$0.2 & 6.6$\pm$
0.5 &   1.4\\
      31 & 3$\pm$2 & 1.2$\pm$0.9 & 22$\pm$1& 9$\pm$4 & 0.4$\pm$0.2 & 17$\pm$1
 &   1.4\\
      32 & 3$\pm$3 & 7$\pm$4 & 26$\pm$2 & 2$\pm$1& 0.5$\pm$0.3 & 6.8$\pm$0.7
 &   0.6\\
      33 & 1.2$\pm$0.8 & 7$\pm$3 & 21.5$\pm$0.8 & 5$\pm$2 & 0.6$\pm$0.2 & 5.9
$\pm$0.4 &   2.3\\
      34 & 0.6$\pm$0.4 & 8$\pm$3 & 21$\pm$1& 8$\pm$4 & 0.6$\pm$0.2 & 7.3$\pm$
0.7 &   0.5\\
      35 & 0.2$\pm$0.1 & 8$\pm$2 & 18$\pm$1& 20$\pm$10 & 0.6$\pm$0.3 & 7.5
$\pm$0.6 &   4.5\\
      36 & 3$\pm$2 & 3$\pm$2 & 22.8$\pm$0.7 & 9$\pm$3 & 0.4$\pm$0.1 & 17.9$\pm$
0.6 &   1.3\\
      37 & 0.3$\pm$0.1 & 9$\pm$1& 19.6$\pm$0.1 & 14$\pm$3 & 0.5$\pm$0.3 & 
9.1$\pm$0.5 &   4.0\\
      38 & 1.2$\pm$0.8 & 7$\pm$3 & 21.6$\pm$0.7 & 16$\pm$6 & 0.6$\pm$0.2 & 21
$\pm$1&   0.9\\
      39 & 6$\pm$5 & 3$\pm$3 & 25$\pm$3 & 2$\pm$2 & 0.5$\pm$0.4 & 7$\pm$1& 
  0.8\\
      40 & 2$\pm$1& 4$\pm$2 & 22$\pm$1& 6$\pm$3 & 0.3$\pm$0.2 & 7.2$\pm$0.5
 &   0.6\\
      41 & 2$\pm$1& 3$\pm$2 & 21.3$\pm$0.8 & 6$\pm$3 & 0.3$\pm$0.1 & 7.2$\pm$
0.6 &   0.9\\
      42 & 1.2$\pm$0.9 & 7$\pm$3 & 22$\pm$2 & 4$\pm$3 & 0.7$\pm$0.3 & 6.0$\pm$
0.7 &   1.4\\
      43 & 2$\pm$1& 6$\pm$2 & 24.7$\pm$0.5 & 5$\pm$1& 0.4$\pm$0.2 & 14.1
$\pm$0.6 &   0.4\\
      44 & 0.3$\pm$0.2 & 9$\pm$2 & 18.9$\pm$0.8 & 13$\pm$6 & 0.8$\pm$0.3 & 7.5
$\pm$0.8 &   1.6\\
      45 & 8$\pm$3 & 2$\pm$1& 27$\pm$1& 8$\pm$3 & 0.16$\pm$0.1 & 48$\pm$2
 &   2.9\\
      46 & 0.5$\pm$0.3 & 8$\pm$2 & 20.7$\pm$0.9 & 12$\pm$5 & 0.5$\pm$0.4 & 9.0
$\pm$0.8 &   0.4\\
      47 & 0.6$\pm$0.3 & 9$\pm$1& 21.5$\pm$0.9 & 15$\pm$6 & 0.11$\pm$0.1 & 
13.6$\pm$0.9 &   2.1\\
      48 & 0.5$\pm$0.2 & 4$\pm$2 & 18.2$\pm$0.3 & 25$\pm$6 & 0.04$\pm$0.03 & 9.8
$\pm$0.3 &   3.9\\
      49 & 0.6$\pm$0.3 & 5$\pm$2 & 19.9$\pm$0.6 & 14$\pm$5 & 0.41$\pm$0.1 & 9.6
$\pm$0.5 &   0.1\\
      50 & 4$\pm$2 & 0.5$\pm$0.4 & 22.1$\pm$0.7 & 10$\pm$3 & 0.05$\pm$0.04 & 17
$\pm$1&   4.4\\
      51 & 0.55$\pm$0.1 & 6$\pm$2 & 20.03$\pm$0.06 & 15$\pm$2 & 0.3$\pm$0.2 & 
10.7$\pm$0.3 &   0.6\\
      52 & 2$\pm$1& 2$\pm$1& 22.1$\pm$0.7 & 7$\pm$1& 0.43$\pm$0.1 & 13.3
$\pm$0.5 &   1.1\\
      53 & 2$\pm$1& 3$\pm$2 & 21$\pm$1& 4$\pm$2 & 0.5$\pm$0.2 & 4.8$\pm$0.4
 &   4.3\\
      54 & 4$\pm$3 & 6$\pm$5 & 23$\pm$3 & 5$\pm$4 & 0.6$\pm$0.4 & 6$\pm$1& 
  0.2\\
      55 & 2$\pm$1& 7$\pm$4 & 23$\pm$1& 6$\pm$3 & 0.3$\pm$0.3 & 10$\pm$1& 
  0.6\\
      56 & 2$\pm$1& 3.$\pm$1& 22.9$\pm$0.7 & 6.9$\pm$0.9 & 0.21$\pm$0.08 & 
12.9$\pm$0.2 &   1.0\\
      57 & 1.2$\pm$0.9 & 7$\pm$3 & 22$\pm$2 & 3$\pm$2 & 0.6$\pm$0.3 & 3.9$\pm$
0.5 &   2.9\\
      58 & 0.5$\pm$0.2 & 7$\pm$2 & 19.8$\pm$0.2 & 12$\pm$2 & 0.03$\pm$0.02 & 
7.45$\pm$0.04 &   3.1\\
      59 & 0.4$\pm$0.2 & 7$\pm$3 & 19.2$\pm$0.4 & 19$\pm$6 & 0.06$\pm$0.05 & 9.0
$\pm$0.5 &   2.0\\
      60 & 1.2$\pm$0.9 & 5$\pm$3 & 20$\pm$1& 9$\pm$4 & 0.4$\pm$0.4 & 7.1$\pm$
1.0 &   1.5\\
      61 & 1.2$\pm$0.9 & 8$\pm$2 & 22$\pm$2 & 8$\pm$5 & 0.5$\pm$0.4 & 9$\pm$1
 &   0.3\\
      62 & 1.2$\pm$0.9 & 7$\pm$4 & 22$\pm$1& 8$\pm$4 & 0.5$\pm$0.4 & 9$\pm$1
 &   0.4\\
      63 & 0.5$\pm$0.3 & 8$\pm$3 & 20.1$\pm$0.9 & 10$\pm$4 & 0.2$\pm$0.2 & 6.4
$\pm$0.6 &   0.5\\
      64 & 1.3$\pm$0.7 & 5$\pm$2 & 21.1$\pm$0.6 & 8$\pm$2 & 0.3$\pm$0.2 & 8.1
$\pm$0.6 &   0.6\\
      65 & 0.5$\pm$0.4 & 7$\pm$3 & 19$\pm$1& 12$\pm$7 & 0.3$\pm$0.3 & 5.5$\pm$
0.8 &   0.5\\
      66 & 0.6$\pm$0.4 & 7$\pm$3 & 21.1$\pm$0.8 & 11$\pm$4 & 0.14$\pm$0.1 & 9.1
$\pm$0.8 &   0.8\\
      67 & 3$\pm$2 & 6$\pm$4 & 24$\pm$2 & 4$\pm$2 & 0.4$\pm$0.4 & 6.5$\pm$0.9 & 
  2.2\\
      68 & 1.2$\pm$0.8 & 7$\pm$3 & 21.8$\pm$0.8 & 3$\pm$1& 0.3$\pm$0.3 & 3.7
$\pm$0.4 &   0.7\\
      69 & 0.20$\pm$0.1 & 8$\pm$2 & 18.1$\pm$0.4 & 33$\pm$8 & 0.12$\pm$0.1 & 
10.2$\pm$0.7 &   1.0\\
      70 & 2$\pm$1& 6$\pm$4 & 22$\pm$2 & 5$\pm$4 & 0.4$\pm$0.4 & 4.1$\pm$0.7
 &   0.3\\
      71 & 0.5$\pm$0.3 & 7$\pm$3 & 19.7$\pm$0.9 & 13$\pm$6 & 0.2$\pm$0.2 & 7.6
$\pm$0.8 &   0.5\\
      72 & 0.4$\pm$0.3 & 7$\pm$3 & 19$\pm$1& 17$\pm$8 & 0.6$\pm$0.4 & 10$\pm$
1&   0.1\\
      73 & 2$\pm$1& 3$\pm$3 & 20$\pm$2 & 8$\pm$5 & 0.8$\pm$0.2 & 7.0$\pm$0.6
 &   1.5\\
      74 & 0.4$\pm$0.3 & 7$\pm$4 & 19$\pm$1& 11$\pm$6 & 0.4$\pm$0.4 & 4.6$\pm$
0.7 &   0.5\\
      75 & 0.6$\pm$0.4 & 7$\pm$4 & 19.7$\pm$0.9 & 9$\pm$4 & 0.3$\pm$0.3 & 5.5
$\pm$0.5 &   1.3\\
      76 & 0.3$\pm$0.2 & 8$\pm$2 & 18.9$\pm$0.8 & 15$\pm$6 & 0.6$\pm$0.3 & 7.2
$\pm$0.7 &   0.7\\
      77 & 0.4$\pm$0.2 & 8$\pm$3 & 19.6$\pm$0.8 & 9$\pm$4 & 0.09$\pm$0.08 & 4.9
$\pm$0.5 &   0.8\\
      78 & 1.2$\pm$0.9 & 8$\pm$2 & 22$\pm$1& 6$\pm$3 & 0.12$\pm$0.1 & 5.3$\pm$
0.6 &   2.0\\
      79 & 1.1$\pm$0.9 & 6$\pm$4 & 20$\pm$1& 7$\pm$4 & 0.3$\pm$0.3 & 4.0$\pm$
0.7 &   0.6\\
      80 & 0.3$\pm$0.2 & 7$\pm$3 & 18.8$\pm$0.8 & 10$\pm$4 & 0.2$\pm$0.2 & 4.1
$\pm$0.5 &   1.6\\
      81 & 1.3$\pm$0.8 & 4$\pm$2 & 20.1$\pm$0.5 & 4$\pm$2 & 0.10$\pm$0.09 & 3.0
$\pm$0.2 &   0.2\\
      82 & 1.0$\pm$1.0 & 6$\pm$4 & 19$\pm$2 & 10$\pm$7 & 0.5$\pm$0.3 & 5.6$\pm$
0.8 &   2.0\\
      83 & 1.3$\pm$0.8 & 3$\pm$2 & 19.5$\pm$0.4 & 7$\pm$2 & 0.21$\pm$0.1 & 4.3
$\pm$0.2 &   0.5\\
      84 & 0.6$\pm$0.4 & 0.4$\pm$0.3 & 16$\pm$1& 12$\pm$6 & 0.6$\pm$0.2 & 3.7
$\pm$0.2 &   0.9\\
      85 & 1.2$\pm$0.9 & 3$\pm$2 & 19$\pm$1 & 6$\pm$3 & 0.4$\pm$0.2 & 2.8
$\pm$0.3 &   1.3\\
      86 & 0.3$\pm$0.2 & 7$\pm$3 & 17$\pm$1& 20$\pm$10 & 0.2$\pm$0.2 & 4.5
$\pm$0.8 &   0.5\\
      87 & 0.6$\pm$0.3 & 4$\pm$2 & 19.2$\pm$0.7 & 11$\pm$4 & 0.3$\pm$0.2 & 6.1
$\pm$0.4 &   0.1\\
      88 & 1.2$\pm$0.9 & 5$\pm$3 & 19.9$\pm$1.0 & 4$\pm$2 & 0.07$\pm$0.06 & 2.5
$\pm$0.3 &   2.8\\
      89 & 0.4$\pm$0.3 & 7$\pm$4 & 18.6$\pm$0.9 & 11$\pm$5 & 0.06$\pm$0.05 & 4.0
$\pm$0.4 &   6.3\\
      90 & 0.20$\pm$0.1 & 6$\pm$4 & 16$\pm$1& 20$\pm$10 & 0.8$\pm$0.2 & 6.6
$\pm$0.7 &   6.6\\

\end{longtable}
\end{center}

\begin{center}
\small
\begin{longtable}{cccccccccc}
\caption{LB measurement source fluxes and positions}\\
\label{LBfluxes}\\

\hline
 & RA & $\delta$ & F$_{8\mu m}$ & F$_{24\mu m}$ & F$_{70\mu m}$ & F$_{160\mu m}$ & F$_{250\mu m}$ & F$_{350\mu m}$ & F$(H_\alpha)$\\
 & [deg] & [deg]& [mJy] & [mJy] & [mJy] & [mJy] & [mJy] & [mJy] & [$\mu$Jy] \\
\hline
\endfirsthead
\multicolumn{3}{c}%
{\tablename\ \thetable{} -- continued from previous page}\\
\hline
 & RA & $\delta$ & F$_{8\mu m}$ & F$_{24\mu m}$ & F$_{70\mu m}$ & F$_{160\mu m}$ & F$_{250\mu m}$ & F$_{350\mu m}$ & F$(H_\alpha)$\\
 & [deg] & [deg]& [mJy] & [mJy] & [mJy] & [mJy] & [mJy] & [mJy] & [$\mu$Jy] \\
\hline
\endhead

       1 &        204.27867 &       -29.826725 & 132 $\pm$ 9.0 & 302 $\pm$ 17 & 
3.17 $\pm$ 0.17 & 5.43 $\pm$ 0.55&2.63 $\pm$ 0.67&1.37 $\pm$ 0.34&70 $\pm$ 7.1\\
       2 &        204.26433 &       -29.900421 & 63 $\pm$ 6.9 & 197 $\pm$ 12 & 
1.69 $\pm$ 0.11 & 2.59 $\pm$ 0.40&1.36 $\pm$ 0.54&0.49 $\pm$ 0.32&52 $\pm$ 2.2\\
       3 &        204.22128 &       -29.858999 & 163 $\pm$ 12 & 283 $\pm$ 18 & 
3.18 $\pm$ 0.20 & 5.03 $\pm$ 0.84&2.25 $\pm$ 0.72&0.70 $\pm$ 0.33&133 $\pm$ 3.2
\\
       4 &        204.28271 &       -29.854724 & 114 $\pm$ 10. & 166 $\pm$ 11 & 
2.48 $\pm$ 0.16 & 4.40 $\pm$ 0.67&1.93 $\pm$ 0.72&0.87 $\pm$ 0.35&83 $\pm$ 3.7\\
       5 &        204.23187 &       -29.830846 & 142 $\pm$ 10. & 127 $\pm$ 8.7
 & 2.41 $\pm$ 0.15 & 4.04 $\pm$ 0.49&1.70 $\pm$ 0.57&0.65 $\pm$ 0.28&
106 $\pm$ 3.9\\
       6 &        204.22652 &       -29.884745 & 126 $\pm$ 9.8 & 192 $\pm$ 14 & 
2.49 $\pm$ 0.17 & 2.93 $\pm$ 0.71&1.26 $\pm$ 0.58&0.48 $\pm$ 0.31&205 $\pm$ 3.6
\\
       7 &        204.22894 &       -29.885796 & 84 $\pm$ 8.3 & 144 $\pm$ 12 & 
2.32 $\pm$ 0.17 & 1.87 $\pm$ 0.68&0.87 $\pm$ 0.59&--&131 $\pm$ 2.9\\
       8 &        204.22172 &       -29.882527 & 65 $\pm$ 7.6 & 146 $\pm$ 11 & 
2.08 $\pm$ 0.15 & 3.38 $\pm$ 0.65&2.64 $\pm$ 0.80&1.34 $\pm$ 0.37&79 $\pm$ 3.1\\
       9 &        204.21930 &       -29.863860 & 268 $\pm$ 17 & 386 $\pm$ 25 & 
4.94 $\pm$ 0.31 & 7.7 $\pm$ 1.2&2.60 $\pm$ 0.83&1.57 $\pm$ 0.44&238 $\pm$ 3.7\\
      10 &        204.22046 &       -29.880233 & 175 $\pm$ 14 & 264 $\pm$ 19 & 
2.29 $\pm$ 0.15 & 5.24 $\pm$ 0.80&3.18 $\pm$ 0.85&1.34 $\pm$ 0.37&153 $\pm$ 3.7
\\
      11 &        204.21971 &       -29.852660 & 93 $\pm$ 8.5 & 106 $\pm$ 11 & 
1.56 $\pm$ 0.13 & 3.21 $\pm$ 0.51&1.54 $\pm$ 0.69&0.42 $\pm$ 0.31&91 $\pm$ 3.6\\
      12 &        204.26957 &       -29.849437 & 159 $\pm$ 13 & 190 $\pm$ 13 & 
3.09 $\pm$ 0.21 & 5.41 $\pm$ 0.98&2.16 $\pm$ 0.87&1.05 $\pm$ 0.40&210 $\pm$ 3.7
\\
      13 &        204.28487 &       -29.869891 & 144 $\pm$ 13 & 125 $\pm$ 9.6 & 
1.79 $\pm$ 0.13 & 3.34 $\pm$ 0.54&1.37 $\pm$ 0.67&0.46 $\pm$ 0.31&101 $\pm$ 3.1
\\
      14 &        204.22800 &       -29.882334 & 116 $\pm$ 9.4 & 132 $\pm$ 11 & 
2.17 $\pm$ 0.16 & 4.14 $\pm$ 0.86&3.37 $\pm$ 0.80&1.27 $\pm$ 0.40&34 $\pm$ 2.5\\
      15 &        204.28634 &       -29.858535 & 70 $\pm$ 8.4 & 106 $\pm$ 11 & 
0.796 $\pm$ 0.080 & 2.93 $\pm$ 0.52&1.26 $\pm$ 0.67&0.69 $\pm$ 0.33&20 $\pm$ 3.6
\\
      16 &        204.28083 &       -29.852275 & 260 $\pm$ 20 & 422 $\pm$ 32 & 
5.04 $\pm$ 0.34 & 8.1 $\pm$ 1.1&5.5 $\pm$ 1.4&1.92 $\pm$ 0.46&193 $\pm$ 4.5\\
      17 &        204.16814 &       -29.855863 & 47 $\pm$ 2.9 & 48 $\pm$ 3.0 & 
0.665 $\pm$ 0.037 & 1.39 $\pm$ 0.19&0.85 $\pm$ 0.22&0.45 $\pm$ 0.12&
62 $\pm$ 0.93\\
      18 &        204.22554 &       -29.844806 & 51 $\pm$ 5.8 & 93 $\pm$ 7.2 & 
0.918 $\pm$ 0.085 & 2.19 $\pm$ 0.45&0.73 $\pm$ 0.62&--&83 $\pm$ 3.2\\
      19 &        204.21463 &       -29.883342 & 58 $\pm$ 7.4 & 91 $\pm$ 9.7 & 
1.06 $\pm$ 0.11 & 1.70 $\pm$ 0.42&2.07 $\pm$ 0.73&0.82 $\pm$ 0.32&55 $\pm$ 3.6\\
      20 &        204.28397 &       -29.881946 & 91 $\pm$ 10.0 & 76 $\pm$ 7.8 & 
1.200 $\pm$ 0.10 & 2.47 $\pm$ 0.44&1.11 $\pm$ 0.60&0.31 $\pm$ 0.29&76 $\pm$ 3.7
\\
      21 &        204.20849 &       -29.878634 & 77 $\pm$ 8.1 & 79 $\pm$ 7.3 & 
1.21 $\pm$ 0.12 & 3.75 $\pm$ 0.52&2.03 $\pm$ 0.74&1.14 $\pm$ 0.36&47 $\pm$ 4.1\\
      22 &        204.26997 &       -29.823576 & 97 $\pm$ 9.1 & 81 $\pm$ 8.0 & 
1.39 $\pm$ 0.12 & 3.10 $\pm$ 0.45&1.94 $\pm$ 0.62&0.78 $\pm$ 0.28&135 $\pm$ 7.3
\\
      23 &        204.29186 &       -29.858028 & 42 $\pm$ 5.9 & 38 $\pm$ 7.0 & 
0.619 $\pm$ 0.076 & 1.00 $\pm$ 0.40&0.93 $\pm$ 0.65&0.51 $\pm$ 0.33&98 $\pm$ 4.7
\\
      24 &        204.25244 &       -29.905010 & 66 $\pm$ 9.6 & 40 $\pm$ 4.0 & 
0.606 $\pm$ 0.072 & 1.62 $\pm$ 0.35&0.86 $\pm$ 0.48&0.29 $\pm$ 0.29&54 $\pm$ 2.3
\\
      25 &        204.21223 &       -29.844608 & 186 $\pm$ 12 & 165 $\pm$ 11 & 
2.82 $\pm$ 0.18 & 5.53 $\pm$ 0.58&2.50 $\pm$ 0.61&1.37 $\pm$ 0.33&184 $\pm$ 4.1
\\
      26 &        204.21877 &       -29.863393 & 74 $\pm$ 7.2 & 147 $\pm$ 13 & 
1.36 $\pm$ 0.13 & 2.17 $\pm$ 0.74&1.96 $\pm$ 0.72&--&155 $\pm$ 2.8\\
      27 &        204.28831 &       -29.849784 & 50 $\pm$ 5.9 & 51 $\pm$ 7.4 & 
0.849 $\pm$ 0.085 & 0.97 $\pm$ 0.40&0.80 $\pm$ 0.60&0.65 $\pm$ 0.30&67 $\pm$ 4.2
\\
      28 &        204.24629 &       -29.907226 & 38 $\pm$ 6.5 & 26 $\pm$ 3.3 & 
0.474 $\pm$ 0.071 & 0.93 $\pm$ 0.32&0.47 $\pm$ 0.45&--&49 $\pm$ 2.3\\
      29 &        204.18835 &       -29.879014 & 32 $\pm$ 3.9 & 32 $\pm$ 2.6 & 
0.473 $\pm$ 0.045 & 1.21 $\pm$ 0.30&0.83 $\pm$ 0.46&0.23 $\pm$ 0.19&
104 $\pm$ 3.5\\
      30 &        204.18091 &       -29.872982 & 22 $\pm$ 3.8 & 41 $\pm$ 3.6 & 
0.459 $\pm$ 0.049 & 0.75 $\pm$ 0.28&0.72 $\pm$ 0.42&--&71 $\pm$ 4.5\\
      31 &        204.23124 &       -29.845585 & 117 $\pm$ 11 & 105 $\pm$ 8.6 & 
1.066 $\pm$ 0.092 & 2.37 $\pm$ 0.56&0.78 $\pm$ 0.59&0.31 $\pm$ 0.30&29 $\pm$ 2.5
\\
      32 &        204.29158 &       -29.819528 & 31 $\pm$ 4.6 & 32 $\pm$ 3.1 & 
0.620 $\pm$ 0.070 & 1.25 $\pm$ 0.31&0.93 $\pm$ 0.25&0.81 $\pm$ 0.21&
109 $\pm$ 4.3\\
      33 &        204.22354 &       -29.813422 & 20 $\pm$ 3.3 & 22 $\pm$ 2.2 & 
0.328 $\pm$ 0.036 & 0.63 $\pm$ 0.23&0.48 $\pm$ 0.39&0.28 $\pm$ 0.23&19 $\pm$ 2.8
\\
      34 &        204.29732 &       -29.831076 & 49 $\pm$ 5.3 & 40 $\pm$ 3.3 & 
0.709 $\pm$ 0.063 & 2.64 $\pm$ 0.39&2.04 $\pm$ 0.42&1.12 $\pm$ 0.27&76 $\pm$ 3.3
\\
      35 &        204.27795 &       -29.823031 & 36 $\pm$ 4.8 & 51 $\pm$ 6.0 & 
0.518 $\pm$ 0.063 & 3.38 $\pm$ 0.45&1.79 $\pm$ 0.57&0.95 $\pm$ 0.30&18 $\pm$ 7.1
\\
      36 &        204.24819 &       -29.810471 & 84 $\pm$ 9.8 & 67 $\pm$ 5.0 & 
1.110 $\pm$ 0.086 & 2.30 $\pm$ 0.37&1.14 $\pm$ 0.50&0.42 $\pm$ 0.21&87 $\pm$ 7.0
\\
      37 &        204.26738 &       -29.899060 & 38 $\pm$ 5.8 & 46 $\pm$ 5.7 & 
0.586 $\pm$ 0.069 & 1.80 $\pm$ 0.40&0.91 $\pm$ 0.50&--&84 $\pm$ 2.0\\
      38 &        204.23665 &       -29.880081 & 102 $\pm$ 12 & 106 $\pm$ 8.4 & 
1.69 $\pm$ 0.16 & 4.3 $\pm$ 1.0&2.56 $\pm$ 0.70&1.03 $\pm$ 0.38&36 $\pm$ 2.4\\
      39 &        204.21163 &       -29.866269 & 42 $\pm$ 6.8 & 39 $\pm$ 6.3 & 
0.616 $\pm$ 0.096 & 1.59 $\pm$ 0.38&1.42 $\pm$ 0.54&0.72 $\pm$ 0.29&42 $\pm$ 4.3
\\
      40 &        204.31140 &       -29.843049 & 46 $\pm$ 4.3 & 29 $\pm$ 2.3 & 
0.458 $\pm$ 0.041 & 1.92 $\pm$ 0.32&1.15 $\pm$ 0.29&0.68 $\pm$ 0.19&42 $\pm$ 15
\\
      41 &        204.25832 &       -29.925232 & 42 $\pm$ 3.7 & 30 $\pm$ 2.2 & 
0.447 $\pm$ 0.032 & 1.27 $\pm$ 0.23&0.72 $\pm$ 0.25&0.39 $\pm$ 0.16&66 $\pm$ 1.5
\\
      42 &        204.28258 &       -29.887080 & 35 $\pm$ 5.3 & 41 $\pm$ 4.7 & 
0.561 $\pm$ 0.065 & 1.95 $\pm$ 0.32&1.11 $\pm$ 0.39&0.82 $\pm$ 0.32&46 $\pm$ 3.3
\\
      43 &        204.21262 &       -29.876963 & 86 $\pm$ 8.3 & 93 $\pm$ 7.8 & 
1.64 $\pm$ 0.14 & 2.97 $\pm$ 0.50&1.72 $\pm$ 0.72&0.77 $\pm$ 0.29&65 $\pm$ 3.8\\
      44 &        204.27939 &       -29.857758 & 29 $\pm$ 5.6 & 44 $\pm$ 6.3 & 
0.483 $\pm$ 0.076 & 3.18 $\pm$ 0.72&1.25 $\pm$ 0.62&0.81 $\pm$ 0.35&35 $\pm$ 3.5
\\
      45 &        204.28841 &       -29.857795 & 321 $\pm$ 28 & 271 $\pm$ 26 & 
4.76 $\pm$ 0.34 & 6.6 $\pm$ 1.1&1.59 $\pm$ 0.88&0.63 $\pm$ 0.34&161 $\pm$ 6.5\\
      46 &        204.27239 &       -29.821631 & 57 $\pm$ 6.5 & 51 $\pm$ 6.7 & 
0.782 $\pm$ 0.085 & 2.94 $\pm$ 0.44&1.79 $\pm$ 0.59&0.90 $\pm$ 0.29&
130 $\pm$ 7.0\\
      47 &        204.23190 &       -29.881713 & 69 $\pm$ 7.6 & 64 $\pm$ 8.1 & 
1.22 $\pm$ 0.12 & 2.92 $\pm$ 0.83&1.94 $\pm$ 0.67&1.09 $\pm$ 0.38&68 $\pm$ 2.2\\
      48 &        204.17584 &       -29.875390 & 77 $\pm$ 7.8 & 27 $\pm$ 3.2 & 
0.490 $\pm$ 0.053 & 1.94 $\pm$ 0.33&1.24 $\pm$ 0.50&0.20 $\pm$ 0.17&22 $\pm$ 4.4
\\
      49 &        204.26047 &       -29.832599 & 46 $\pm$ 5.1 & 35 $\pm$ 2.9 & 
0.446 $\pm$ 0.054 & 2.36 $\pm$ 0.40&1.75 $\pm$ 0.62&0.80 $\pm$ 0.31&--\\
      50 &        204.26316 &       -29.827777 & 140 $\pm$ 15 & 66 $\pm$ 6.4 & 
0.87 $\pm$ 0.11 & 2.59 $\pm$ 0.50&2.08 $\pm$ 0.73&0.42 $\pm$ 0.27&43 $\pm$ 3.1\\
      51 &        204.26779 &       -29.925076 & 53 $\pm$ 4.1 & 38 $\pm$ 2.9 & 
0.635 $\pm$ 0.045 & 2.16 $\pm$ 0.26&1.13 $\pm$ 0.29&0.49 $\pm$ 0.17&63 $\pm$ 3.3
\\
      52 &        204.28075 &       -29.909675 & 80 $\pm$ 6.2 & 64 $\pm$ 4.1 & 
0.854 $\pm$ 0.057 & 2.38 $\pm$ 0.31&1.14 $\pm$ 0.30&0.52 $\pm$ 0.21&
135 $\pm$ 3.2\\
      53 &        204.32062 &       -29.886212 & 26 $\pm$ 2.9 & 19 $\pm$ 1.9 & 
0.294 $\pm$ 0.029 & 0.72 $\pm$ 0.17&0.84 $\pm$ 0.31&0.35 $\pm$ 0.17&36 $\pm$ 1.1
\\
      54 &        204.23495 &       -29.885641 & 38 $\pm$ 6.8 & 42 $\pm$ 7.1 & 
0.551 $\pm$ 0.088 & 1.33 $\pm$ 0.70&0.88 $\pm$ 0.51&0.32 $\pm$ 0.27&15 $\pm$ 2.6
\\
      55 &        204.28509 &       -29.878952 & 65 $\pm$ 8.0 & 41 $\pm$ 6.7 & 
0.838 $\pm$ 0.089 & 1.55 $\pm$ 0.41&0.70 $\pm$ 0.63&--&61 $\pm$ 3.9\\
      56 &        204.27136 &       -29.804832 & 51 $\pm$ 9.2 & 36 $\pm$ 3.6 & 
0.753 $\pm$ 0.097 & 1.31 $\pm$ 0.41&0.60 $\pm$ 0.41&0.25 $\pm$ 0.16&46 $\pm$ 6.9
\\
      57 &        204.24438 &       -29.801254 & 15 $\pm$ 3.8 & 17 $\pm$ 2.5 & 
0.257 $\pm$ 0.043 & 0.73 $\pm$ 0.26&0.86 $\pm$ 0.48&0.36 $\pm$ 0.19&18 $\pm$ 7.1
\\
      58 &        204.22839 &       -29.926435 & 37 $\pm$ 3.0 & 18 $\pm$ 1.5 & 
0.385 $\pm$ 0.030 & 1.44 $\pm$ 0.22&0.81 $\pm$ 0.28&0.30 $\pm$ 0.16&
15 $\pm$ 0.82\\
      59 &        204.17916 &       -29.878426 & 54 $\pm$ 5.6 & 27 $\pm$ 3.4 & 
0.561 $\pm$ 0.060 & 1.79 $\pm$ 0.35&1.04 $\pm$ 0.48&0.25 $\pm$ 0.19&
139 $\pm$ 3.6\\
      60 &        204.21423 &       -29.871364 & 40 $\pm$ 6.1 & 31 $\pm$ 5.4 & 
0.440 $\pm$ 0.082 & 1.58 $\pm$ 0.48&1.52 $\pm$ 0.68&0.43 $\pm$ 0.29&--\\
      61 &        204.26419 &       -29.850760 & 53 $\pm$ 6.9 & 66 $\pm$ 9.3 & 
1.01 $\pm$ 0.12 & 2.89 $\pm$ 0.82&1.13 $\pm$ 0.76&0.71 $\pm$ 0.37&98 $\pm$ 4.0\\
      62 &        204.26217 &       -29.847377 & 53 $\pm$ 7.6 & 48 $\pm$ 7.5 & 
0.79 $\pm$ 0.12 & 1.75 $\pm$ 0.73&0.80 $\pm$ 0.78&0.37 $\pm$ 0.30&175 $\pm$ 6.1
\\
      63 &        204.30113 &       -29.838620 & 47 $\pm$ 4.4 & 28 $\pm$ 2.3 & 
0.514 $\pm$ 0.052 & 2.46 $\pm$ 0.39&2.09 $\pm$ 0.44&1.18 $\pm$ 0.28&39 $\pm$ 2.0
\\
      64 &        204.21281 &       -29.835625 & 54 $\pm$ 5.1 & 33 $\pm$ 3.1 & 
0.585 $\pm$ 0.058 & 1.80 $\pm$ 0.29&1.07 $\pm$ 0.40&0.60 $\pm$ 0.23&61 $\pm$ 3.2
\\
      65 &        204.23465 &       -29.825908 & 35 $\pm$ 4.9 & 20 $\pm$ 4.4 & 
0.369 $\pm$ 0.064 & 1.31 $\pm$ 0.33&0.89 $\pm$ 0.48&0.27 $\pm$ 0.24&--\\
      66 &        204.27451 &       -29.895053 & 46 $\pm$ 9.2 & 22 $\pm$ 4.1 & 
0.54 $\pm$ 0.11 & 1.32 $\pm$ 0.42&0.84 $\pm$ 0.54&--&49 $\pm$ 2.0\\
      67 &        204.20733 &       -29.869728 & 40 $\pm$ 6.3 & 32 $\pm$ 4.9 & 
0.651 $\pm$ 0.087 & 1.57 $\pm$ 0.39&1.36 $\pm$ 0.50&0.73 $\pm$ 0.29&60 $\pm$ 4.2
\\
      68 &        204.28385 &       -29.911307 & 22 $\pm$ 2.7 & 17 $\pm$ 1.7 & 
0.296 $\pm$ 0.026 & 1.05 $\pm$ 0.17&0.84 $\pm$ 0.27&0.43 $\pm$ 0.19&11 $\pm$ 1.8
\\
      69 &        204.27882 &       -29.887547 & 72 $\pm$ 10. & 28 $\pm$ 6.5 & 
0.441 $\pm$ 0.069 & 2.16 $\pm$ 0.44&1.62 $\pm$ 0.66&0.61 $\pm$ 0.32&
116 $\pm$ 2.9\\
      70 &        204.29993 &       -29.850675 & 38 $\pm$ 4.3 & 20 $\pm$ 2.5 & 
0.399 $\pm$ 0.050 & 1.80 $\pm$ 0.36&2.00 $\pm$ 0.50&1.05 $\pm$ 0.27&44 $\pm$ 2.4
\\
      71 &        204.21230 &       -29.881552 & 64 $\pm$ 7.5 & 48 $\pm$ 7.0 & 
0.699 $\pm$ 0.095 & 2.29 $\pm$ 0.44&1.74 $\pm$ 0.71&0.78 $\pm$ 0.32&42 $\pm$ 3.5
\\
      72 &        204.22276 &       -29.853089 & 57 $\pm$ 6.5 & 65 $\pm$ 9.1 & 
0.750 $\pm$ 0.099 & 2.44 $\pm$ 0.54&1.06 $\pm$ 0.63&0.43 $\pm$ 0.31&15 $\pm$ 3.2
\\
      73 &        204.21339 &       -29.841575 & 61 $\pm$ 5.2 & 63 $\pm$ 5.2 & 
0.685 $\pm$ 0.065 & 2.01 $\pm$ 0.33&1.63 $\pm$ 0.51&0.65 $\pm$ 0.23&74 $\pm$ 3.4
\\
      74 &        204.29638 &       -29.826215 & 35 $\pm$ 5.1 & 22 $\pm$ 3.0 & 
0.346 $\pm$ 0.050 & 1.78 $\pm$ 0.35&1.39 $\pm$ 0.32&1.18 $\pm$ 0.27&29 $\pm$ 3.7
\\
      75 &        204.24362 &       -29.804870 & 32 $\pm$ 5.3 & 21 $\pm$ 3.1 & 
0.346 $\pm$ 0.051 & 1.34 $\pm$ 0.29&1.16 $\pm$ 0.52&0.57 $\pm$ 0.22&36 $\pm$ 7.0
\\
      76 &        204.28500 &       -29.866274 & 38 $\pm$ 5.7 & 45 $\pm$ 7.2 & 
0.549 $\pm$ 0.075 & 2.46 $\pm$ 0.52&0.87 $\pm$ 0.66&0.34 $\pm$ 0.30&52 $\pm$ 2.5
\\
      77 &        204.30424 &       -29.860591 & 32 $\pm$ 3.4 & 15 $\pm$ 2.0 & 
0.325 $\pm$ 0.038 & 1.80 $\pm$ 0.30&1.59 $\pm$ 0.45&0.77 $\pm$ 0.25&22 $\pm$ 1.4
\\
      78 &        204.28876 &       -29.852195 & 32 $\pm$ 5.0 & 23 $\pm$ 5.8 & 
0.503 $\pm$ 0.070 & 1.52 $\pm$ 0.43&0.74 $\pm$ 0.59&0.63 $\pm$ 0.30&68 $\pm$ 3.5
\\
      79 &        204.28907 &       -29.846556 & 29 $\pm$ 4.3 & 19 $\pm$ 5.8 & 
0.306 $\pm$ 0.053 & 0.74 $\pm$ 0.39&0.97 $\pm$ 0.66&0.60 $\pm$ 0.30&87 $\pm$ 3.7
\\
      80 &        204.19144 &       -29.880882 & 23 $\pm$ 3.6 & 14 $\pm$ 1.9 & 
0.221 $\pm$ 0.032 & 0.94 $\pm$ 0.29&0.77 $\pm$ 0.46&0.32 $\pm$ 0.22&12 $\pm$ 3.3
\\
      81 &        204.32204 &       -29.864861 & 24 $\pm$ 2.1 & 12 $\pm$ 1.1 & 
0.200 $\pm$ 0.018 & 1.19 $\pm$ 0.20&0.97 $\pm$ 0.25&0.66 $\pm$ 0.14&19 $\pm$ 1.2
\\
      82 &        204.20639 &       -29.859928 & 27 $\pm$ 6.2 & 26 $\pm$ 4.2 & 
0.340 $\pm$ 0.064 & 1.05 $\pm$ 0.39&1.20 $\pm$ 0.64&0.85 $\pm$ 0.29&34 $\pm$ 3.8
\\
      83 &        204.19682 &       -29.818118 & 26 $\pm$ 1.7 & 16 $\pm$ 1.1 & 
0.225 $\pm$ 0.017 & 1.00 $\pm$ 0.13&0.70 $\pm$ 0.19&0.43 $\pm$ 0.12&
36 $\pm$ 0.54\\
      84 &        204.29686 &       -29.926976 & 31 $\pm$ 3.4 & 24 $\pm$ 1.9 & 
0.170 $\pm$ 0.017 & 0.62 $\pm$ 0.17&0.44 $\pm$ 0.23&0.18 $\pm$ 0.13&39 $\pm$ 1.4
\\
      85 &        204.29044 &       -29.908584 & 22 $\pm$ 3.2 & 14 $\pm$ 1.6 & 
0.133 $\pm$ 0.023 & 0.71 $\pm$ 0.16&0.82 $\pm$ 0.29&0.47 $\pm$ 0.20&
7.5 $\pm$ 1.3\\
      86 &        204.24101 &       -29.839948 & 37 $\pm$ 6.2 & 13 $\pm$ 2.8 & 
0.239 $\pm$ 0.055 & 1.45 $\pm$ 0.49&0.90 $\pm$ 0.85&0.36 $\pm$ 0.34&
167 $\pm$ 2.6\\
      87 &        204.27353 &       -29.917803 & 44 $\pm$ 4.0 & 31 $\pm$ 2.7 & 
0.429 $\pm$ 0.043 & 1.58 $\pm$ 0.25&1.05 $\pm$ 0.31&0.44 $\pm$ 0.18&
190 $\pm$ 2.6\\
      88 &        204.23122 &       -29.914030 & 18 $\pm$ 2.2 & 7.6 $\pm$ 1.0 & 
0.178 $\pm$ 0.025 & 0.84 $\pm$ 0.22&0.35 $\pm$ 0.25&--&7.30 $\pm$ 0.75\\
      89 &        204.23179 &       -29.802769 & 21 $\pm$ 4.7 & 7.1 $\pm$ 1.9 & 
0.123 $\pm$ 0.051 & 0.58 $\pm$ 0.28&0.64 $\pm$ 0.39&0.36 $\pm$ 0.21&--\\
      90 &        204.20338 &       -29.873538 & 30 $\pm$ 5.7 & 74 $\pm$ 7.4 & 
0.380 $\pm$ 0.078 & 1.74 $\pm$ 0.39&1.15 $\pm$ 0.53&0.66 $\pm$ 0.30&25 $\pm$ 3.7
\\

\end{longtable}
\end{center}

\begin{center}
\small
\begin{longtable}{cccccccc}
\caption{LB measurement SED fitting parameters}\\
\label{LB_sed_table}\\

\hline
 & $\chi_{\text{UV}}$ & $\chi_{\text {col}}$ & T$_{\text {dust}}$ & M$_{\text {dust}}$ & F$_{24}$ & L$_{\text {dust}}$ & $\chi^2_{\text{FIT}}$ \\
 &                   &                     &    [K]   &       [$10^4~M_\odot$]  &     &   [$10^{40}$erg/s]  &   \\     
\hline
 \endfirsthead
\multicolumn{3}{c}%
{\tablename\ \thetable{} -- continued from previous page}\\
\hline
& $\chi_{\text{UV}}$ & $\chi_{\text {col}}$ & T$_{\text {dust}}$ & M$_{\text {dust}}$ & F$_{24}$ & L$_{\text {dust}}$ & $\chi^2_{\text{FIT}}$ \\
 &                   &                     &    [K]   &       [$10^4~M_\odot$]  &     &   [$10^{40}$erg/s]  &   \\    
\hline
\endhead

       1 & 0.7$\pm$0.3 & 4$\pm$2 & 19.8$\pm$0.7 & 40$\pm$20 & 0.8$\pm$0.2 & 67
$\pm$2 &   1.3\\
       2 & 0.5$\pm$0.3 & 4$\pm$3 & 18.1$\pm$1.0 & 30$\pm$10 & 0.9$\pm$0.1 & 36
$\pm$1&   1.7\\
       3 & 3$\pm$2 & 2$\pm$1& 22.3$\pm$0.2 & 20$\pm$5 & 0.8$\pm$0.1 & 65$\pm$
1&   1.9\\
       4 & 2$\pm$1& 4$\pm$2 & 22.1$\pm$0.9 & 30$\pm$10 & 0.7$\pm$0.1 & 49
$\pm$2 &   1.4\\
       5 & 3$\pm$1& 2.$\pm$1& 23.5$\pm$0.7 & 17$\pm$2 & 0.40$\pm$0.08 & 
44.8$\pm$0.5 &   1.5\\
       6 & 6$\pm$4 & 2$\pm$1& 25$\pm$2 & 8$\pm$5 & 0.8$\pm$0.1 & 45$\pm$3 & 
  0.8\\
       7 & 7$\pm$4 & 4$\pm$2 & 28$\pm$2 & 4$\pm$2 & 0.7$\pm$0.2 & 36$\pm$2 & 
  0.4\\
       8 & 0.5$\pm$0.3 & 9$\pm$1& 20.7$\pm$0.9 & 30$\pm$10 & 0.8$\pm$0.2 & 41
$\pm$2 &   2.3\\
       9 & 3$\pm$2 & 2$\pm$1& 22.8$\pm$0.7 & 30$\pm$10 & 0.8$\pm$0.1 & 97
$\pm$2 &   2.3\\
      10 & 1.3$\pm$0.8 & 1$\pm$1 & 18$\pm$1& 70$\pm$30 & 0.7$\pm$0.1 & 59
$\pm$2 &   1.6\\
      11 & 2$\pm$1& 2$\pm$1& 22.1$\pm$0.7 & 15$\pm$2 & 0.6$\pm$0.2 & 32
$\pm$1&   1.5\\
      12 & 2$\pm$1& 3$\pm$1& 22.9$\pm$0.7 & 23$\pm$4 & 0.6$\pm$0.2 & 58
$\pm$3 &   1.1\\
      13 & 4$\pm$2 & 1.2$\pm$0.8 & 23$\pm$1 & 14$\pm$5 & 0.5$\pm$0.1 & 38
$\pm$2 &   1.0\\
      14 & 1.3$\pm$0.7 & 5$\pm$2 & 20.8$\pm$0.4 & 37$\pm$9 & 0.7$\pm$0.1 & 46
$\pm$1&   1.6\\
      15 & 0.5$\pm$0.3 & 1.1$\pm$0.9 & 16.8$\pm$0.9 & 50$\pm$20 & 0.7$\pm$0.1
 & 24$\pm$1&   4.2\\
      16 & 2$\pm$1& 3$\pm$2 & 21$\pm$1& 50$\pm$20 & 0.8$\pm$0.2 & 106$\pm$5
 &   2.3\\
      17 & 1.5$\pm$0.5 & 3$\pm$2 & 20.2$\pm$0.4 & 14$\pm$3 & 0.61$\pm$0.08 & 
15.6$\pm$0.2 &   2.7\\
      18 & 1.2$\pm$0.8 & 3$\pm$2 & 19$\pm$1 & 20$\pm$8 & 0.9$\pm$0.2 & 21.6
$\pm$0.9 &   2.2\\
      19 & 1.2$\pm$0.9 & 3$\pm$2 & 19$\pm$2 & 30$\pm$10 & 0.8$\pm$0.2 & 23$\pm$
1&   2.8\\
      20 & 3$\pm$2 & 2$\pm$1& 22.8$\pm$0.7 & 12$\pm$5 & 0.4$\pm$0.2 & 25$\pm$2
 &   1.5\\
      21 & 0.6$\pm$0.3 & 5$\pm$2 & 19.5$\pm$0.6 & 40$\pm$20 & 0.5$\pm$0.1 & 30
$\pm$1&   0.4\\
      22 & 2$\pm$1& 3$\pm$2 & 21.3$\pm$0.8 & 23$\pm$8 & 0.5$\pm$0.2 & 30$\pm$
1&   1.1\\
      23 & 3$\pm$2 & 2$\pm$2 & 22$\pm$1& 8$\pm$5 & 0.5$\pm$0.3 & 12.7$\pm$0.9
 &   1.0\\
      24 & 3$\pm$2 & 1.2$\pm$0.9 & 21.5$\pm$0.9 & 11$\pm$4 & 0.2$\pm$0.2 & 15
$\pm$1&   0.2\\
      25 & 2$\pm$1& 2$\pm$1& 22.1$\pm$0.7 & 32$\pm$4 & 0.45$\pm$0.08 & 
58.1$\pm$0.7 &   1.2\\
      26 & 2$\pm$2 & 1$\pm$1 & 20$\pm$3 & 20$\pm$10 & 0.9$\pm$0.2 & 30$\pm$2
 &   3.1\\
      27 & 4$\pm$3 & 3$\pm$2 & 24$\pm$2 & 6$\pm$3 & 0.6$\pm$0.2 & 16$\pm$1& 
  2.4\\
      28 & 5$\pm$4 & 2$\pm$1& 24$\pm$2 & 4$\pm$3 & 0.3$\pm$0.3 & 10$\pm$1& 
  0.4\\
      29 & 1.3$\pm$0.7 & 4$\pm$2 & 20.2$\pm$0.3 & 11$\pm$3 & 0.6$\pm$0.1 & 11.2
$\pm$0.6 &   0.6\\
      30 & 2$\pm$1& 4$\pm$3 & 20$\pm$2 & 11$\pm$8 & 0.9$\pm$0.2 & 10.0$\pm$0.9
 &   0.6\\
      31 & 3$\pm$2 & 0.3$\pm$0.2 & 21$\pm$1& 15$\pm$6 & 0.5$\pm$0.2 & 27$\pm$
1&   1.7\\
      32 & 0.6$\pm$0.4 & 7$\pm$3 & 20.6$\pm$0.9 & 15$\pm$6 & 0.5$\pm$0.1 & 13.4
$\pm$0.8 &   4.7\\
      33 & 2$\pm$1& 3$\pm$2 & 21$\pm$1& 6$\pm$4 & 0.6$\pm$0.2 & 7.2$\pm$0.7
 &   0.5\\
      34 & 0.4$\pm$0.2 & 6$\pm$2 & 18.7$\pm$0.2 & 40$\pm$10 & 0.4$\pm$0.1 & 
20.1$\pm$0.9 &   1.2\\
      35 & 0.15$\pm$0.05 & 9$\pm$2 & 16.5$\pm$0.2 & 80$\pm$20 & 0.7$\pm$0.1 & 
18.9$\pm$0.9 &   4.2\\
      36 & 2$\pm$1& 2$\pm$1& 22.1$\pm$0.7 & 13$\pm$2 & 0.45$\pm$0.08 & 
23.7$\pm$0.3 &   0.7\\
      37 & 1.2$\pm$0.9 & 4$\pm$3 & 20$\pm$1& 19$\pm$9 & 0.7$\pm$0.2 & 15$\pm$
1&   1.2\\
      38 & 1.3$\pm$0.8 & 4$\pm$2 & 20.6$\pm$0.6 & 40$\pm$10 & 0.57$\pm$0.1 & 38
$\pm$1&   0.3\\
      39 & 1.2$\pm$0.9 & 5$\pm$3 & 20$\pm$1& 20$\pm$10 & 0.5$\pm$0.2 & 15$\pm$
1&   1.2\\
      40 & 0.55$\pm$0.1 & 3.$\pm$1& 18.8$\pm$0.3 & 27$\pm$5 & 0.3$\pm$0.1 & 
14.1$\pm$0.2 &   0.3\\
      41 & 2$\pm$1& 2$\pm$1& 20.3$\pm$0.8 & 12$\pm$5 & 0.4$\pm$0.1 & 11.7
$\pm$0.9 &   1.3\\
      42 & 0.4$\pm$0.3 & 6$\pm$3 & 18.7$\pm$0.6 & 30$\pm$10 & 0.6$\pm$0.2 & 15.2
$\pm$0.9 &   0.6\\
      43 & 2$\pm$1& 4$\pm$2 & 22.1$\pm$0.9 & 20$\pm$8 & 0.5$\pm$0.1 & 32$\pm$
2 &   1.1\\
      44 & 0.20$\pm$0.1 & 8$\pm$3 & 17.1$\pm$0.9 & 50$\pm$20 & 0.7$\pm$0.2 & 16
$\pm$1&   2.8\\
      45 & 8$\pm$2 & 1.3$\pm$0.7 & 26.4$\pm$0.8 & 14$\pm$4 & 0.4$\pm$0.2 & 83
$\pm$4 &   2.7\\
      46 & 0.5$\pm$0.3 & 5$\pm$2 & 19.0$\pm$0.6 & 40$\pm$10 & 0.5$\pm$0.2 & 22
$\pm$1&   0.3\\
      47 & 1.2$\pm$0.8 & 5$\pm$2 & 20.7$\pm$0.7 & 30$\pm$10 & 0.5$\pm$0.2 & 28
$\pm$2 &   0.6\\
      48 & 2$\pm$1& 1.2$\pm$0.8 & 20.8$\pm$0.3 & 12$\pm$2 & 0.02$\pm$0.02 & 
14.2$\pm$0.7 &   7.5\\
      49 & 0.4$\pm$0.2 & 4$\pm$2 & 17.8$\pm$0.1 & 40$\pm$10 & 0.4$\pm$0.1 & 
15.8$\pm$0.9 &   0.3\\
      50 & 3$\pm$2 & 0.4$\pm$0.3 & 21.3$\pm$1.0 & 20$\pm$8 & 0.09$\pm$0.08 & 26
$\pm$1&   2.0\\
      51 & 1.4$\pm$0.6 & 3$\pm$1& 20.2$\pm$0.3 & 20$\pm$5 & 0.33$\pm$0.1 & 
16.8$\pm$0.5 &   1.1\\
      52 & 2$\pm$1& 1.4$\pm$0.6 & 21.0$\pm$0.1 & 15$\pm$2 & 0.40$\pm$0.08 & 
21.3$\pm$0.4 &   1.7\\
      53 & 1.4$\pm$0.6 & 2$\pm$1& 19.7$\pm$0.2 & 10$\pm$2 & 0.4$\pm$0.2 & 7.6
$\pm$0.3 &   2.4\\
      54 & 2$\pm$1& 3$\pm$2 & 20$\pm$2 & 10$\pm$10 & 0.6$\pm$0.2 & 13$\pm$2 & 
  0.2\\
      55 & 6$\pm$4 & 2$\pm$1& 25$\pm$1& 6$\pm$3 & 0.2$\pm$0.2 & 16$\pm$2 & 
  1.0\\
      56 & 3$\pm$2 & 3$\pm$2 & 23.8$\pm$0.9 & 6$\pm$3 & 0.3$\pm$0.2 & 14$\pm$1.
 &   0.1\\
      57 & 0.5$\pm$0.4 & 7$\pm$4 & 19$\pm$1& 14$\pm$8 & 0.7$\pm$0.2 & 6.6$\pm$
0.8 &   1.0\\
      58 & 1.4$\pm$0.6 & 3.$\pm$1& 20.2$\pm$0.3 & 13$\pm$3 & 0.09$\pm$0.08 & 
10.5$\pm$0.3 &   0.9\\
      59 & 2$\pm$1& 3$\pm$2 & 21.3$\pm$0.8 & 13$\pm$6 & 0.08$\pm$0.07 & 14
$\pm$1&   2.3\\
      60 & 1.2$\pm$0.9 & 3$\pm$2 & 18.9$\pm$1.0 & 20$\pm$10 & 0.5$\pm$0.3 & 13
$\pm$1&   0.6\\
      61 & 1.2$\pm$0.9 & 6$\pm$3 & 20$\pm$1& 20$\pm$10 & 0.7$\pm$0.2 & 23$\pm$
2 &   0.5\\
      62 & 2$\pm$2 & 3$\pm$2 & 22$\pm$2 & 12$\pm$7 & 0.5$\pm$0.2 & 17$\pm$2 & 
  0.1\\
      63 & 0.3$\pm$0.1 & 5$\pm$1& 18.1$\pm$0.5 & 47$\pm$6 & 0.23$\pm$0.10
 & 17.1$\pm$0.3 &   1.2\\
      64 & 1.4$\pm$0.6 & 3$\pm$2 & 20.1$\pm$0.5 & 20$\pm$5 & 0.26$\pm$0.1 & 15.7
$\pm$0.8 &   0.9\\
      65 & 1.3$\pm$0.8 & 3$\pm$2 & 19.8$\pm$0.8 & 15$\pm$6 & 0.3$\pm$0.3 & 10
$\pm$1&   0.6\\
      66 & 3$\pm$2 & 3$\pm$2 & 22.8$\pm$0.7 & 6$\pm$2 & 0.10$\pm$0.09 & 11.3
$\pm$0.9 &   0.5\\
      67 & 1.2$\pm$0.8 & 6$\pm$3 & 20.5$\pm$0.9 & 17$\pm$7 & 0.4$\pm$0.3 & 15
$\pm$1&   1.5\\
      68 & 0.5$\pm$0.3 & 5$\pm$2 & 19.1$\pm$0.6 & 16$\pm$6 & 0.4$\pm$0.2 & 8.2
$\pm$0.6 &   0.9\\
      69 & 1.3$\pm$0.8 & 2$\pm$1& 18.6$\pm$0.6 & 30$\pm$10 & 0.11$\pm$0.1 & 
16$\pm$1&   1.2\\
      70 & 0.3$\pm$0.2 & 6$\pm$3 & 17.9$\pm$0.5 & 40$\pm$20 & 0.15$\pm$0.1 & 
13.7$\pm$0.9 &   3.1\\
      71 & 1.2$\pm$0.8 & 3$\pm$2 & 19.2$\pm$0.8 & 30$\pm$10 & 0.4$\pm$0.2 & 19
$\pm$1&   0.5\\
      72 & 1.2$\pm$0.8 & 3$\pm$2 & 19.3$\pm$0.9 & 30$\pm$10 & 0.6$\pm$0.2 & 20
$\pm$1&   1.3\\
      73 & 0.7$\pm$0.3 & 2.$\pm$1& 18.6$\pm$0.7 & 30$\pm$10 & 0.6$\pm$0.2 & 
19.4$\pm$0.9 &   0.7\\
      74 & 0.3$\pm$0.2 & 6$\pm$3 & 17.6$\pm$0.5 & 50$\pm$10 & 0.3$\pm$0.2 & 12.6
$\pm$0.8 &   2.7\\
      75 & 0.6$\pm$0.4 & 4$\pm$2 & 19.0$\pm$0.9 & 20$\pm$10 & 0.3$\pm$0.2 & 10.5
$\pm$0.8 &   0.7\\
      76 & 0.5$\pm$0.3 & 5$\pm$3 & 18.6$\pm$1.0 & 30$\pm$10 & 0.7$\pm$0.2 & 15
$\pm$1&   2.7\\
      77 & 0.3$\pm$0.2 & 6$\pm$2 & 18.1$\pm$0.4 & 40$\pm$10 & 0.07$\pm$0.06 & 
11.4$\pm$0.6 &   1.5\\
      78 & 1.2$\pm$0.9 & 6$\pm$3 & 20$\pm$1& 17$\pm$8 & 0.4$\pm$0.4 & 12$\pm$
1&   0.7\\
      79 & 2$\pm$1& 3$\pm$2 & 20$\pm$2 & 10$\pm$10 & 0.4$\pm$0.4 & 8$\pm$1& 
  1.9\\
      80 & 1.2$\pm$0.9 & 3$\pm$2 & 19$\pm$1 & 14$\pm$7 & 0.3$\pm$0.2 & 7.1
$\pm$0.8 &   0.1\\
      81 & 0.3$\pm$0.2 & 5$\pm$2 & 17.4$\pm$0.3 & 30$\pm$10 & 0.10$\pm$0.09 & 
8.0$\pm$0.5 &   2.2\\
      82 & 0.4$\pm$0.3 & 6$\pm$4 & 17$\pm$2 & 40$\pm$30 & 0.6$\pm$0.3 & 10$\pm$
1&   2.1\\
      83 & 0.7$\pm$0.3 & 3$\pm$2 & 18.8$\pm$0.4 & 14$\pm$4 & 0.27$\pm$0.10 & 7.6
$\pm$0.4 &   1.2\\
      84 & 0.6$\pm$0.4 & 0.2$\pm$0.1 & 17$\pm$1 & 19$\pm$9 & 0.5$\pm$0.2 & 
6.8$\pm$0.3 &   2.6\\
      85 & 0.4$\pm$0.3 & 2$\pm$1& 16.5$\pm$0.8 & 30$\pm$10 & 0.4$\pm$0.2 & 5.7
$\pm$0.5 &   1.1\\
      86 & 1.2$\pm$0.9 & 2$\pm$2 & 18.7$\pm$0.8 & 20$\pm$10 & 0.07$\pm$0.06 & 9
$\pm$1&   1.3\\
      87 & 1.3$\pm$0.7 & 2$\pm$1& 19.4$\pm$0.5 & 19$\pm$6 & 0.36$\pm$0.1 & 
12.8$\pm$0.4 &   0.3\\
      88 & 1.2$\pm$0.8 & 4$\pm$2 & 19.8$\pm$0.8 & 8$\pm$3 & 0.07$\pm$0.06 & 5.1
$\pm$0.4 &   1.8\\
      89 & 1.1$\pm$0.9 & 3$\pm$2 & 18$\pm$2 & 20$\pm$10 & 0.10$\pm$0.09 & 4.6
$\pm$0.7 &   1.4\\
      90 & 0.2$\pm$0.1 & 4$\pm$3 & 14.9$\pm$0.7 & 70$\pm$30 & 0.9$\pm$0.2 & 15
$\pm$1&   7.4\\

\end{longtable}
\end{center}

\end{document}